\documentclass[aps,10pt,prl,twocolumn,showpacs,groupedaddress]{revtex4}

\usepackage{graphicx}  
\usepackage{dcolumn}   
\usepackage{bm}        
\usepackage{amssymb}   
\usepackage{amsmath}

\usepackage{bm}
\DeclareBoldMathCommand{\bfmu}{\mu}

\begin{document}

\title{Material Science for Quantum Computing with Atom Chips}
\author{Ron Folman}
	\email{folman@bgu.ac.il}

	\affiliation{Department of Physics, Ben-Gurion University of the Negev, Be'er Sheva 84105, Israel}
\date{\today}

\begin{abstract}
In its most general form, the atom chip is a device in which neutral or charged particles are positioned in an isolating environment such as vacuum (or even a carbon solid state lattice) near the chip surface. The chip may then be used to interact in a highly controlled manner with the quantum state. I outline the importance of material science to quantum computing (QC) with atom chips, where the latter may be utilized for many, if not all, suggested implementations of QC. Material science is important both for enhancing the control coupling to the quantum system for preparation and manipulation as well as measurement, and for suppressing the uncontrolled coupling giving rise to low fidelity through static and dynamic effects such as potential corrugations and noise. As a case study, atom chips for neutral ground state atoms are analyzed and it is shown that nanofabricated wires will allow for more than $10^4$ gate operations when considering spin-flips and decoherence. The effects of fabrication imperfections and the Casimir-Polder force are also analyzed. In addition, alternative approaches to current-carrying wires are briefly described. Finally, an outlook of what materials and geometries may be required is presented, as well as an outline of directions for further study.
\end{abstract}

\maketitle

Whatever the eventual physical realization is, quantum computing (QC) will most probably require fabrication technology, analogous but not identical to what has been done for classical computers. This is so also for the neutral particles described in this special issue.

Scalability will require highly repeatable architectures. Fidelity will require extremely low levels of DC fluctuations (e.g. electron or photon scattering due to imperfections such as surface roughness of wires and waveguides) and AC fluctuations in the form of electro-magnetic noise (the eventual numbers would be determined by the allowed infidelity or error rate, e.g. \cite{error1,error2}). The variety of essential technologies for the different operations of a quantum computer such as isolation through vacuum, trapping, cooling, transporting, state preparation, manipulation, measurement and electronic readout, will require the accurate and complex integration capabilities which only monolithic fabrication methods exhibit. More so, it is plausible that the eventual realization will necessitate hybrid systems in which atoms serve, for example, as memory while superconducting circuits make up the logic gates. This integration between different quantum systems will most probably again require chip technology.

We may then talk of an atom chip for QC hosting any eventual desired particle and interaction, and where the atom chip may be very different from its form today ~\cite{RonRev, ReichelRev, fortagh}. Indeed, already now diverse fabrication techniques are advancing at a rate which allows us to assume that any eventual scheme for the quantum computer may be put on an atom chip. This may include on one extreme cold atoms in an optical lattice within a miniature vacuum chamber inside a substrate, and on the other extreme, solid-state atoms or atom-like systems (e.g. NV centers in diamond nano crystals) embedded in a mesh of guides for photons and electrons fabricated on a surface. Our definition of the atom chip is therefore broadened to include all quantum objects close to a "classical" surface. This somewhat arbitrary division does not include, for example, quantum dots and superconducting qubits which are an inherent part of the surface. The main focus of this paper will be on an atom chip having a room temperature surface.

When discussing a quantum system close to a macroscopic body one should think of a classical environment which is continuously attempting to lower the fidelity of the quantum operations by all means at its disposal. Hence, while the controlled coupling between this environment and our quantum system should be sufficient to enable necessary operations, we must suppress the uncontrolled coupling. The eventual figure of merit would then be the ratio between the controlled and the uncontrolled coupling, or more simply put, how many logic gate operations may be done with high fidelity. Due to the necessity of chip technology, as explained above, I believe that material science and technology will have just as much effect on this ratio as will the actual physical nature of the chosen quantum system and the experimental tools used to interact with it.

As the nature of the quantum system used changes, so does the nature of the coupling with the chip. For example, charged particles have exhibited a very strong dependence of the heating rate on the distance to the chip \cite{baruch, dubessy, ion_heating} (most probably caused by contaminations or inhomogeneities on the surface \cite{chuang}). Polar molecules \cite{Meijer} and Rydberg atoms \cite{Ryd_Spreeuw,Ryd_Pfau,Ryd_Scheel, Ryd_Martin, Ryd_Carsten}, while exhibiting advantageous features concerning logic gates due to strong coupling, will also have enhanced atom-surface interaction relative to neutral atoms, as their permanent or induced dipole is much stronger.

In this paper we shall focus on ground state neutral atoms as a case study which should be followed for any system eventually used. The current-carrying wire will be used as a specific example of the problem of controlled vs. uncontrolled coupling. A brief discussion of an alternative atom chip in the form of an optical chip ends this paper.

\section{Advances in fabrication}

Before discussing as a case study the fundamental limitations of the current-carrying wire, let us briefly note the state of the art in fabrication.

\begin{figure}[b]%
\includegraphics[width=\columnwidth]{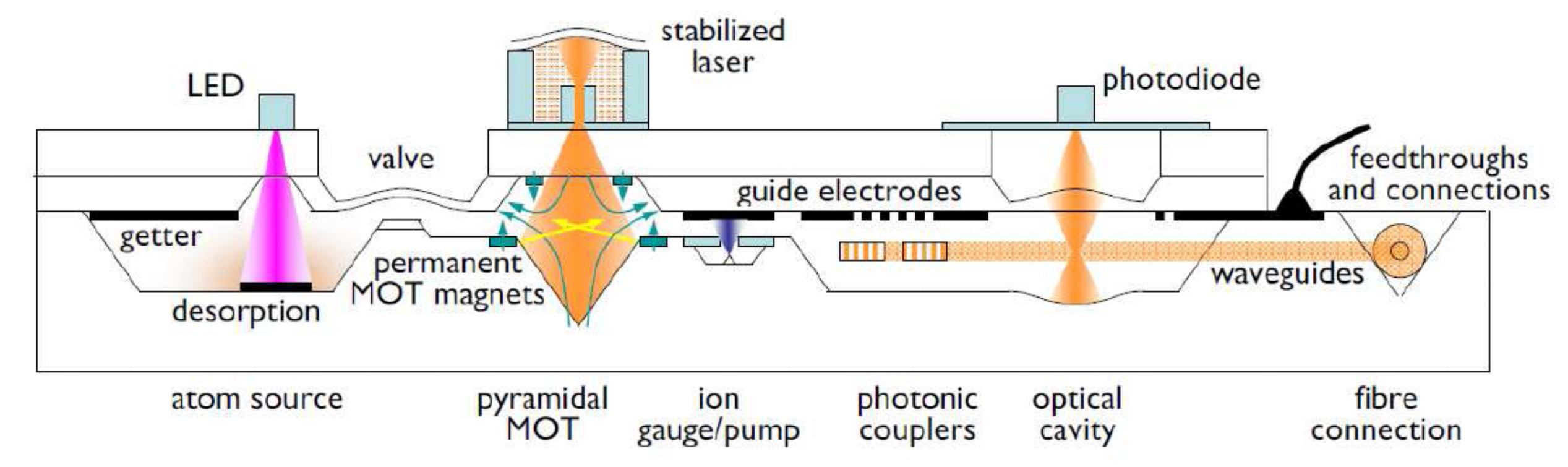}%
\caption{(Color Online)
A futuristic vision of the atom chip (courtesy of Tim Freegarde). Here all the concepts of miniature vacuum, miniature light and particle sources, photonics and micro magnetic traps, all come as an integrated device on top of one substrate. Eventually, also readout and pre-amplification electronics could be put on the same chip. Machines such as wafer bonders enable such structures. Taken from \cite{chapter}. Copyright Wiley-VCH Verlag GmbH $\&$ Co. KGaA. Reproduced with permission.
}%
\label{freegarde}%
\end{figure}

Since its conception, the vision of the atom chip has been to create a platform for any required atomic and molecular optics tool. In Fig. \ref{freegarde}, a general view of such a vision is presented.

Advances in this direction have recently been made. For example, in the group of Dana Anderson, miniature vacuum chambers inside substrates are already functioning, and are poised to enable one of the best, if not the best, optical resolution to date; in the group of Ed Hinds an integrated pyramid magneto optical trap has been etched into the substrate; in the group of Robert Spreeuw, permanent magnet lattices have been fabricated; and in the group of Philipp Treutlein, multi-layer chips, including micro-wave guides, have been developed.

The vision for a quantum computer would be to create a dense lattice of traps which may be individually addressed and between which controlled coupling may be initiated. Such a lattice may have an optical periodicity if optical far fields are used, or sub-optical periodicity if near fields are used, or field sources such as current-carrying wires or permanent magnets, where the size limitation comes from e-beam lithography resolution, which today stands at a few tens of nano-meters. A procedure to optimize the chip design for such lattices has been defined \cite{roman}. Another limitation comes from the smallest sustainable particle-surface distance (e.g. due to van der Waals forces) and this will be discussed to some extent in the following. Near every lattice site there will be another field source, such as a charged dot, or optical sub wavelength aperture (including plasmonics for enhanced transmission and focus), in order to facilitate single qubit rotations with single site addressability.

\begin{figure}[b]%
\includegraphics[width=\columnwidth]{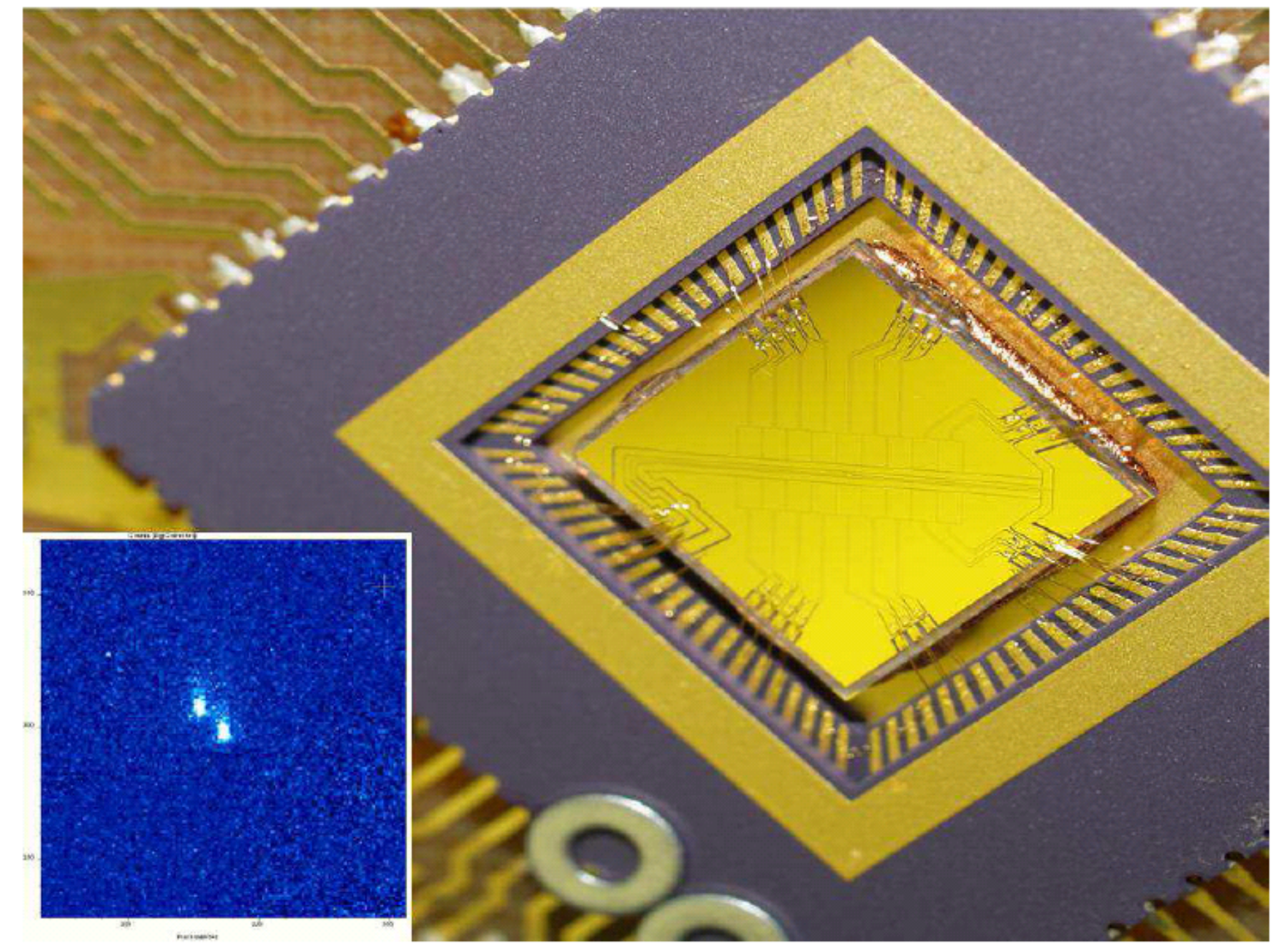}%
\caption{(Color Online)
IonChip produced in Ben-Gurion University before it is put into the vacuum chamber in Mainz. Inset: first two ions trapped on the chip. Courtesy of Ferdinand Schmidt-Kaler.
}%
\label{ionchip}%
\end{figure}

To facilitate rapid advances in the field of material science for atom chips, and with the aim of providing research and development services to the quantum optics community, we have constructed a fabrication facility at Ben-Gurion University with emphasis on quantum optics. This facility with a resolution of 10 nm and with the ability to integrate numerous materials and processes, has already given services to numerous laboratories in Europe and the United-States. Projects include current-carrying chips, permanent magnet chips, ion chips, chips for cold electrons, chips for diamond NV centers, photonics chips and so on. In Fig. \ref{ionchip}, an ion chip developed for the Mainz group of Ferdinand Schmidt-Kaler is presented as an example.

To conclude, these and other examples from the atom chip community \cite{chapter}, as well as overwhelming advances in fabrication techniques in general, allow us the optimistic point of view that any eventual realization of a quantum computer may be put in the future on an atom chip. The main challenge will be to maximize the ratio of controlled to uncontrolled coupling, and in the following the current-carrying wire is analyzed as a case study.

\section{Atom-surface distance}

For any system eventually used, it seems reasonable to assume that small system-environment distances will eventually be required. This is due to physical needs such as resolution and density of traps, as well as technical needs such as low power consumption.

In the neutral atom example, decreasing the atom-surface distance should increase trap gradients sufficiently to construct tunneling barriers with widths on the order of the atomic de Broglie wavelength. In the context of QC, this will allow for higher fidelity gates when they are potential dependent. The effect of small atom-surface distances will be quantified in the following section. High gradients also mean that qubits are better localized and this would be important for any 2-qubit interaction which is sensitive to the distance. Better localization is also important for single qubit rotation, as fixed elements contributing to such a rotation (e.g. charged dots) will be better spatially aligned with the qubit. Such high trap gradients may also allow for improved atom-light interactions such as probing without heating in the Lamb-Dicke regime, or side-band cooling.

At small atom-surface distances, interactions with the nearby surface become important. For example, spatial and temporal magnetic field fluctuations, due to electron scattering and Johnson noise, respectively, determine the minimal atom-surface distance, as they cause potential corrugations, spin-flips, and decoherence. There have been several experiments utilizing cold atoms to study these interactions~\cite{tubingen_frag, Hinds_spinflip, orsay1, MagScanN, simonSc, emmert}, and many suggestions on how to overcome their damaging effects~\cite{Valery, OrsayFragSup, fermani, YoniPRB, TalAni,peter}. Also becoming prominent for small atom-surface distances is the Casimir-Polder~(CP) force \cite{Milton}; being attractive, it reduces the magnetic barrier and allows atoms to tunnel to the surface, as already observed \cite{VuleticCP,cornell}. Several accurate measurements of the van der Waals, Casimir and CP forces between atoms and surfaces have been made \cite{hinds_vdw, vigue}. These topics will be elaborated upon in the following.

As atom-surface proximity grows so will the need for suppressing hindering surface effects by material science and engineering. The final accepted levels of material infidelity would be defined by the final allowed gate infidelity which in turn depends on error correction thresholds.

Finally, let us note that from the above it is evident that achieving small atom-surface distances would not only be advantageous for atom optics and QC, but would also contribute to the fundamental study of atom-surface and surface phenomena. However, the topic of the atom as a probe is beyond the scope of this chapter (see for example \cite{simonSc,YoniPRB,atom-probe,carsten_probe}).

Let me also note that there are numerous ideas aimed at bringing atoms closer to a surface or a macroscopic object~\cite{Ovchinnikov, Schmiedmayer, Shevchenko, rosenblit1, Ricci, Bender, Gillen, Chang2009}, all of which are, however, not based on the current-carrying wire, the latter being the focus of this chapter. Some additional ideas for small atom-surface distances will be noted in the photonics section.

\subsection{Control of Atom-Atom distance}

As mentioned earlier, small atom-surface distances may be advantageous to QC devices because of different reasons such as power consumption. However, as noted, one of the most important features of a small atom-surface distance are the high potential gradients made available ~\cite{RonRev}. These will enable not only high density lattices for atoms - as required by the demand for scalability, and a high degree of alignment - as required for single qubit rotations, but also high fidelity in the control of barriers between atom trap sites, as required for example by the collision gate (see first chapter of this special issue).

As an example of the fidelity awarded by a small atom-surface distance, let us briefly examine the technical stability of a single barrier between two qubit sites. Tunneling may be used as a figure of merit for this stability. As noted previously, stable control of high gradients will most likely be relevant at some level to any QC scheme with isolated particles, and so let us examine in the following the stability of a barrier, not only with the collision gate in mind.

To quantify such tunneling, let us construct a simple potential which has no direct connection to QC. Consider a waveguide formed by a single atom chip wire (in the $\hat{x}-$direction) and an external bias field; current through a second atom chip wire in the $\hat{y}-$direction, namely in a right-angle~``X'' wire configuration~\cite{ReichelIFM}, is added to generate a simple potential barrier. The configuration used here is, incidentally, exactly opposite to the ``dimple'' configuration recently used for compressing atom chip traps~\cite{Horikoshi, Anderson-poster}.

\begin{figure}[ht]
	\includegraphics[width=7cm]{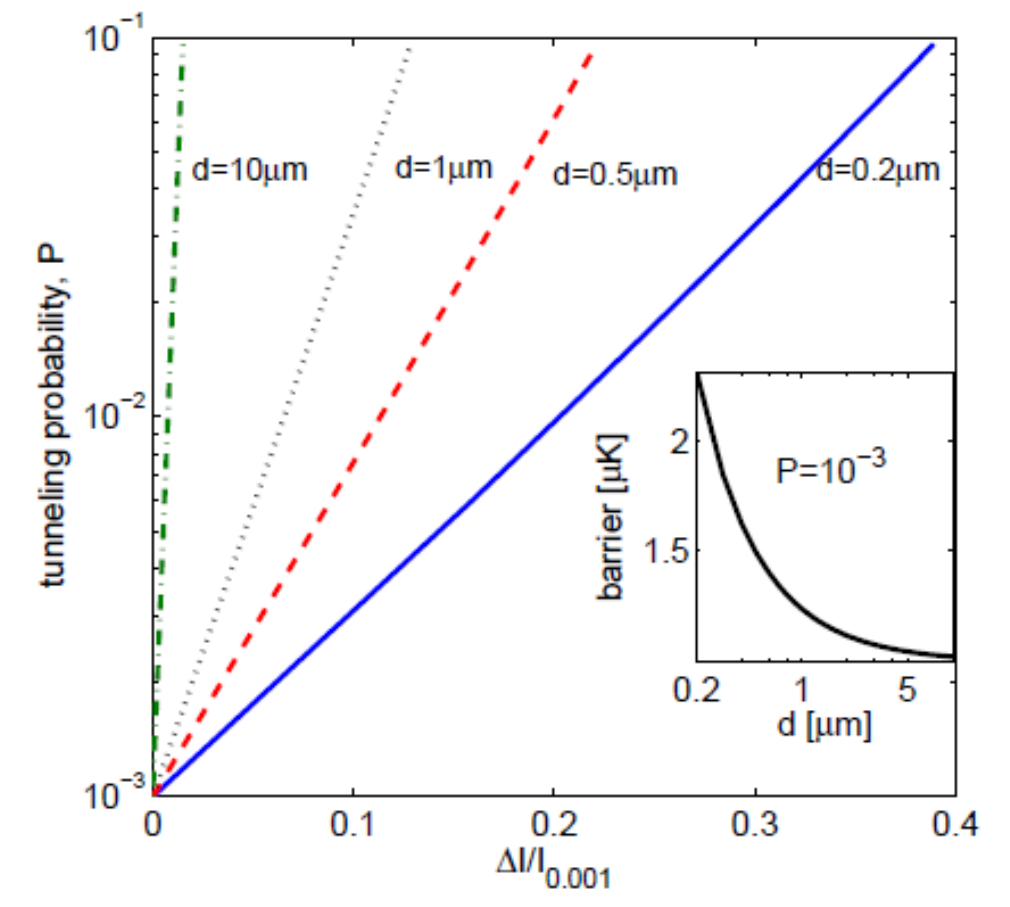}
		\caption{(Color Online) Tunneling probability through a barrier at several heights~$d$ as a function of the change in the current~$\Delta I$ through the control wire, relative to the current~$I_{0.001}$ for a probability~$P(I)=0.001$. A kinetic energy of~$\rm1\,\mu$K is assumed for a single atom of~$\rm^{87}$Rb. For this X-trap model, changing the current by a few percent causes a drastic change in the tunneling probability for~$d=\rm10\,\mu$m. Good control over the tunneling probability requires the trap height to be about~$d\lesssim\rm1-2\,\mu$m. The inset shows the potential barrier required to maintain a tunneling probability of~$0.001$ as a function of the atom-surface separation~$d$: for smaller~$d$, a higher barrier is required so better tunneling control is attained. Taken from \cite{Salem2010} Copyright IOP Publishing Ltd. Reproduced with permission.}	
		\label{fig:barrier}
\end{figure}

For a specific atom-surface distance~$z$, the magnetic potential in the~$\hat{x}-$direction, generated by the crossing wire, is given by:
\begin{equation}
	V(x)=\mu_A B_0+\frac{\mu_A \mu_0 I}{2\pi}\ \frac{z}{z^2+x^2},
	\label{eq:barrier}
\end{equation}
where $\mu_0$ is the permeability constant,~$\mu_A$ is the atomic magnetic dipole moment along the direction of the trap bottom (Ioffe-Pritchard) field~$B_0$, and $I$ is the current in the crossing wire. A 1D single-particle tunneling probability through the barrier can then be calculated in the~WKB approximation as
\begin{equation}
  P=\exp\left(-\frac{1}{\hbar}\int_{-x_E}^{x_E}dx \sqrt{2m[V(x)-E]}\right),
 	\label{eq:Tunnel}
\end{equation}
where~$E$ is the kinetic energy of the atom and~$V(\pm x_E)=E$. Assuming a kinetic energy of~$E=1\,\mu$K for a~$\rm^{87}Rb$ atom (corresponding to a free-particle de Broglie wavelength of~$\approx\rm0.33\,\mu$m) in the~$\left|F=2,m_F=2\right\rangle$ state, we may then easily calculate the change required in the current~$I$ in order to cause a given proportional change in the tunneling probability, as a function of the atom-surface distance~$z$. The results of this calculation are shown in Fig.~\ref{fig:barrier} for changing the tunneling probability from~0.001 to some higher probability. The calculation suggests that control over the tunneling probability requires a distance~$z$ on the order of~$\rm1-2\,\mu$m for experimentally reasonable values of current control. In the simple model of Eq.~(\ref{eq:barrier}), this corresponds to a barrier half-width of~$\rm2-4\,\mu$m, comparable to experiments that have observed interference between adjacent wells with the addition of non-static fields~\cite{JoergIFM,Ketterle_int}. Thus, the desired static magnetic potentials can be generated only if atoms can be brought down to micron or sub-micron distances above the wires on the atom chip surface, at which point the tunneling rate can be tuned over a significant dynamic range by adjusting the current in the crossing wire, or in other words the high gradient potential is much more stable against current noise.

As another brief example of the enhanced trapping potential engineering resolution as the atom-surface distance is decreased, let us examine the creation of a 1D lattice by a "snake"-shaped current-carrying wire.

Shaping the wire edges may be used for creating potential variations desired for manipulating atoms near the atom chip surface. In the following section we shall characterize the dependence of magnetic field variations on wire edge imperfections. Based on this dependence we may now discuss quantitatively the deliberate ``tailoring'' of magnetic trapping potentials by engineering wire edge profiles. Using again tunneling as a figure of merit for the purposes of this brief demonstration, we are particularly interested in potentials with sufficient variation so that tunneling barriers can be controlled. We may now wish to determine the highest ``potential resolution'', namely, the smallest distinguishable distance between adjacent wells separated by static tunneling barriers, as a function of~$d$.

As a test case for quantifying this potential resolution, let us consider a configuration in which a thin wire is curved with a certain periodicity~$\lambda$ that corresponds to a wave-vector $k=2\pi/\lambda$. If the amplitude of this curvature is small with respect to the wavelength, then, as will be explained in the next section, the magnetic field above the wire is given by a single~$|k|$ component in Eq.~(\ref{eq:dBx}), and then~$V(x)=V_0\cos kx$, where $V_0=\mu_A\mu_0 Ik^2\delta y_{\rm center}{\rm K}_1(kz)$ (see Appendix A for details).
\begin{figure}[ht]
	\includegraphics[width=0.45\textwidth]{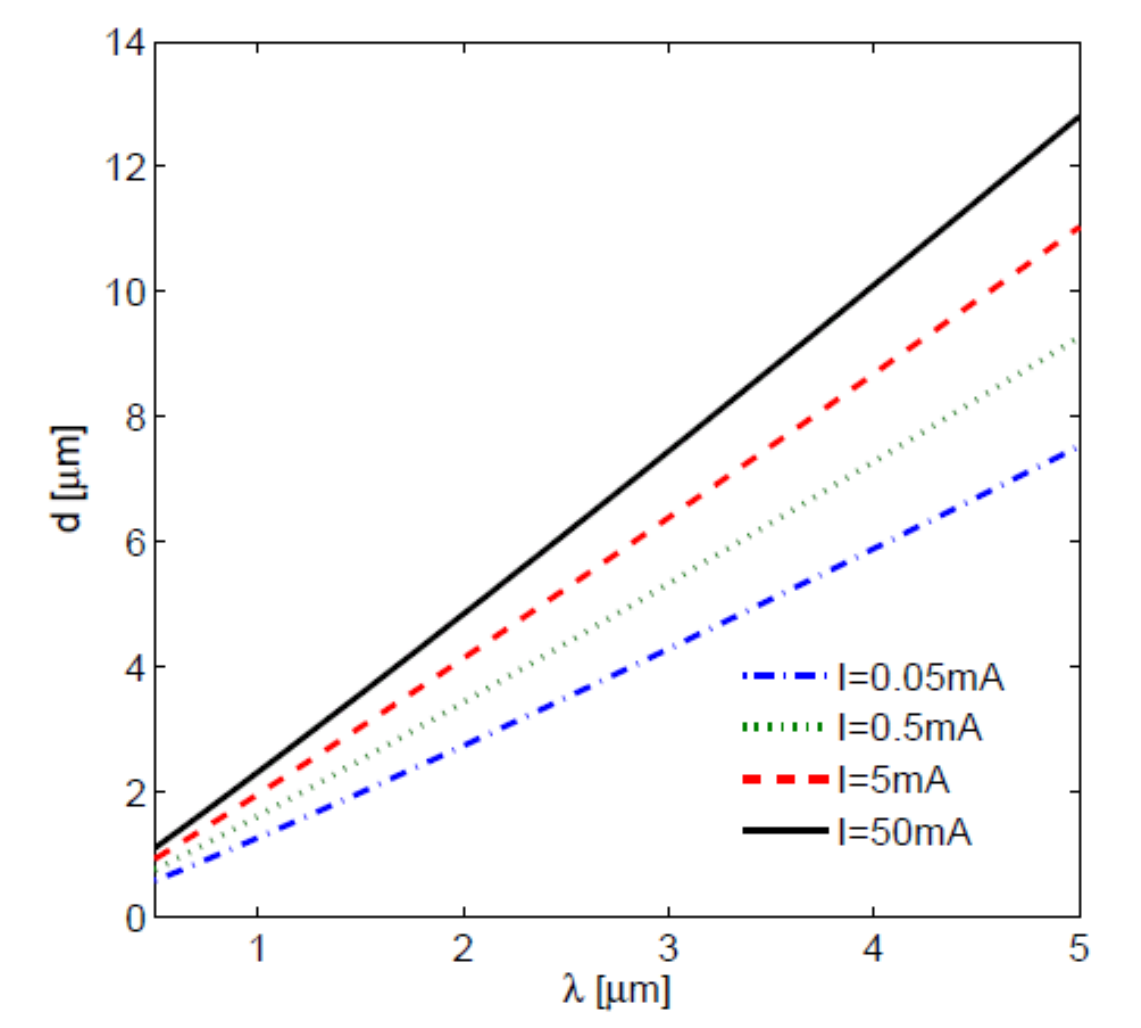}
\caption{(Color Online) Potential spatial resolution achievable with wire currents in the range~$\rm0.05-50$ mA. Presented is the maximum atom-surface distance~$d$ for which the longitudinal barrier between two adjacent minima in a periodic potential is at least twice the energy of the longitudinal ground state. Obtaining static magnetic potential features with a resolution on the order of the de Broglie wavelength, i.e. for a potential periodicity on the order of~$\lambda\approx\rm1\,\mu$m, requires the atom-surface distance to be~$d\lesssim\rm2\,\mu$m. Wire currents of~$0.5$ and~$\rm5$ mA are the maximum currents that can be sustained through~$20$ and~$\rm100$ nm atom chip wires, respectively [Fig.~\ref{fig:rho}(c)]. Increasing the current by three orders of magnitude to~$\rm50$ mA serves to increase the required height by just a factor of two, despite being well beyond a safe atom chip current even for a~$\rm200$ nm nanowire. As explained in the following section on corrugations, the spatial resolution achieved with such a wire goes down exponentially with atom-surface distance. Taken from \cite{Salem2010} Copyright IOP Publishing Ltd. Reproduced with permission.}
  \label{fig:PotRes}
\end{figure}

At the minima of such a periodic potential, the longitudinal frequency is~$\omega=\sqrt{V_0 k^2/m}$, where~$m$ is the atomic mass. In order to engineer potential barriers between adjacent minima higher than the single-atom ground state energy by a factor of say~$\eta$, we require~$2V_0>\frac{1}{2}\eta\hbar\omega$, or~$V_0>(\eta^2/16)\hbar^2k^2/m$. In Fig.~\ref{fig:PotRes} we plot the maximum atom-surface distance for which a longitudinal barrier with~$\eta=2$ can be obtained. These curves show that the maximum atom-surface distance still allowing tunneling control is on the order of the potential periodicity~$\lambda$. Designing the edges of a wire as the sum of different modulations therefore allows engineering of any periodic potential up to a resolution determined by the atom-surface distance. Consequently, atom-surface distances of~$\rm1-2\,\mu$m (or sub-micron distances in some cases) will be required to fully exploit the potential of an atom chip based on static magnetic fields.

Let us note that it stands to reason that a quantum information processing device based on ground state neutral atoms and current-carrying wires will also require static electric fields (e.g. for Stark shifts useful for single site addressability or barrier control), RF fields and MW fields (e.g. for controlling barriers, see for example \cite{MWgate,Treulein_MW}). All these forms of radiation address the atom in the near field and so they may all be viewed as static fields. Hence the above discussion of a static barrier is relevant also to these fields. Namely, the closer the atom to the radiation source, the higher the gradients and the possibility to act locally and with greater immunity to instabilities.

As an example of quantitative considerations concerning gradients, let us utilize a square cross section wire. If $d$ is the atom-surface distance and we want to increase the lattice density by decreasing the trap-trap distance linearly with $d$, to maintain the same trap depth the gradients will obviously need to grow as $1/d$. However, to avoid finite size effects, the wire width should also go down linearly with $d$. This means that that the cross section will go down as $1/d^2$. Maintaining a constant current density, this also means that the current goes down as $1/d^2$ canceling the natural gradient growth which is proportional to $d^2$. We thus see that in order to reap the benefits of a small atom-surface distance in terms of high density lattices, one requires growing current densities, and indeed it has been shown that smaller cross section wires enable higher current densities \cite{groth} (see also Fig. \ref{fig:rho}c).

Finally, it is noted above that in order to avoid finite size effects which degrade the trap gradient, the wire width should be on the same scale as the atom-surface distance, i.e. for the above noted heights of $d\lesssim\rm1-2\,\mu$m (see Fig.~\ref{fig:barrier}) one requires a micron-scale wire. As will be shown in the following, it is advantageous to utilize even smaller wire dimensions, namely nanowires. This will enable improving operational parameters at the above heights, or decreasing the atom-surface distance even further without hindering effects. If, in the future, vdW and CP polder are manipulated efficiently, atom-surface distances may become on the order of hundreds or even tens of nano-meters, and to exploit the full potential of this, state-of-the-art fabrication will have to be utilized.

\section{Potential corrugations}

From the early days of atom chips and to this day, the current-carrying wire has been the 'work-horse' of this apparatus. It is therefore fitting that we use it as a case study in order to present how conducting research into material science and material engineering in the context of any required coupling, may result in the suppression of unwanted effects.

Every field source has its inherent inhomogeneities, whether AC or DC, which are bound to cause infidelities. When considering a current-carrying wire as the trapping potential, let us first note the effect of static potential corrugations caused by inhomogeneous electron trajectories, i.e. fluctuating current directions \cite{lukin-frag,hinds-frag}. Such potential corrugations may warp the potential and cause infidelity in potential dependent qubit rotations and logic gates.

Such current irregularities are produced by wire imperfections, namely, geometrical properties (wire edge roughness and surface roughness), and internal bulk inhomogeneities. Since atom chip traps are formed by canceling the magnetic field~$B_y$ generated by the current density~$J^0_x$ at a specific distance from the wire~$d$, the minimum of the trapping potential lies along the wire direction~$\hat{x}$. Variations in this potential~$\delta B_x(x)$ are then directly related to changes in the direction of the magnetic field generated by the wire imperfections.

In previous work~\cite{simonSc,YoniPRB}, we concluded that internal bulk inhomogeneities play a minor role in thin wires~(thickness $h<\rm250$ nm). For wide wires, surface roughness dominates the potential corrugation, but as we show, edge roughness dominates for narrow wires. Consequently, as we are interested in narrow wires in order to avoid finite size effects as discussed above, here we need to consider only current deviations due to edge roughness since the wires considered are thinner and narrower than~$h\approx w<\rm250$ nm. In Appendix A the theory for narrow wires is detailed. A general theory for all wire dimensions is presented in \cite{YoniPRB}.

\begin{figure}[ht]
		\includegraphics[width=0.45\textwidth]{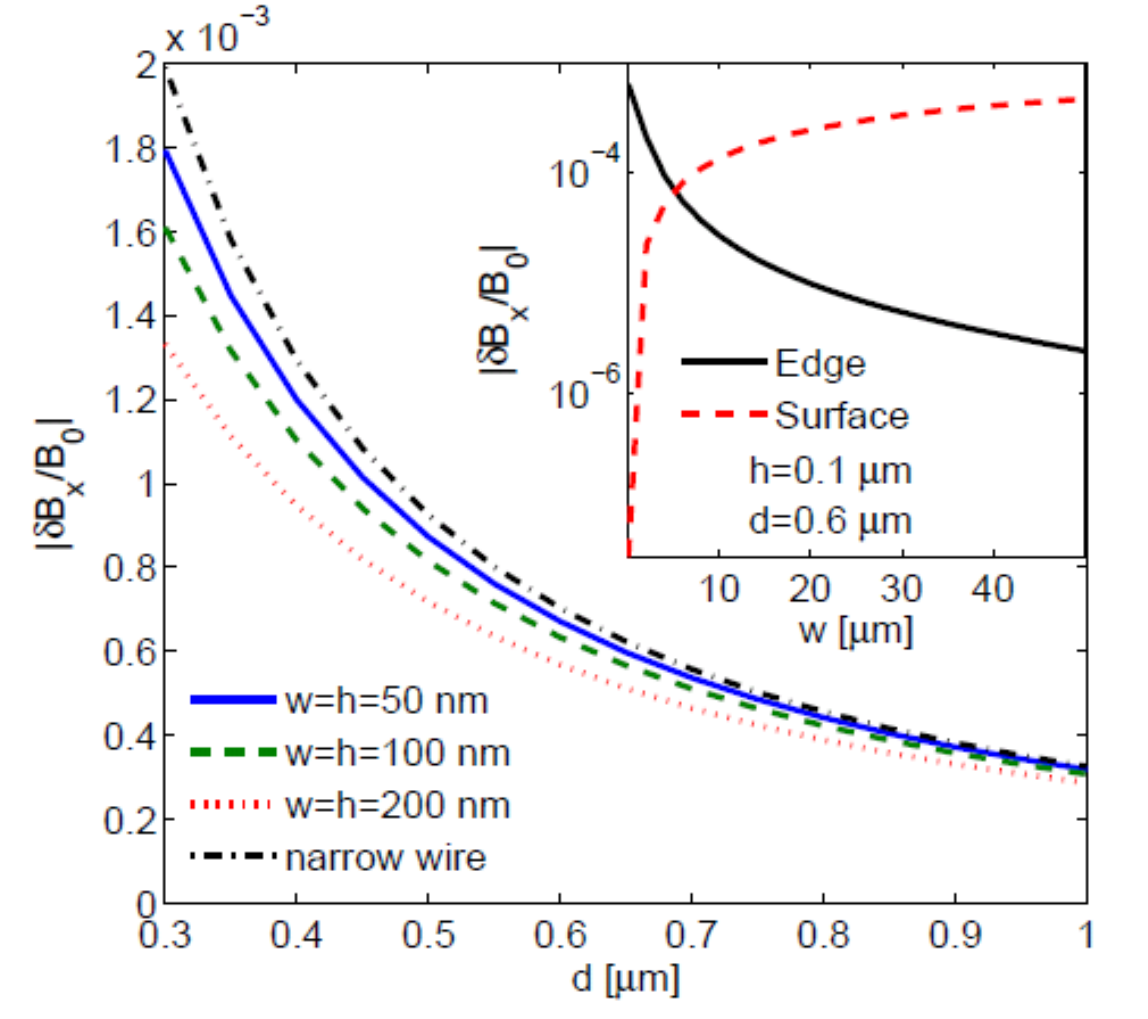}
	\caption{(Color Online) Directional variation of the magnetic fields $|\delta B_x / B_0|$, calculated from Eq.~(\ref{eq:BS}) as a function of the atom-surface distance~$d$. We consider wires with square cross-sections of~$\rm50-200$ nm and the narrow-wire approximation presented in Eq.~(\ref{eq:fragrms}). The same edge roughness is used for calculating the magnetic variations for all the wires. The small differences amongst the wires, despite relatively higher edge roughness of the narrower wires, corresponds to Eq.~(\ref{eq:fragrms}) in which only the absolute quantity~$\delta y_c^{rms}$ appears. These differences are smallest for~$d\gg w$ and become larger as~$d$ approaches~$w$. The inset shows the directional variation of the magnetic field at a fixed height of~$d=\rm0.6\,\mu$m and for a fixed wire thickness of~$h=\rm0.1\,\mu$m, where we plot the influence of edge roughness (solid curve) and surface roughness (dashed curve) over a wide range of wire widths~$w$. The effect of the surface roughness drops strongly for narrower wires, since long wavelengths of the magnetic corrugations are suppressed~\cite{YoniPRB}. Hence, for nanowires magnetic variations are completely dominated by edge roughness. Taken from \cite{Salem2010} Copyright IOP Publishing Ltd. Reproduced with permission.}
  \label{fig:fragmentation}
\end{figure}

In Fig.~\ref{fig:fragmentation}, one may view the calculated directional variations of the magnetic field~$|\delta B_x/B_0|$, generated by the trapping wire, as a function of the atom-surface distance~$d$ for several wire cross-sections. The edge roughness amplitude is measured from our fabricated wires and was found to be frequency-independent [$\alpha=0$ in Eq.~(\ref{eq:fragrms})] with a measured root-mean-square deviation of~$\rm2$ nm between~$\rm100-800$ nm~[Let us note that in the case of edge roughness with~$1\over f$ power spectrum ($\alpha=1$), the directional variations of the magnetic field~$\delta B_x/B_0$ will be an order of magnitude higher~($8\times10^{-3}$ compared to~$7\times10^{-4}$ at~$d=\rm0.6\,\mu$m), and will lead to significantly larger density perturbations.]. In accordance with Eq.~(\ref{eq:fragrms}), we see that, for a given edge roughness~$\delta y_c^{rms}$ (where $y_c$ is the position of the center of the wire), smaller wires  produce only slightly larger magnetic corrugations. We also see that the influence of the surface roughness~$\delta z_{\pm}(x,y)$ is negligible for the narrow wires discussed in this work due to the suppression of long wavelengths in the magnetic corrugations~\cite{YoniPRB}.

We thus find that as the atom-surface distance becomes closer, the corrugation amplitude increases. Furthermore, as the distance decreases shorter and shorter wavelengths affect the potential, and hence on the scale of a single qubit site or in the distance between two qubit sites, significant potential changes may arise.

\subsection{Material engineering for potential corrugations}

\begin{figure}[b]%
\includegraphics[width=\columnwidth]{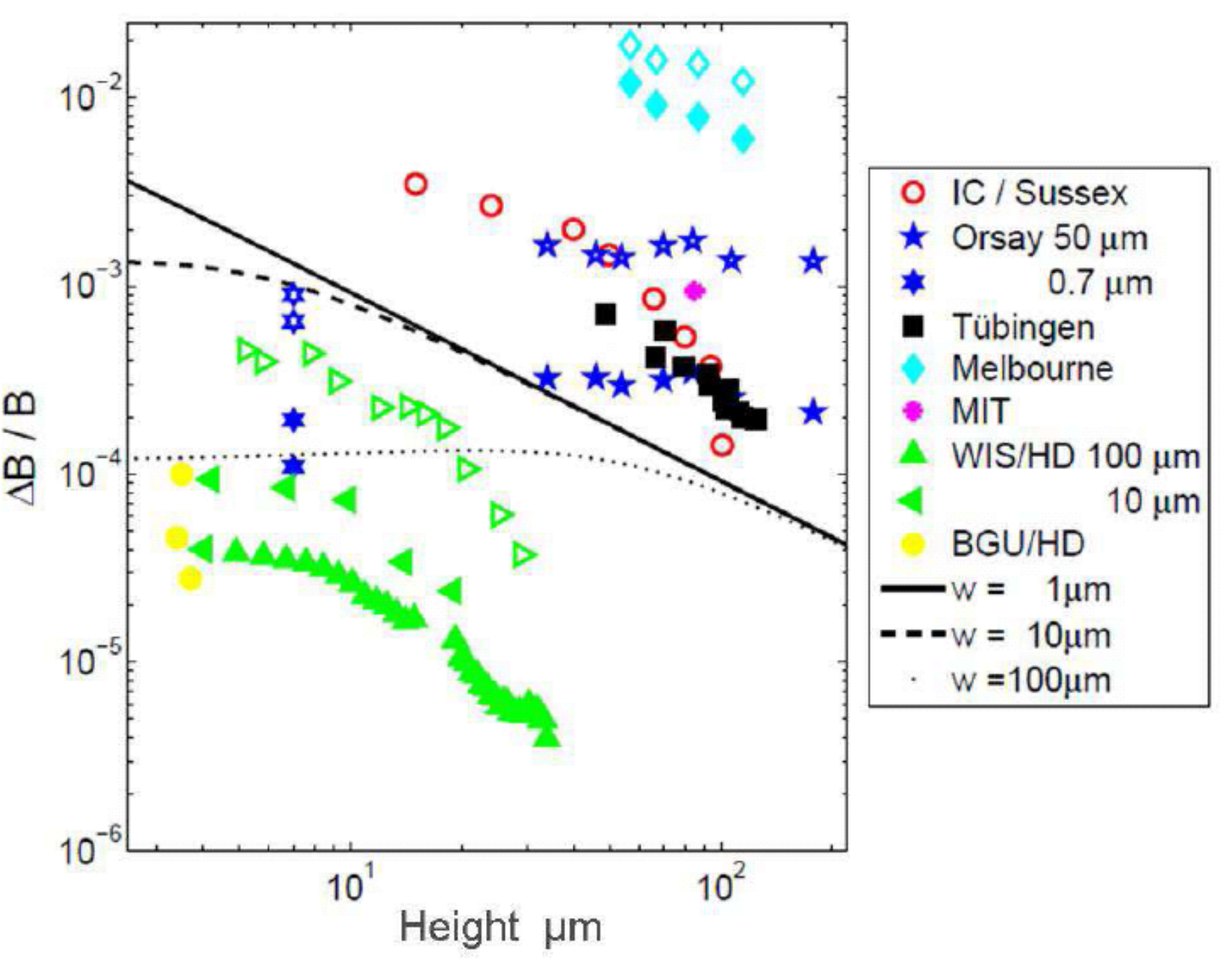}%
\caption{(Color Online)
Comparison of different atom chip magnetic potential roughness measurements.
    Data is taken from figures in the published literature and PhD theses, and displayed with color codes according to the experimental group.
    Filled symbols denote \emph{rms} values, and data displayed as open symbols are peak to valley maximum height of the roughness.
    \emph{Sussex}: data from a gold coated copper wire \cite{Hinds_spinflip};
    \emph{Orsay}: data from an electro plated wire \cite{orsay1}, and data from a 5 wire setup on an evaporated gold chip;
    \emph{T{\"u}bingen}: data from electroplated wires (PhD thesis, J. Fortagh);
    \emph{Melbourne}: data from a permanent magnet atom chip \cite{Whi06};
    \emph{MIT}: data from electro plated wires;
    \emph{WIS/Heidelberg (HD)}: chips fabricated at the Weizmann Institute of Science (WIS): a 100 $\mu$m wire \cite{FAB:Kruger07},
    2003 data from various 10 $\mu$m wires \cite{Krueger2004c,Krueger2004}.
    The open triangles give the peak to valley of the worst ever roughness observed in HD: a 10 $\mu$m wire (PhD thesis, L. Della Pietra).
    \emph{BGU/HD}: Data from 3 different wires fabricated at Ben-Gurion University (BGU) and analyzed in HD \cite{simonSc}. The lines show the limits of maximal
roughness allowed in order to be able to reach the one-dimensional regime
of $\mu < \hbar\omega_{\perp}$for different wire widths \cite{FAB:Kruger07}. Taken from \cite{chapter} Copyright Wiley-VCH Verlag GmbH $\&$ Co. KGaA. Reproduced with permission.
}%
\label{edge_comparison}%
\end{figure}

One way to solve the problem is through better fabrication of the standard gold wires. In Fig. \ref{edge_comparison}, comparative data is presented showing that different methods may deliver very different corrugation magnitudes. The lift-off method used by BGU seems to be the best to date. However, more work is required to improve edge roughness.

\begin{figure}
\includegraphics[width=\columnwidth]{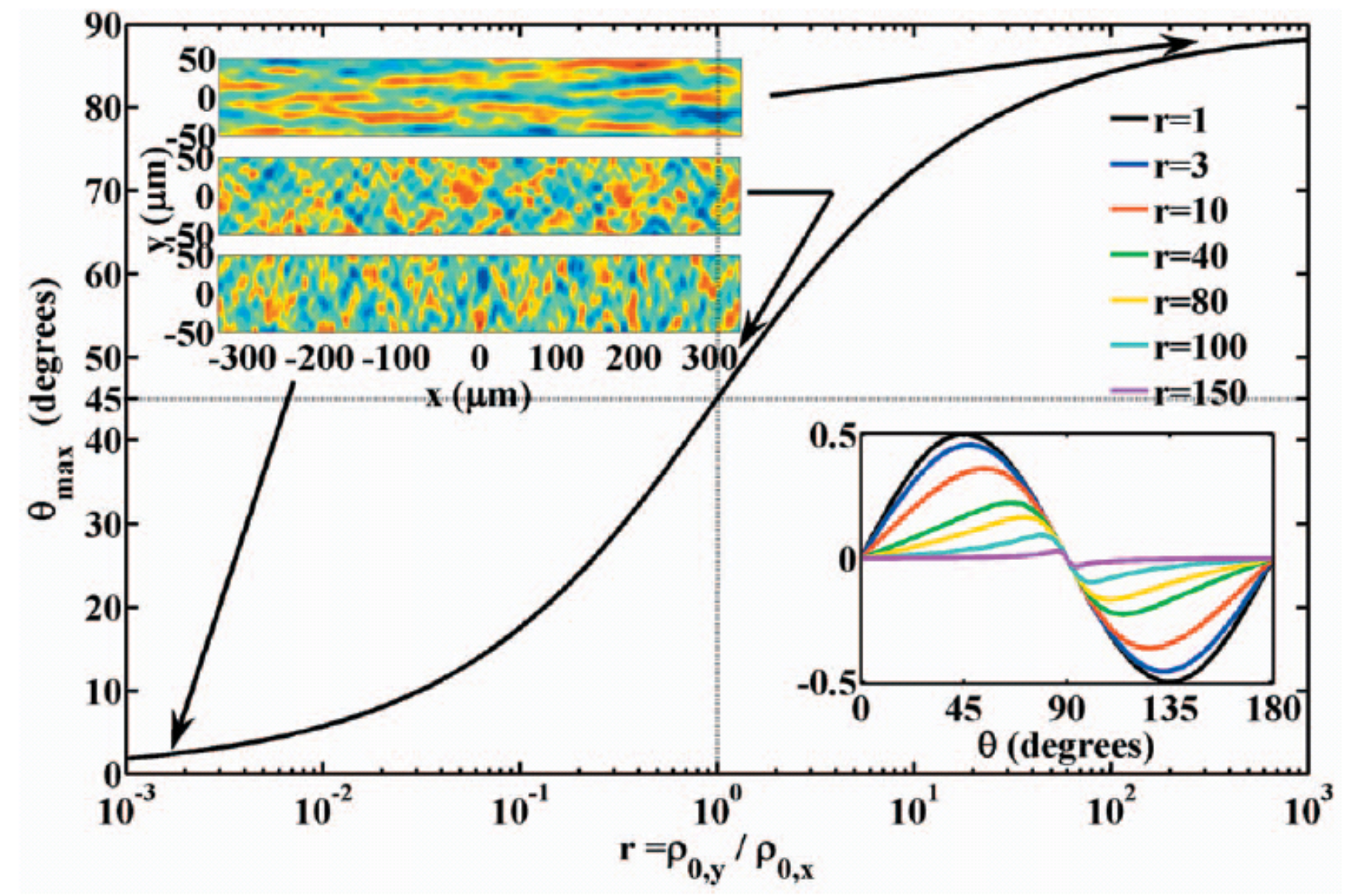}
\caption{(Color Online)
 For a current flowing through an electrically
anisotropic wire, the perturbation wavefront angle $\theta$, giving rise
to maximum transverse electron scattering, will depend on the ratio $r$ between the transverse and longitudinal resistivities. The two-dimensional maps show the predicted atomic density above a wire in the extreme cases
$\theta=0$ ($r\ll 1$), $\theta=90$ ($r\gg 1$), and the isotropic case $\theta=45$ ($r= 1$).
Bottom inset: the magnetic corrugation amplitude as a function of
$\theta$, which is suppressed when $r$ is greater than one. Taken from \cite{YoniPRB}. See Ref. \cite{TalAni} for additional details.
}%
\label{aniso_corrugations}%
\end{figure}

Another route to minimizing static corrugations is by different choices of wire material. An example of a class materials that may reduce electron scattering are electrically anisotropic materials~\cite{YoniPRB}. Here, low transverse conductance will suppress scattering into the transverse direction. One should remember that it is exactly this kind of scattering which causes potential corrugations. In Fig. \ref{aniso_corrugations} a calculation is presented showing that not only do these materials change the pattern of scattering, but they also suppress their amplitude considerably.

\section{Johnson noise}

Let us now continue with the analysis of our case study concerning the benefits awarded by material science and engineering. As current-carrying wires require metallic layers, let us analyze possible sources for AC fluctuations, i.e. noise. Such fluctuations can arise from technical noise in the wires or thermally activated electron currents, and may cause trap loss (spin-flip), climbing the vibrational level ladder (heating) and dephasing (decoherence).

The theory describing the affect of thermal noise \cite{Carsten_thermal, Sidles} on ultra cold atoms is now well established for isotropic materials (e.g. \cite{RonRev,HenkelFundamentalLimits,HenkelPotting,HenkelPottingWilkens,ScheelSpatialDecoherence}). Much of the existing theory was developed and explained by Carsten Henkel. Other models, based for example on non-local electrodynamics, have also been developed \cite{baruch}.

The method typically used, incoherently adds up elementary current sources in the material, within the quasi-static approximation~\cite{HenkelPotting}.
Although approximate (up to corrections on the order of a factor 2), this approach can be used to obtain closed-form solutions
for many wire geometries. This is in contrast to methods based on Green functions, which are accurate but limited to simple geometries such as a half-space or
a laterally infinite layer. It should be noted that the quasi-static approximation is valid when the skin depth $\delta=\sqrt{2/ \sigma_{0} \mu_0 \omega}$ (where $\sigma_0$, $\mu_0$, and $\omega$ are the electrical conductivity, permeability of free space, and the radiation frequency, respectively) is much
larger than both the atom-surface distance $d$ and the wire thickness
$h$. At room-temperature the skin depth of Au at $1$ MHz
is on the order of $70~\mu$m. The electrically anisotropic materials, which will be discussed in the following, usually have a much
higher resistivity, and their skin depths are orders of magnitude larger at the same frequencies (e.g. for graphite $\delta\approx 1$ mm).
The quasi-static approximation thus applies even better
to the latter. This approximation has been corroborated
to a high degree of accuracy in experiment~\cite{Valery,VuleticCP}
and theory~\cite{Zhang07}.

Let us briefly review the basic theory. The thermal radiation couples
to the atoms' magnetic moment by the Zeeman interaction $V\left(\vec{x},t\right)=-
\bfmu \cdot \vec{B}\left(\vec{x},t\right)$, where $\bfmu$ is
the magnetic dipole moment operator.  This operator can
be written for convenience as $\bfmu = \mu_B g_F {\bf F}$, with $\mu_{\rm
B}$ Bohr's magneton, $g_F$ the Land\'e factor of the appropriate
hyperfine level, and ${\bf F}$ the total spin operator.  As the magnetic field
noise is random in time and space it averages to zero, and its effect
is expressed through its two-point correlation function $\left\langle
V\left(\vec{x}_1,t\right) V\left(\vec{x}_2,t'\right) \right\rangle$,
between different points in time $t,t'$ and space
$\vec{x}_1,\vec{x}_2$.

The field and magnetic moment being vectors, one actually needs the
cross-correlations between their components to characterize
the different processes induced by the noise (spin-flips, heating,
and decoherence). It follows from time-dependent perturbation theory
that the rate for a transition from an initial state
$\left|0\right\rangle$ to a final state $\left|f\right\rangle$ of the
system is given by \cite{LandauLifshitz}
\begin{eqnarray}
\Gamma_{0 \to f} & = & \frac{1}{\hbar^2}\sum_{i,j}\int\!{\rm d}(t - t')\,{\rm e}^{{\rm i}\omega_{0f}(t - t')}\times
\label{Eq:Fermi} \\
& \times & \left\langle \langle 0 | \mu_i B_i( \vec{x}, t ) | f \rangle \langle f | \mu_j B_j( \vec{x}, t' ) | 0 \rangle \right\rangle \nonumber
\end{eqnarray}
where the indices $i, j$ label the Cartesian components of the
magnetic field. The argument $\vec{x}$ of the magnetic field in Eq.
(\ref{Eq:Fermi}) is the atomic position. The matrix elements
thus involve spatial (overlap) integrals over the spatial wave
function and magnetic field (e.g.\ see Eq. (\ref{Eq:heatingRate})
below). These integrals can be written in terms of the
two-point correlation function of the magnetic field. The integral of
this correlation function over the time difference $(t - t')$ is
related to the cross-spectral density $S_B^{ij}( \vec{x}_1, \vec{x}_2; \omega_{0f} )$ of the magnetic fluctuations at positions $\vec{x}_1, \vec{x}_2$, and at the transition frequency $\omega_{0f}$ \cite{HenkelFundamentalLimits,Mandel}. This
is defined as
\begin{eqnarray}
\lefteqn{S^{ij}_B\left(\vec{x}_1,\vec{x}_2;\omega\right)\equiv}
\label{Eq:PowerSpectrumDefinition} \\
&& \equiv \int_{-\infty}^{\infty} {\rm d} \left(t-t'\right) e^{-{\rm
i} \omega(t-t')}\left\langle
B_i\left(\vec{x}_1,t\right)B_j\left(\vec{x}_2,t'\right)\right\rangle
\nonumber,
\end{eqnarray}
or equivalently by
\begin{equation}
\left\langle
B_i^*\left(\vec{x}_1,\omega\right)B_j\left(\vec{x}_2,\omega
'\right)\right\rangle = 2 \pi \delta \left(\omega - \omega
'\right)S_B^{ij}\left(\vec{x}_1,\vec{x}_2;\omega\right).
\label{Eq:spectralDensity}
\end{equation}

In cases where the spatial degrees of freedom can be considered
classically, $\vec{x}$ in Eq. (\ref{Eq:Fermi}) can be taken as the coordinate of the
atoms, giving rise to a position-dependent transition rate between
spin states. Instead
of the wave function overlap integrals, the coordinates are then
taken as $\vec{x}_1=\vec{x}_2$. For calculating at the trap center
one takes $\vec{x}_1=\vec{x}_2=\vec{x}$ \cite{Note:TrapInhomogeneity}.

The noise power spectrum $S_B^{ij}( \vec{x}_1, \vec{x}_2; \omega_{0f}
)$ is quite flat at all relevant frequencies which correspond for
example to transitions between Zeeman magnetic sub-levels (radio-frequency), or
the trap vibrational states (Hz to kHz range). Hence a low-frequency
limit $\omega_{0f} \rightarrow 0$ can be taken. If this spectrum is
not flat in the relevant frequency range, the time dependence of loss
and decoherence processes is more complicated and does not reduce to
a simple rate \cite{HenkelFundamentalLimits}.

Spin-flips are transitions from trapped ($|0\rangle$) to untrapped
($|f\rangle$) internal states. These states are eigenstates of the
spin operator component parallel to the quantization axis,
defined by the static trapping field at the bottom of the trap (we neglect the magnetic field direction deviations at the edges of the trap).
Due to the form of the dipole moment operator $\mu_i$,
the Zeeman interaction can induce a spin-flip
only when the direction of the magnetic field fluctuation is
perpendicular to the quantization axis.
Hence, in this case, we can rewrite (\ref{Eq:Fermi}) as
\begin{equation}
\gamma_{{\rm spin~ flip}} = \sum_{l,m=\perp} \frac{\mu_l \mu_m
}{\hbar^2}S_{B}^{lm}\left(\vec{x};\omega_{0f}\right),
\label{Eq:spinfliprate}
\end{equation}
where we sum over the perpendicular components $l,m$ only, and take
the matrix elements $\mu_l = \mu_B g_F \langle 0 | F_l | f \rangle$
of the dipole moment between the states $|0\rangle$ and $|f\rangle$.
The spin-flip rate can be measured from the lifetime of a magnetic
micro-trap, and by varying the trap distance one can discriminate loss due to surface induced magnetic field fluctuations, against trap loss of different origin~\cite{Hinds_spinflip,VuleticCP,Cornell2,TubingenNoise,ZhangNoise}.

Decoherence or dephasing of a quantum superposition state can occur without
changing the occupation of the states involved in the superposition,
affecting only the phase. We distinguish between spin decoherence and
spatial decoherence. The former involves the change of the relative phase
of two internal states ($|1\rangle$ and $|2\rangle$) in a
superposition at the same spatial location $\vec{x}$, while the latter involves
the change of the relative phase of two spatially separated components of
the atom cloud (positions $\vec{x}_{1}$ and $\vec{x}_{2}$),
trapped in the same internal state.
Here we shall consider only `classical dephasing' in which the phase
change arises from the fluctuations in the Zeeman shift. Such shifts occur, to lowest order, for magnetic field fluctuations parallel to the atom's magnetic
dipole moment. Hence for both types of decoherence processes we need
only the parallel component of the same power spectrum appearing in
the spin-flip case (Eq. (\ref{Eq:spinfliprate})). The spin
decoherence rate can thus be written as
\begin{equation}
\gamma_{{\rm spin~ decoherence}} = \frac{\Delta
\mu_\|^2}{2\hbar^2}S_\|\left(\vec{x};0\right),
\label{Eq:spinDecoherenceRate}
\end{equation}
with $\Delta \mu_\| = \left\langle 2 \left| \mu_\| \right|
2\right\rangle - \left\langle 1 \left| \mu_\| \right| 1
\right\rangle$ being the differential magnetic
moment of the two trapped states, and $S_\|\left(\vec{x};0\right)$
the low-frequency limit parallel component of the
noise spectrum, at the trap center $\vec{x}$. This decoherence rate
can be measured from the reduced contrast
of interference fringes in a series of Ramsey spectroscopy
experiments, performed as a function of the atom-surface distance.
In fact, long coherence times have already been measured close to the surface \cite{ReichelClock}. This measurement puts an upper limit on the decoherence rate as other processes may mask the dephasing due to Johnson noise (e.g. inhomogeneous magnetic field) \cite{ReichelRephasing}.

For the case of a spatially separated superposition state, the rate of decoherence of the relative phase of the atomic states between two points $\vec{x}_1,\vec{x}_2$ involves the correlation function of the difference in the magnetic fields $B_\|( \vec{x} _1, t ) -B_\|(\vec{ x}_2, t' )$, and can be written as
\begin{equation}
\gamma_{{\rm spatial~decoherence}} = \frac{\mu_\|^2}{2\hbar^2} \left[S_{11}+S_{22}-2S_{12}\right],
\label{Eq:spatialDecoherenceRate}
\end{equation}
where we denote for convenience $S_{ij}=S_\|(\vec{x}_i,\vec{x}_j;\omega \rightarrow 0)$, and assume a correlation spectrum symmetric in $\vec{x}_1$, $\vec{x}_2$ and flat in frequency. We again take only the field components parallel to the quantization axis, that shift the relative phase between the two parts of the wave function. In Eq. (\ref{Eq:spatialDecoherenceRate}) $\mu_\|$ is the magnetic moment matrix element of the single internal state. The low-frequency limit is valid here as the ``scattering cross section" for transitions involved in the decoherence process is in practice independent of the frequency \cite{HenkelFundamentalLimits,ScheelSpatialDecoherence}.

This decoherence rate can be measured by studying the interference pattern
contrast reduction in double-well experiments \cite{JoergIFM,Reichel2010}, where the atom cloud is separated in two parts, which are then held at a fixed separation
for a given time. The interference pattern is obtained by overlapping
the cloud parts upon release from the trap. The interference contrast
is directly related to the product of split time and decoherence
rate; it can be measured as a function of the atom-surface distance
by repeating the measurement at different heights. It is interesting to note that while quite a lot of theory has been done, spatial dephasing on atom chips has not to this day been measured. This is probably due to the fact that most cold atom experiments utilize a BEC, and in this system phase diffusion due to atom-atom interaction is very fast. This measurement is of great interest as the correlation lengths of the noise need to be examined carefully. This number may have a dominant effect on what the ${\it N}$-qubit dephasing time would be \cite{Nqubit1,Nqubit2,Nqubit3,Nqubit4}.

Finally, heating of the trapped atoms (as a result of exciting
external degrees of freedom while retaining the same internal state)
can also be caused by the coupling to the thermal radiation.
The transition rate of atoms initially
in the ground vibrational state $\left|0\right\rangle$
to higher states $\left|f\right\rangle$ with energy splitting
$\hbar \omega_{0f}$ is of the form
\cite{HenkelFundamentalLimits}
\begin{equation}
\Gamma_{0 \rightarrow f} = \frac{\mu_\|^2}{\hbar^2} \int {\rm
d}\vec{x}_1 {\rm d}\vec{x}_2 M_{f0}^*(\vec{x}_1) M_{f0}(\vec{x}_2)
S_\|\left(\vec{x}_1,\vec{x}_2; \omega_{0f} \right),
\label{Eq:heatingRate}
\end{equation}
where we find once more the spin operator matrix element $\mu_\|$ in
the direction parallel to the quantization axis,
and now also the wave functions of
the levels involved in the transition $M_{f0}( \vec{x} ) =
\psi_f^*(\vec{x})\psi_0(\vec{x})$.
The spatial integration
here provides a probe of the spatial correlation of the magnetic
field noise. In practice, it is
enough to consider transitions from the ground state to either of the
first two excited levels \cite{RonRev}.

Thus, we see that the important term common to all rates is the
spectral density of the magnetic field fluctuations or power spectrum
$S_B^{ij}\left(\vec{x}_1,\vec{x}_2;\omega\right)$. This quantity
holds all of the relevant information about the magnetic field
fluctuations leading to the harmful processes, while the other terms
in each of the rates describe the coupling of the noise to the atoms
through the magnetic dipole moment. Furthermore, we see that a
measurement of either the spin or spatial decoherence rates, or of
the heating rate, will give strong indications to any of the other
two of these three processes, as they all depend on the noise power
of the same field component.

\begin{figure}[b]
\centerline{%
\resizebox{80mm}{!}{%
\includegraphics*{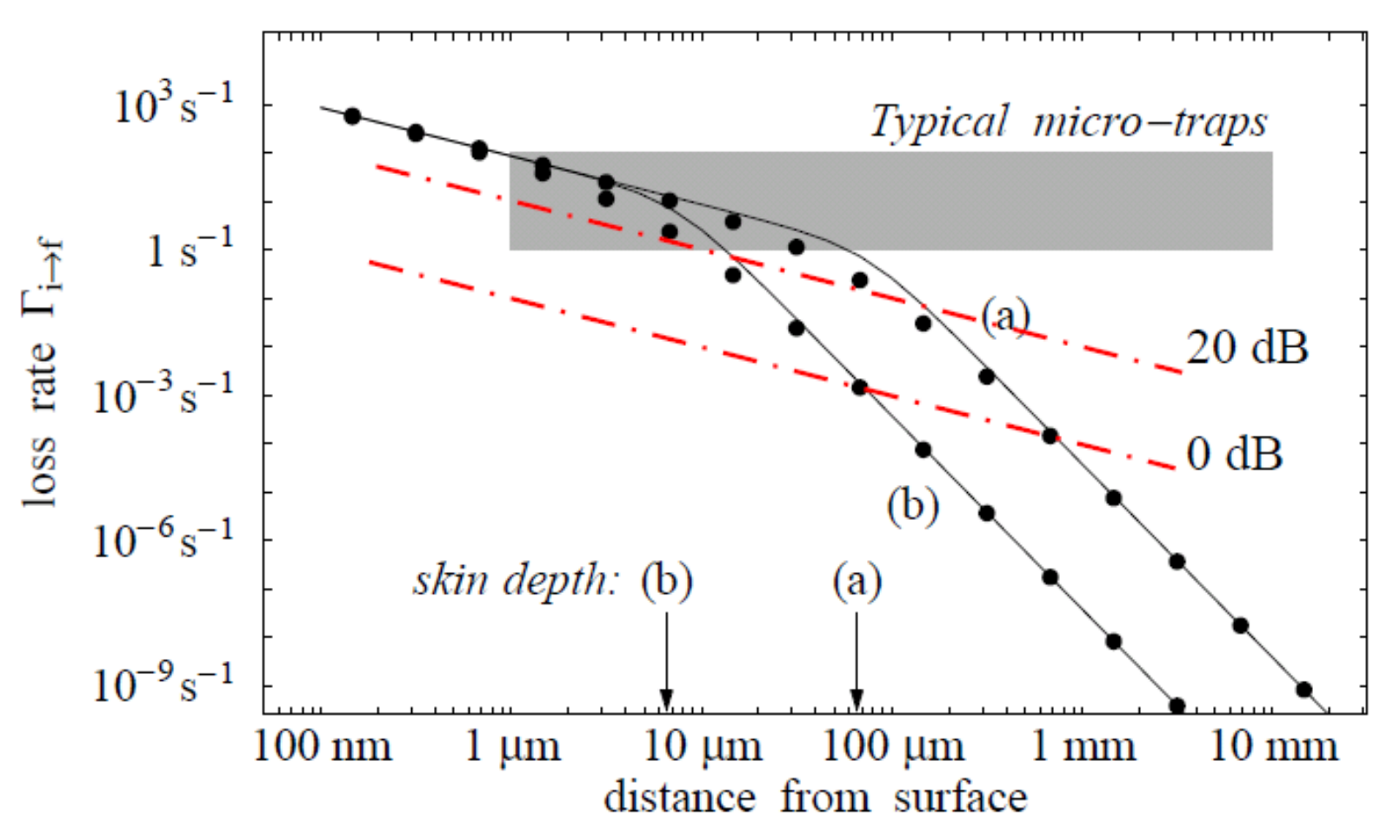}}}
\caption[]{(Color Online) Rate of spin-flips in a microtrap above a planar
metal substrate. The black solid lines are due to the near
field noise
generated by thermally excited currents (Johnson noise) in
the substrate.
The dots are an exact numerical calculation from~\cite{HenkelPottingWilkens}.
Larmor frequencies
$\omega_{L} = \mu_{B} |{\bf B}( {\bf r} )| / \hbar =
1\,{\rm MHz}\times 2\pi$ (a) and $100\,{\rm MHz}\times 2\pi$ (b)
are chosen.
The red dash-dotted lines describe the noise due to technical fluctuations
in the electric current of a side guide. The current noise
is assumed at shot noise level (0 dB) or 20 dB above shot noise.
The guide height $d$ is lowered by ramping down the wire current $I$
with a constant ratio $I/d$ and
at fixed bias field $B_b = 100\,$G. Taken from \cite{HenkelFundamentalLimits}.}
\label{APB}
\end{figure}

One may summarize these results with an example, as presented in Fig. \ref{APB}, showing the increasing spin-flip rate as a function of decreased atom-surface distance. A qualitatively similar behavior is found for heating and decoherence, as the noise amplitude increases with decreasing distance.

\subsection{Material engineering for Johnson noise}

\begin{figure}[b]
\includegraphics[width=\columnwidth] {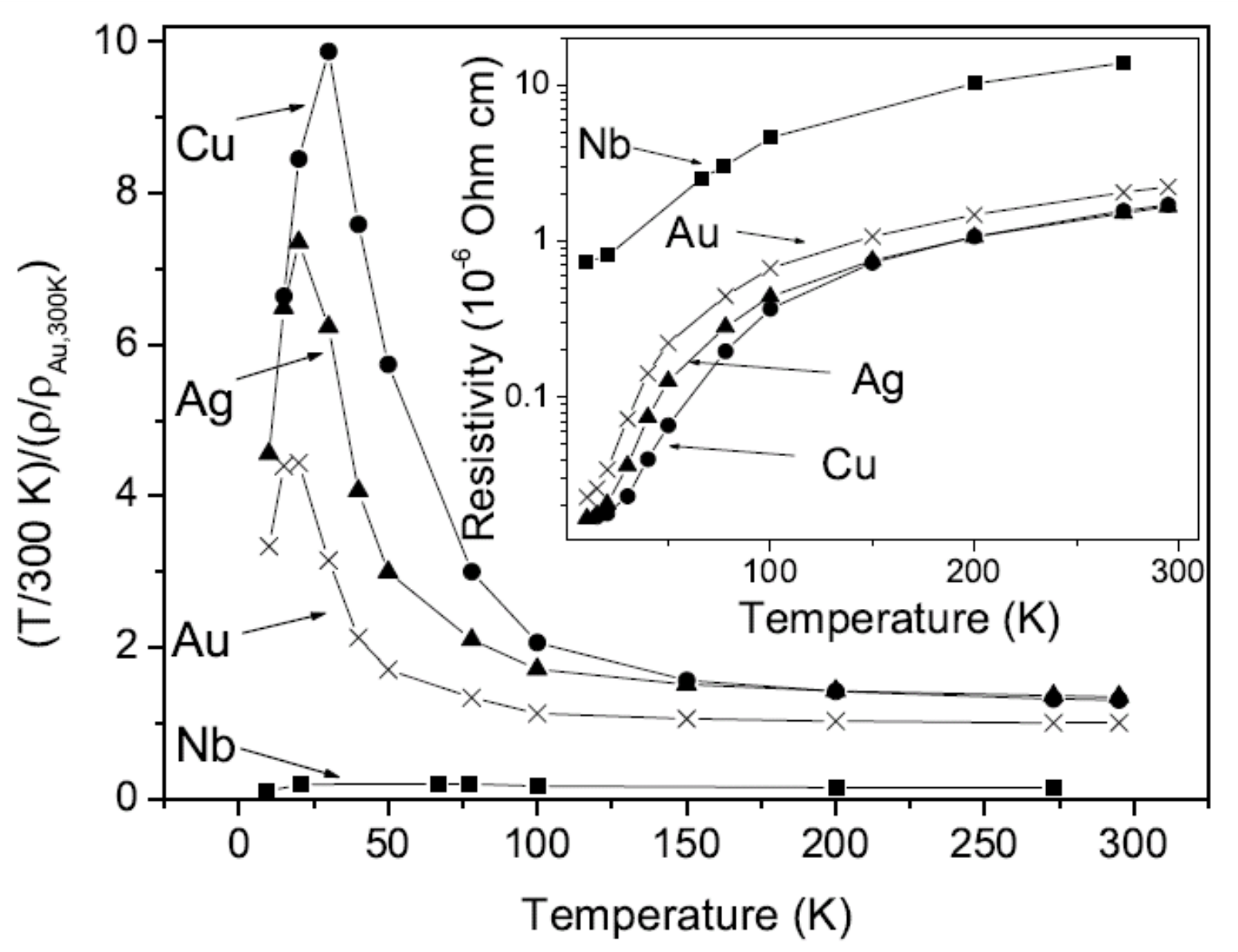}
\caption{ Temperature dependence of the normalized magnetic noise for wires made of copper (circles), silver (triangles) gold (crosses) and niobium
(squares). The noise is normalized to the value for gold at
$T=$300 K. Inset: temperature dependence of the resistivity
(extracted from \cite{Malkov}). Taken from \cite{Valery}.} \label{alloy1}
\end{figure}

Typically, the noise spectrum $S$ is proportional to the ratio $T/\rho$ of the surface temperature and the surface resistivity \cite{HenkelFundamentalLimits}. This seems to indicate that cooling the surface would not be advantageous as typically resistivity is linear with temperature. In fact, as is shown in Fig. \ref{alloy1}, cooling of normal metals even worsens the situation. Although the eventual goal should be to utilize material science so as to realize an operational QC device at room temperature, cooling down to $4$K should still be considered as technologically feasible. Let us therefore analyze the situation at low temperatures for standard conductors.

\begin{figure}[b]
\includegraphics [width=\columnwidth] {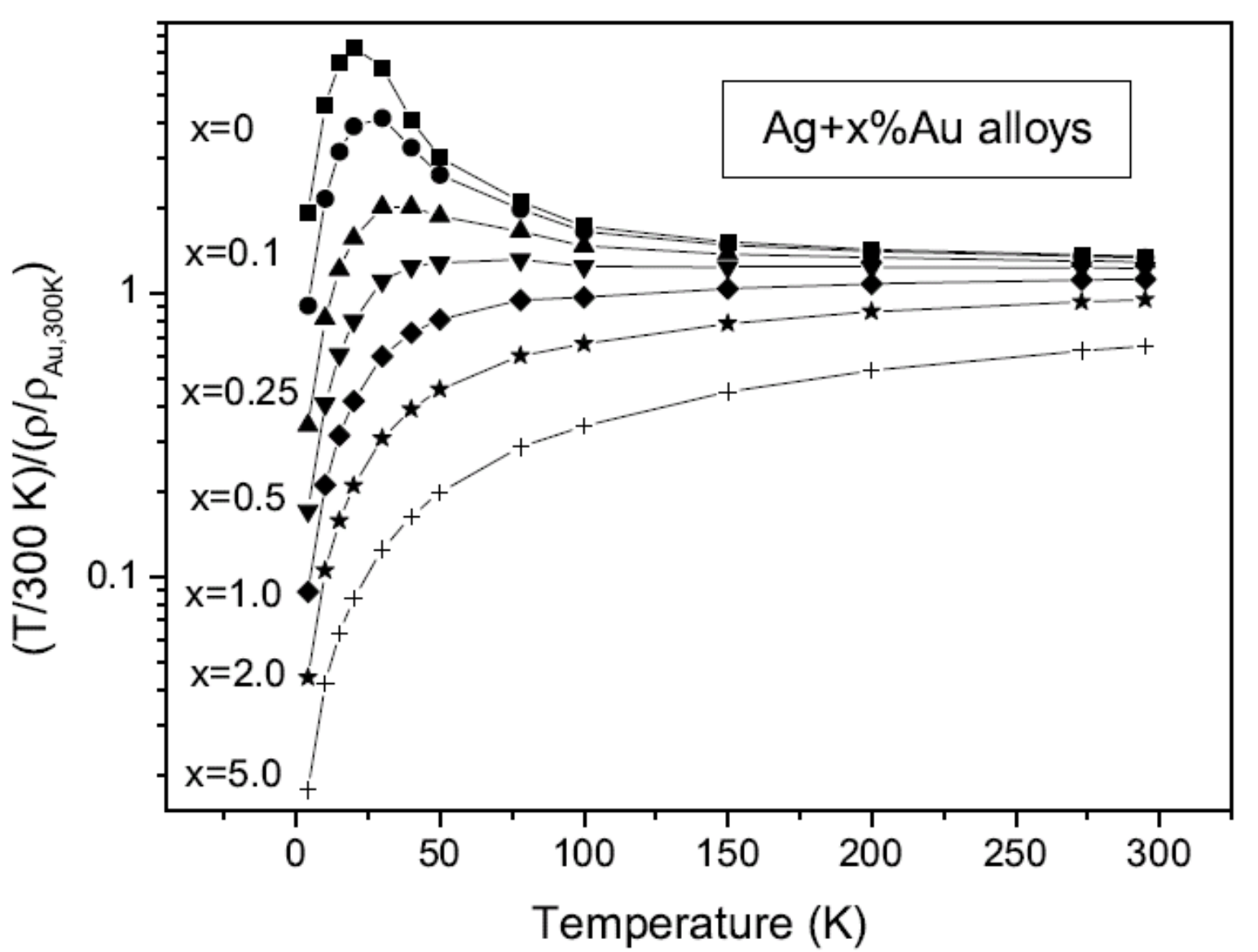}
\caption{Temperature dependence of the ratio $T/\rho$ (normalized
to its value for gold at 300 K) for silver and its alloys with
gold: pure silver (squares), and with 0.1\% gold (circles), 0.25\%
(triangles), 0.5\% (inverted triangles), 1\% (diamonds), 2\%
(stars), and 5\% (crosses). Note the difference compared to Fig. \ref{alloy1}. Taken from \cite{Valery}.}
\label{alloy2}
\end{figure}

To avoid complications of magnetic permeability and
hysteresis of the chip, let us consider here only nonmagnetic metals
(having no long-range magnetic order). The resistivity of these
metals is essentially a sum of two contributions $\rho = \rho_{0}
+ \rho_{\rm ph}$: a temperature independent residual resistivity
$\rho_0$, due to scattering of charge carriers by crystal defects
and impurities and a phonon contribution $\rho_{\rm ph}$ \cite{Valery}. By manipulating the relative strength of these two terms we may engineer the ratio $T/\rho$ as a function of temperature. In Fig. \ref{alloy2} this ratio is presented as a function of the amount of defects introduced into the material.

\begin{figure}[b]
\includegraphics [width=\columnwidth]{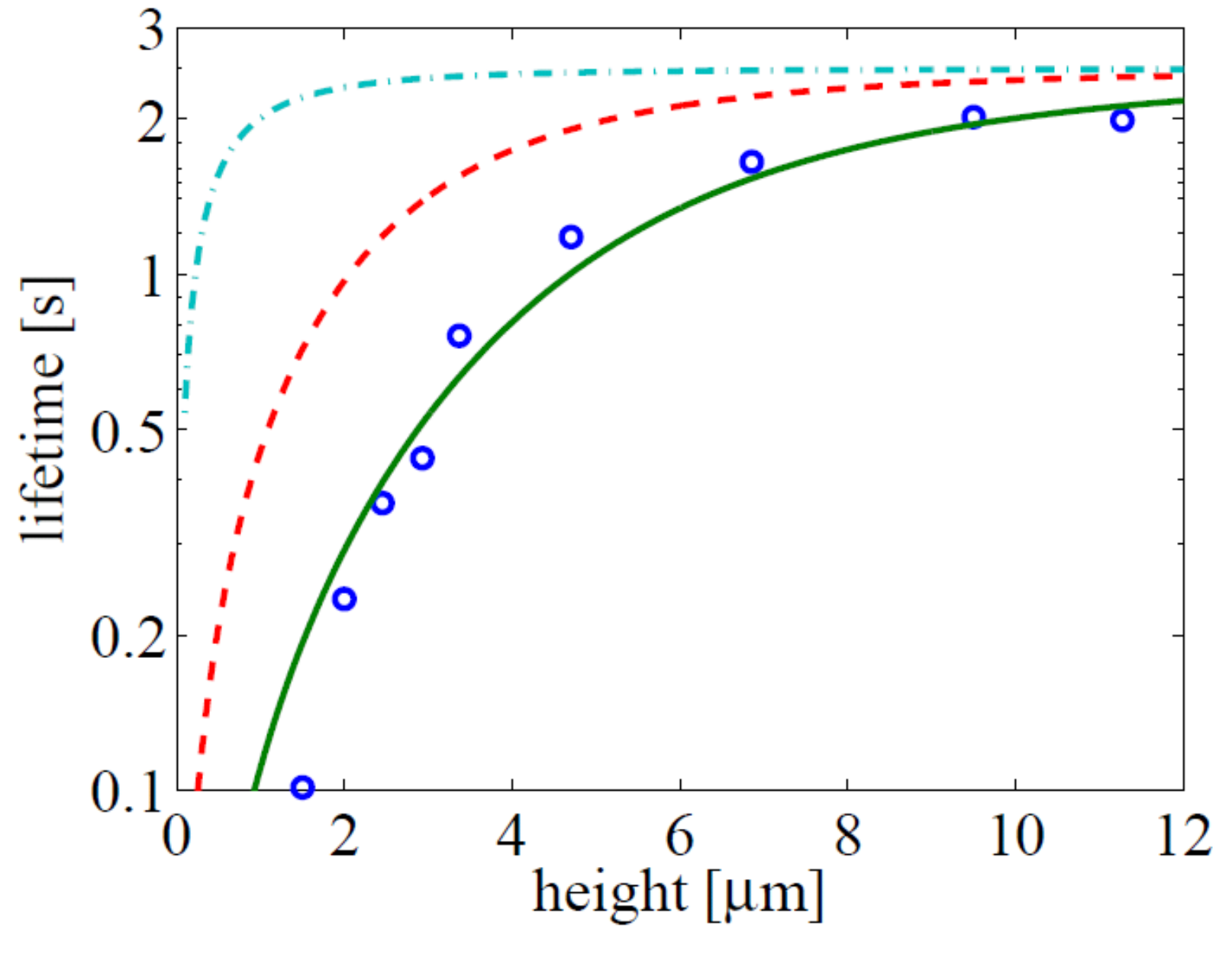}
\caption{(Color Online) Comparison of trapping lifetimes of $^{87}$Rb atoms above
a copper wire on an atom chip~\cite{VuleticCP} with a theoretical
calculation (solid line). Predicted lifetimes are also shown for a
similar wire made of an alloy of Ag with 6\% Au content, cooled
down to $T=$77 K (dashed) and 4.2 K (dash-dotted). In both cases the
resistivity was taken equal to gold resistivity at $T$=300 K,
$\rho_{\rm Au}=2.21~\mu\Omega \cdot$cm (same penetration depth).
The needed gold concentrations are estimated by
$x=5.4$\% and $x=6$\%, respectively. Note that the calculation presented here
differs from the one made in~\cite{VuleticCP}. Van der Waals
forces are not taken into account. Let us also mention that the
maximum noise reduction factor of 70 expected from the calculations is not
visible due to the effect of the technical noise and/or
background gas collisions limiting the lifetime in this experiment
to a maximum of $\tau_{\rm tech}=2.5\,s$~\cite{LinPrivate}. Taken from \cite{Valery}.}
\label{alloy3}
\end{figure}

To conclude this first example of how material engineering may alter the hindering effect of Johnson noise, the expected lifetime as a function of atom-surface distance for different temperatures is presented in Fig. \ref{alloy3}. To verify the theoretical calculation, a comparison to experimental results is also made. As can be seen, more than an order of magnitude improvement in lifetime may be achieved by simply introducing defects into the material.

As a second example for the potential advantages of material engineering for QC let us now calculate the possible suppression of the qubit dephasing by utilizing an anisotropic material instead of a normal metal. Although realistic hyperfine qubits will most likely use states between which only second-order Zeeman shifts and higher exist, for simplicity we shall use two qubit states differentiated by a first-order Zeeman shift.

As presented in the previous section, the noise power spectrum is related to the cross correlation function of the magnetic field fluctuations [Eqs. (\ref{Eq:PowerSpectrumDefinition}),(\ref{Eq:spectralDensity})]. Before calculating this correlation function, we define the coordinate system
such that the wire length $L$ is along the $\hat{x}$ direction, its
width $w$ along the $\hat{y}$ direction, and its thickness $h$ along
$\hat{z}$.  As in typical Ioffe-Pritchard magnetic traps, a non-zero field at the trap bottom polarizes the atoms along the wire, and so the quantization axis in our analysis is along the $\hat{x}$ axis.

We can make a distinction between two types of\linebreak anisotropic
materials. Materials with a `layered conductance' are relatively good conductors
along two axes and have one axis of bad conductance: $\sigma_{yy} \ll
\sigma_{xx} \sim \sigma_{zz}$. We always assume that the
wire is aligned to one axis of good conductance, so that current may
flow easily and create a magnetic trap. Materials that have only one
direction of good conductance, $\sigma_{yy} \sim \sigma_{zz} \ll
\sigma_{xx}$ or $\sigma_{zz} \ll \sigma_{yy} \ll
\sigma_{xx}$, will be denoted as `quasi-1D conductors'. A detailed treatment of the anisotropic material and the noise it induces, is presented in Appendix B.

\begin{figure}
\centering
\includegraphics*[width=\columnwidth]{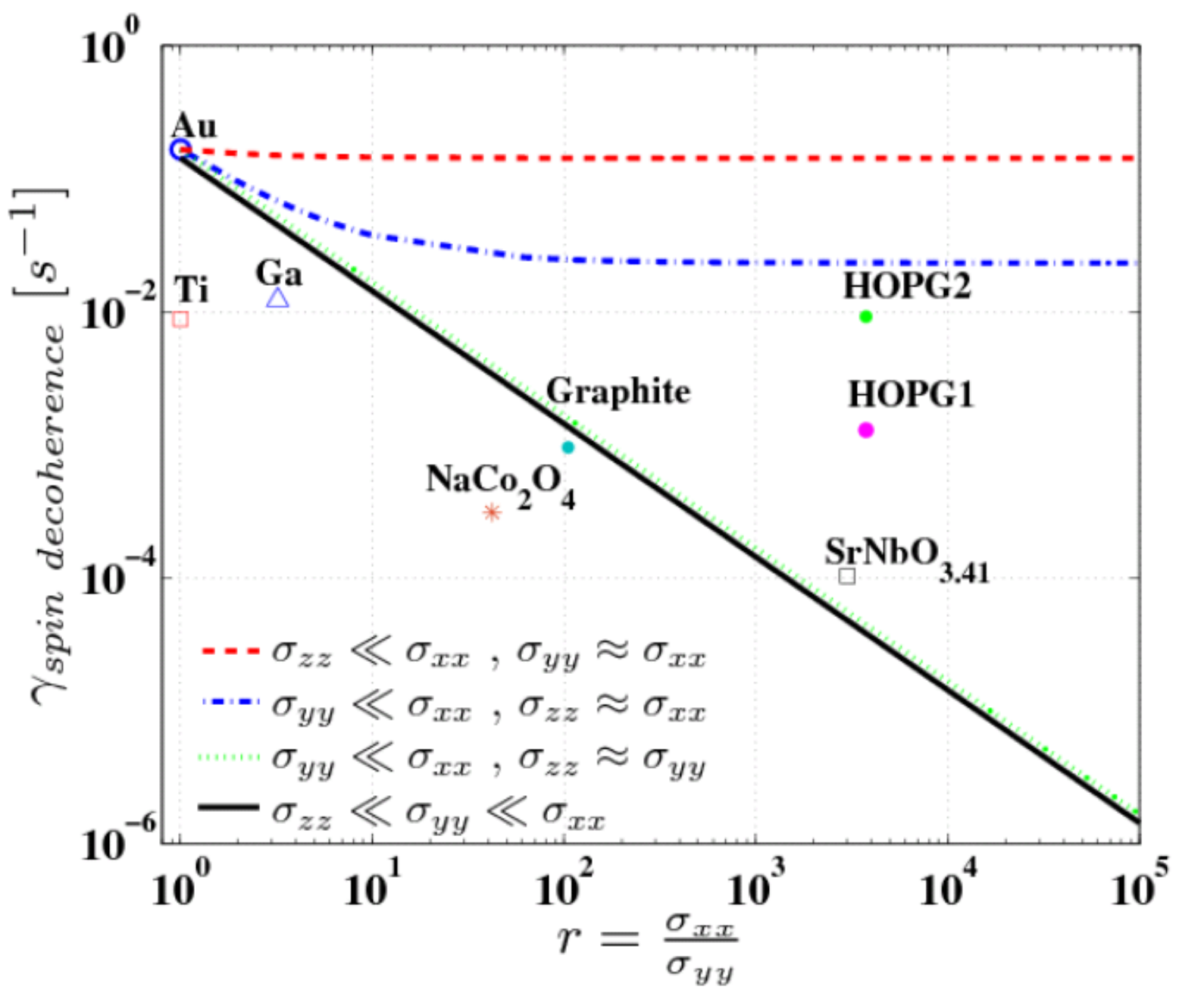}
\caption{(Color Online) Lines: Spin decoherence rate $\gamma_{{\rm spin~ decoherence}}$ as a function of electrical anisotropy $r=\sigma_{xx}/\sigma_{yy}$, for layered and quasi-1D conducting materials. Trap and wire parameters are $d=5~\mu$m, $w=10~\mu$m, $h=2.15~\mu$m, and a surface temperature of $T=300$ K. For these lines, the good conductivity along the wire was assumed to be identical to that of Au. For layered materials having the badly conducting axis along the wire thickness (dashed red), only a slight improvement is gained, and the dependence on the anisotropy is negligible. If the badly conducting axis is along the width of the wire (dashed-dotted blue), the improvement is more pronounced, but still saturates at relatively low anisotropy. For quasi-1D materials (dotted green - quasi-1D with both low conductivity terms of the same order; solid black - the more extreme case having $\sigma_{zz} \ll \sigma_{yy} \ll \sigma_{xx}$, where we assumed $\sigma_{zz} \approx \sigma_{yy} / r$), the suppression is much more significant. For high anisotropy the decoherence rate scales as $1/r$ with the anisotropy. Points: Examples of specific materials, not normalized to Au. Taken from \cite{TalAni}. \label{Fig:spinDecoherence}}
\end{figure}

In contrast to the case of spin-flips, the decoherence and heating rates,
Eqs. (\ref{Eq:spinDecoherenceRate},\ref{Eq:spatialDecoherenceRate},\ref{Eq:heatingRate}),
depend only on the noise in the parallel field component  $B_{xx}$.  This component may be strongly reduced for highly anisotropic materials. To quantify this, consider
the ratio
\begin{equation}
\frac{B_{xx}^{{\rm aniso}}}{B_{xx}^{{\rm iso}}} = \frac{\sigma_{zz}X_{yy} + \sigma_{yy}X_{zz}}{\sigma_{xx}\left(X_{yy} + X_{zz}\right)},  \label{Eq:noiseSuppressionRatio}
\end{equation}
where $\sigma_{ii}$ are the conductivities and
$X_{ii}$ the geometry-dependent factors (see Appendix B), and where the reference level is an isotropic conductivity set to the `good axis', $\sigma_{0} = \sigma_{xx}$.
The possible improvements depend on the relative
magnitudes of the anisotropic conductivities. Four cases can be
distinguished, as illustrated in Fig. \ref{Fig:spinDecoherence}.
For materials with layered conductance, the worst choice is to have
the second good conducting axis in the chip plane, along the wire's
width: $\sigma_{yy} \sim \sigma_{xx} \gg \sigma_{zz}$ (dashed red line).
The ratio~(\ref{Eq:noiseSuppressionRatio}) then tends to
$\left(1 + X_{yy} / X_{zz} \right)^{-1}$ which is not significantly
smaller than unity and where the conductivity anisotropy $\sigma_{xx} / \sigma_{yy}$ actually does not enter.
With the other choice, having the badly conducting axis along the wire
width (dashed-dotted blue), we get a reduction of about one order of magnitude for a small
wire geometry. Materials that are quasi-1D conductive have a much larger potential for noise
suppression: for comparable `bad axes', $\sigma_{yy} \sim \sigma_{zz}
\ll \sigma_{xx}$, Eq. (\ref{Eq:noiseSuppressionRatio}) scales inversely with the
anisotropy ratio $r=\sigma_{xx} / \sigma_{yy}$ which may be very large. The suppression is somewhat
more pronounced in the extreme case
$\sigma_{zz} \ll \sigma_{yy} \ll \sigma_{xx}$.

The reduction of the decoherence and heating rates is illustrated in
Fig. \ref{Fig:spinDecoherence} for the cases discussed above.
We plot the spin decoherence rate~(\ref{Eq:spinDecoherenceRate}) for a
superposition state of the hyperfine levels
$\left| F=2,m_F=2 \right \rangle$ and $\left| F=2,m_F=1 \right \rangle$
in the ground state of $^{87}$Rb.
The anisotropic conductor is chosen such that the largest conductivity value
$\sigma_{xx}$ coincides with the one for Au, an isotropic conductor
taken as reference. The calculated rates for some specific materials are also given.
It can be seen that quasi-1D materials
are much more appealing to suppress heating and decoherence, although
about one order of magnitude can already be gained with layered
materials, even at a relatively small anisotropy.
In addition, for most anisotropic materials, even the best direction
conducts worse than Au; therefore the rates presented in this graph even for layered materials are smaller than for Au.

To conclude, heating and decoherence may be suppressed by several orders of magnitude even at room temperature, by using electrically anisotropic materials for current-carrying structures on atom chips.

\begin{figure}[b]
 	\includegraphics[width=\columnwidth]{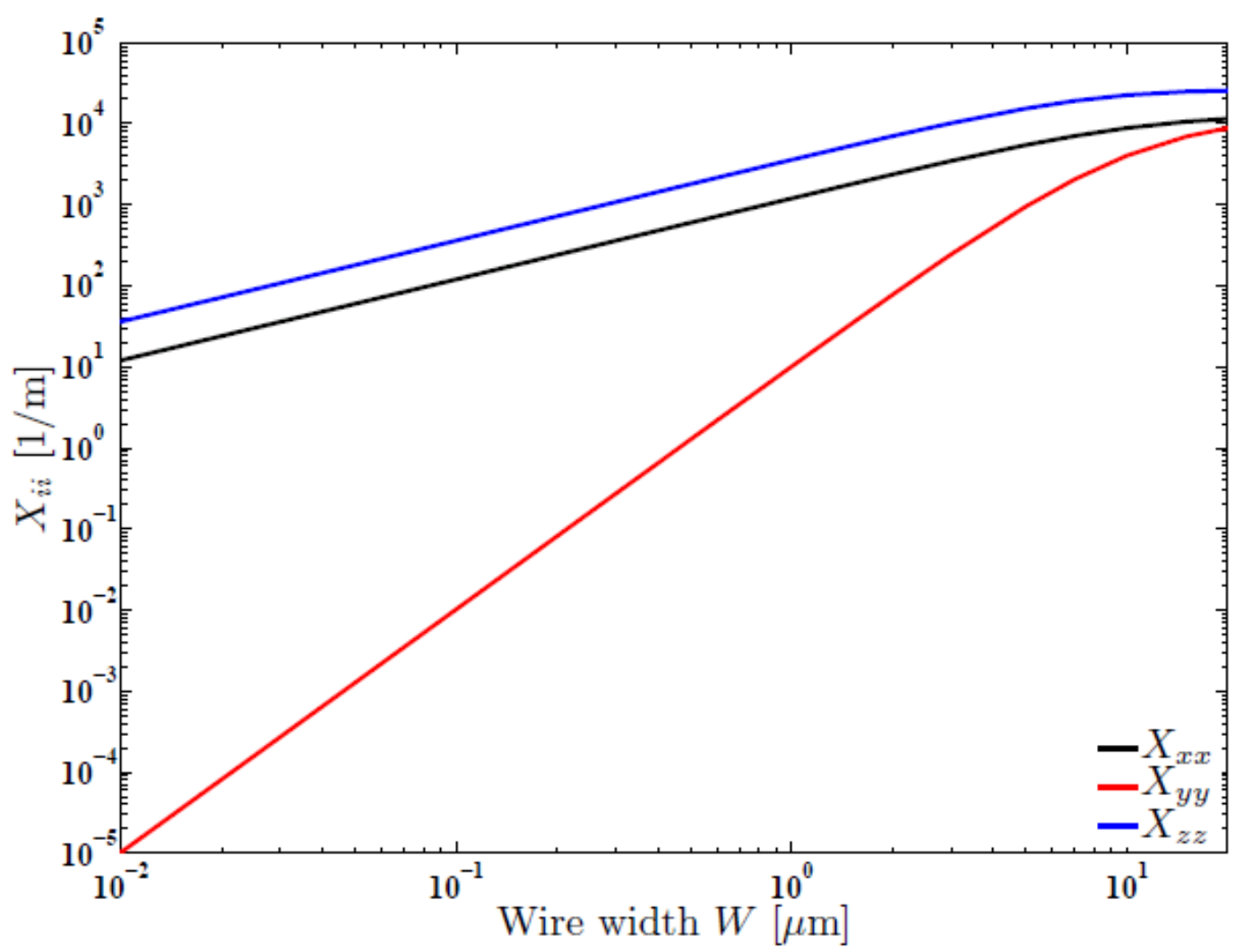}
	\includegraphics[width=\columnwidth]{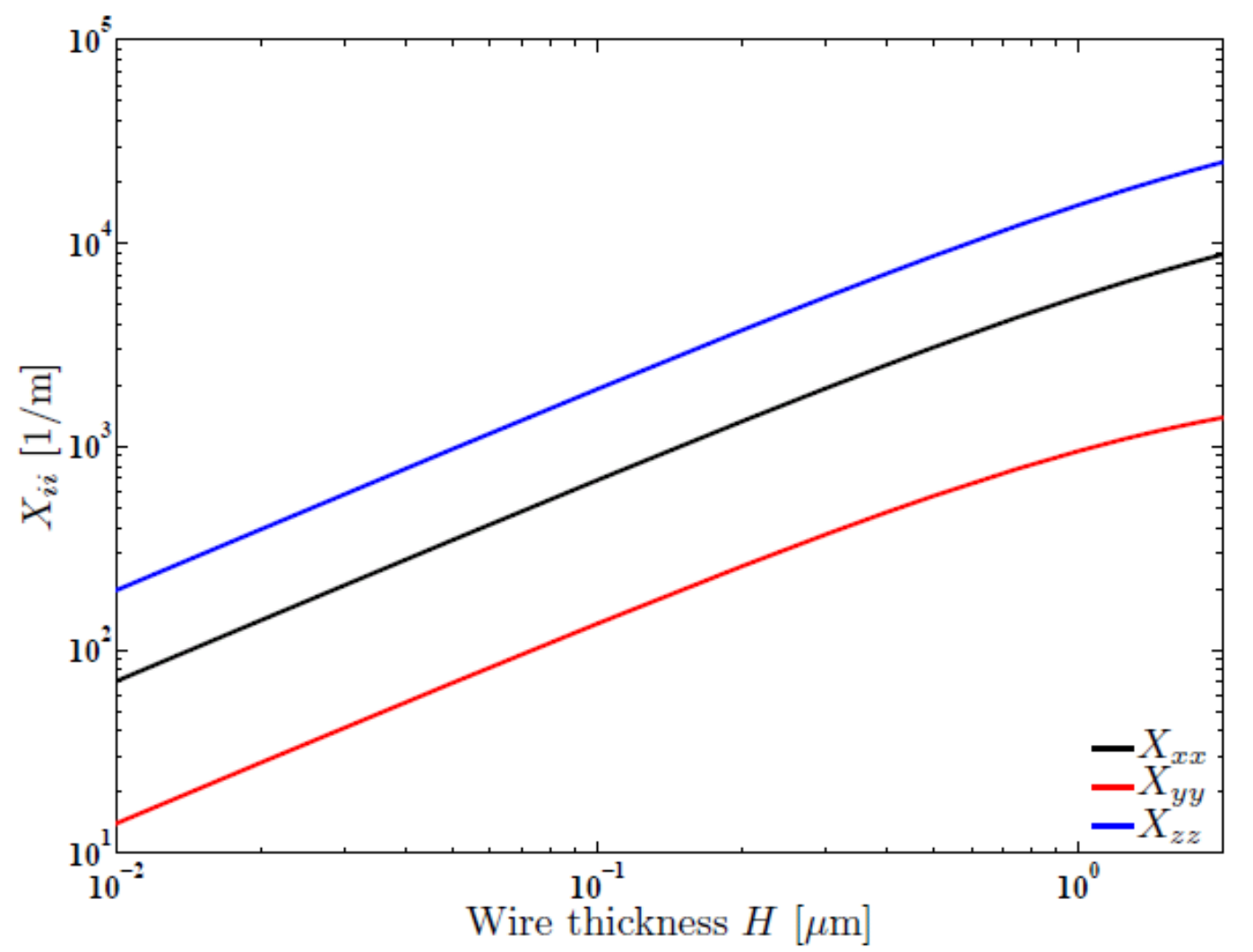}
\caption{(Color Online) Dependence of the geometrical factors $X_{ii}$ on wire width and thickness. Top: Varying the wire width $w$, while $d = 5~\mu$m, and the wire thickness is $h = 1~\mu$m. Only $X_{yy}$ is reduced. Bottom: Varying $h$, while $d = 5~\mu$m, and $w = 5~\mu$m. All three factors are reduced. It can be seen that a situation where both
$X_{yy}$ and $X_{zz}$ become much smaller than $X_{xx}$ is not obtained. This is because all three $X_{ii}$ factors have similar scaling on the wire thickness, thus reducing thickness does not lead to
the desired situation. The trends are the same also for sub-$\mu$m atom-surface distances (not shown). Taken from \cite{TalPhD}.} \label{Geometry_factors}
\end{figure}

It is interesting to note that the anisotropic material hardly helps with the spin-flip rate as for the two perpendicular components $B_{yy}$ and $B_{zz}$, which are responsible for spin-flips, we find that both mix the highly conducting $\sigma_{xx}$ and the low conductivity terms. The geometrical factors $X_{ij}$ have been analyzed in
detail (see appendix A in \cite{Valery}) for the case of
rectangular wires.  From this analysis it emerges that for
any reasonable wire geometry, one cannot obtain the necessary situation in which the $X_{yy}$ and $X_{zz}$ factors are small relative to $X_{xx}$ [see Eq. (\ref{Eq:B_ii})]. This is presented in Fig. \ref{Geometry_factors}. Thus the main difference in the noise components is due to the difference in conductivity terms, where the conductivity $\sigma_{xx}$ is dominant.  Consequently, the improvement to the trap lifetime using anisotropic materials at room temperature is expected to be at most on the order of $\sim\! 2$.

Let us now briefly review some additional alternatives to normal metals:

Carbon Nano Tubes (CNTs) are an example of such a conductor. CNTs are able to hold a current density which is 2-3 orders of magnitude higher than that for gold. Consequently, although they have a very small cross section, they may hold enough current to create stable atom traps \cite{fermani, CNT, peano}. As they are crystalline in nature, they may also reduce the static corrugation. Following from their small amount of material the CP force may also be suppressed, especially if the tubes are made to be suspended. In addition, contrary to normal metals, CNTs do not have a broad absorption spectrum and so they may be put close to optical high finesse elements without hindering their qualities. Let us note that CNTs may also be used for other purposes in the context of atom optics \cite{fortaghCNT,Hau}. See \cite{chapter} for additional details.

Another material potentially advantageous for QC is the superconductor ~\cite{emmert, roux, valery2, mukai, cano, fortagh2010, fortagh2011}. These materials are expected to give rise to orders of magnitude less static scattering and noise. The drawbacks of superconductors include the cooling itself, the sensitivity to fabrication (e.g. the transition temperature is very sensitive to contaminations), DC fields and noise created by vortices, and current inhomogeneity (most of the current is in the wires edges). As trapping atoms in the field of vortices has been achieved \cite{ZhangSC, muller}, one may even contemplate the loading of a lattice of traps above a lattice of vortices. For more details of the state-of-the-art the reader is referred to Ref. \cite{chapter}. As noted previously, the greatest challenge to material engineering is to enable long coherence times at room temperature, while maintaining a fast enough computer clock time.

Other metallic layers go beyond current-carrying structures and may include permanent magnets (which may also be made of non metallic or high resistivity materials) \cite{hinds_permanent, amsterdam, hannaford}. Recent work \cite{amsterdam}, also described in this issue, has realized a dense two dimensional array of trapped atoms and it seems reasonable to assume that controllable gates are feasible. The main advantages of permanent magnets are the lack of corrugations due to electron scattering and the suppressed Johnson noise due to the low conductivity of the material, as well as technical noise due to the lack of current. The main drawback is the demagnetization with temperature and time, and the weaker dynamic control of the fields.

\begin{figure}[b]
\includegraphics[width=\columnwidth] {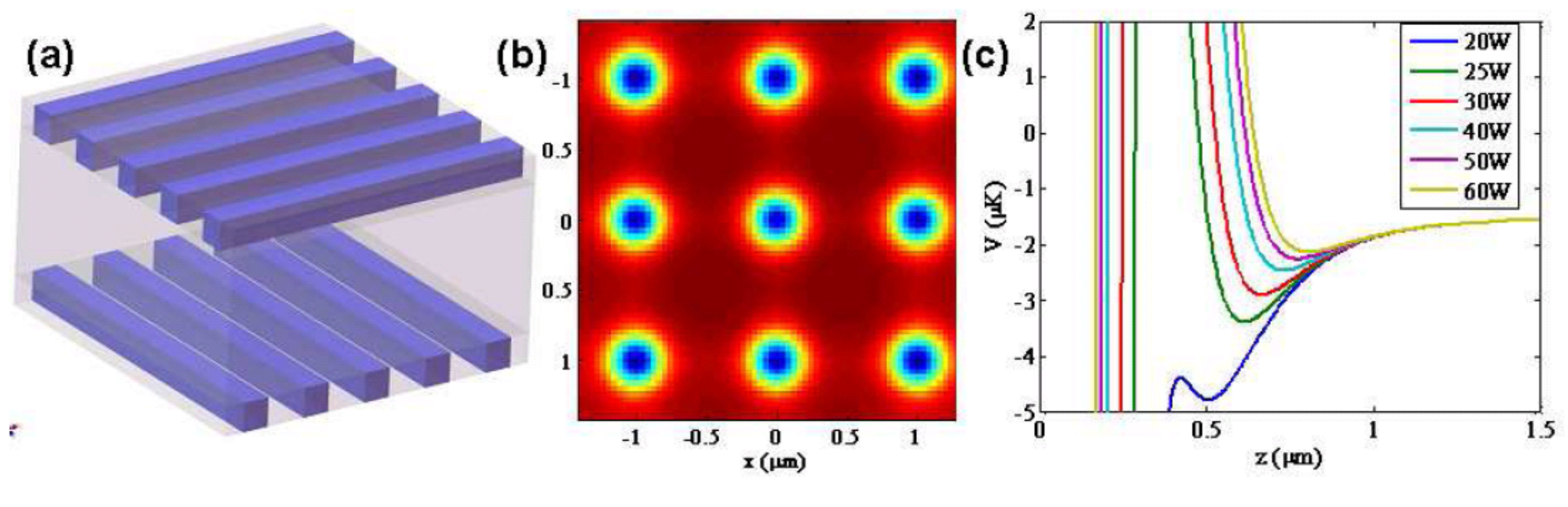}
\caption{(Color Online) The charge dots chip: (a) two layers of crossing wires will provide a 2D array of point capacitors, each layer connected to an opposite voltage. In our simulation the wires are 1 $\mu$m apart while the layers are also 1 $\mu$m apart. The wires are made of ITO and the insulating material between the wires and between the layers is $Si_3N_4$ (the similar index of refraction will suppress diffraction patterns). A prism is located below these two layers and the atoms are above. This particular simulation assumes that each capacitor is charged with one electron in the top layer and one positive hole in the bottom layer. (b-c) Trap potentials calculated with a beam waist of 100 $\mu$m, an incident angle of 65 degrees (the critical angle is 43.45), and $\lambda=778$ nm (2 nm detuning); the decay length of the evanescent wave is 72 nm. The CP force is included. (b) A 2D contour plot of the trap array ($z=644$ nm). The surface laser power is 28 W and the frequencies are about 15 kHz in all directions. The depth is $1.6~\mu$K. The spontaneous emission rate at the center of the trap is 1 Hz. The number of atoms per trap is limited to 20 (TF approximation). (c) The trap as a function of surface light intensity (blue to yellow): 20,25,30,40,50 and 60 W (the actual laser power needed will be much smaller as close to the critical angle there is an enhancement of up to an order of magnitude. A cavity configuration may also be employed). Except for the 20 W curve, the tunneling rates are negligible. Simulation made by Yonathan Japha.} \label{yoni_charged_dots}
\end{figure}

Additional metallic structures not carrying current may include charged dots where the attractive potential is balanced by magnetic fields \cite{folman2003}, a magnetic mirror \cite{joerg1998} or evanescent waves \cite{Shevchenko2004}. As an example of such a lattice, an array of traps made of localized electric fields balanced by an evanescent field is presented in Fig. \ref{yoni_charged_dots}. While this example evokes fabricated electrodes, an advanced chip may utilize self assembled charged dots, where the limiting factor would be the charge diffusion rate.

Finally, we have not touched upon the issue of technical noise, namely, the current fluctuations in the current-carrying wire. The noise spectrum of the most fundamental noise limit, i.e. shot noise, increases as one over the distance square \cite{RonRev,HenkelFundamentalLimits}. This of course implies that going close to the surface may pose a serious problem, in which case techniques aimed at achieving sub-shot noise may have to be employed i.e. creating correlations between electrons (the mesoscopic community uses terminology such as "distribution function of current"). However, long before this limit is reached, "normal" technical noise, namely noise originating from the power supply or "picked-up" by the cables that feed the atom chip \cite{ketterle_noise}, will need to be suppressed. The effect of this noise may be very different from that of the Johnson noise discussed earlier. For example, contrary to the "white noise" scenario utilized for Johnson noise, technical noise may be "colored". Such noise may, for example, cause asymmetric population transfers between the qubit states, if the potential is state dependent, as it is in the case of the collision gate \cite{shimi}. In fact, if the center frequency or the shape of the noise are time dependent, this may cause population fluctuations, i.e. random walks on the ${\it z}$-axis of the Bloch sphere (contrary to the equilibrium which is reached with white noise). To the best of our knowledge, only one measurement compared Johnson and technical noise on the same chip \cite{emmert}.

\section{Nanofabricated wires}

Let us finalize the discussion concerning current-carrying wires by analyzing the case of nanofabricated wires. Analyzing such wires is important when one wishes to answer the question of what is the smallest possible atom-surface distance. The answer to this question may reveal the fidelity limits induced by the current-carrying wire on the single qubit and 2-qubit operations. In this context, nanofabricated wires are important, for example, as noted previously, in order to avoid finite size effects which degrade the produced gradients. Smaller dimensions are also advantageous for reducing the amount of Johnson noise. In the following analysis we will also take into account the CP force.

\begin{figure}[b]
 	\includegraphics[width=\columnwidth]{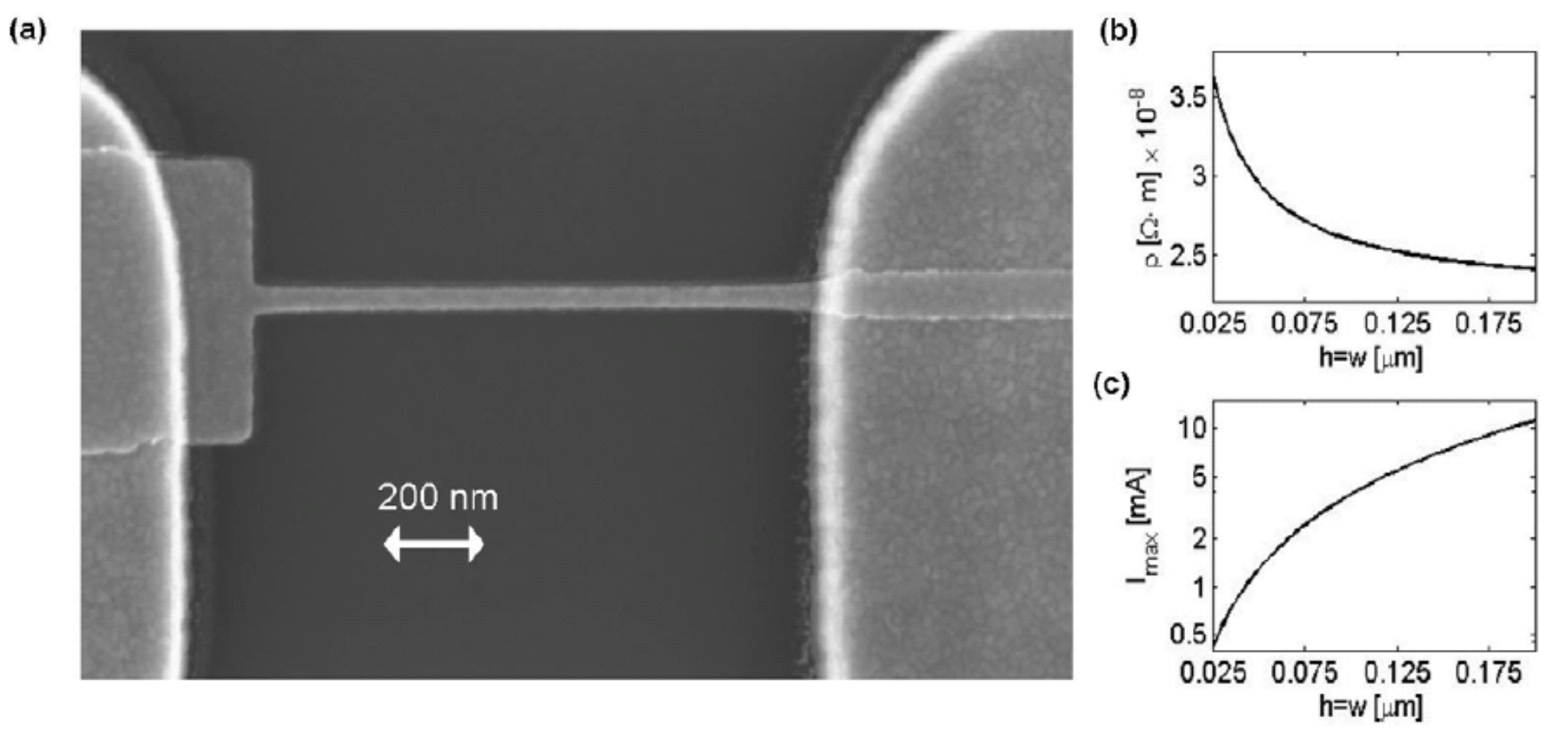}
	\caption{(a)~Scanning electron microscope~(SEM) image of a~$\rm1\,\mu$m long, $\rm20$ nm thick and~$\rm50$ nm wide gold wire. (b)~Calculated dependence of resistivity on wire dimensions (square cross sections), based on the Fuchs-Sondheimer surface scattering model. (c)~Maximum current considered safe for atomchip wire operation, calculated for different wire cross-sections, assuming the nanowire resistivity~$\rho$ shown in~(b) and the temperature coefficient~$\alpha$ for bulk gold. Taken from \cite{Salem2010} Copyright IOP Publishing Ltd. Reproduced with permission.
\label{fig:rho}}
\end{figure}

First, let us familiarize ourselves with how a nanofabricated wire looks like and behaves. A test wire fabricated at our facility and the calculated features of such a wire, are present in Fig. \ref{fig:rho}.

Next, let us take into consideration the CP force. The CP potential between a polarizable atom and dielectric or conducting objects~\cite{casimir} is one of the fundamental outcomes of zero-point vacuum fluctuations. It emerges from the fact that a dielectric or conducting object modifies the modes of the electromagnetic~(EM) field in its vicinity, modes which interact with the atomic polarization. In our case, an attractive CP potential arises from the conducting gold nanowire and from the~Si wafer coated with a~$\rm100$ nm thick SiO$_2$ layer (used to prevent electrical shorts). The CP potential reduces the potential barrier for tunneling to the surface, thereby limiting the possibility of trapping atoms near the surface.

The~EM modes of the combined surface+wire system are not analytically solvable and we will therefore carry out a separate examination of the CP potential emerging from the~Si+SiO$_2$ planar wafer, as discussed also in Ref.~\cite{CNT}, and from a simplified model that takes the wire as a perfectly conducting circular cylinder of a certain diameter. We then take the sum of the two contributions as an estimate for the combined potential as a sort of pairwise additive approximation (PAA). Based on earlier experience from the planar two-layer system, we may anticipate that this approach should at least give the right order of magnitude for the CP potential.

In general, the CP potential may be written in the form
\begin{equation}
U_{CP}(\vec{x})=i\hbar\int_{-\infty}^{\infty} d\omega\ \alpha(\omega)\left[\Gamma(\vec{x},\vec{x},\omega)
-\Gamma_0(\vec{x},\vec{x},\omega)\right],
\label{cp_general}
\end{equation}
where~$\alpha(\omega)$ is the frequency-dependent atomic polarizability and~$\Gamma({\bf r},{\bf r},\omega)$ is the trace over the Green's tensor of the electromagnetic field at the same point~${\bf r}$, with~$\Gamma_0$ being the Green's tensor in empty space, responsible for a space-independent Lamb shift. For distances from the dielectric or conducting object much larger than~$\lambda_0/2\pi$, where~$\lambda_0$ is the wavelength corresponding to the lowest optical transition frequency, the CP potential generated by a planar structure made from a layer of thickness~$t$ with a dielectric constant~$\epsilon_1$ atop an infinitely thick dielectric layer of dielectric constant~$\epsilon_2$ has the form (see Refs. \cite{Salem2010,CNT})
\begin{equation}
U_{CP}(z)=-\ \frac{\hbar c\alpha_0}{2\pi}\ \frac{1}{z^4}\ F(\epsilon_1,\epsilon_2,t/z),
\end{equation}
where~$\alpha_0$ is the static atomic polarizability. The dimensionless function~$F$ takes the single-layer limiting value~$F\sim\frac{3}{4}~\frac{\epsilon-1}{\epsilon+1}~\phi(\epsilon)$ with~$\epsilon=\epsilon_1$ when~$z\ll t$, and with~$\epsilon=\epsilon_2$ when~$z\gg t$, where~$\phi(\epsilon)$ is on the order of unity. $F=\frac{3}{4}$~is obtained in the vicinity of a perfectly conducting thick layer. In our case~$\alpha_0=47.3\times10^{-24}\,\rm{cm^3}$ is the ground state static polarizability of the~$\rm^{87}Rb$ atom, $\epsilon_1=4$ for the~SiO$_2$ layer, and~$\epsilon_2=12$ for the~Si wafer.

\begin{figure}
 	\includegraphics[width=0.45\columnwidth]{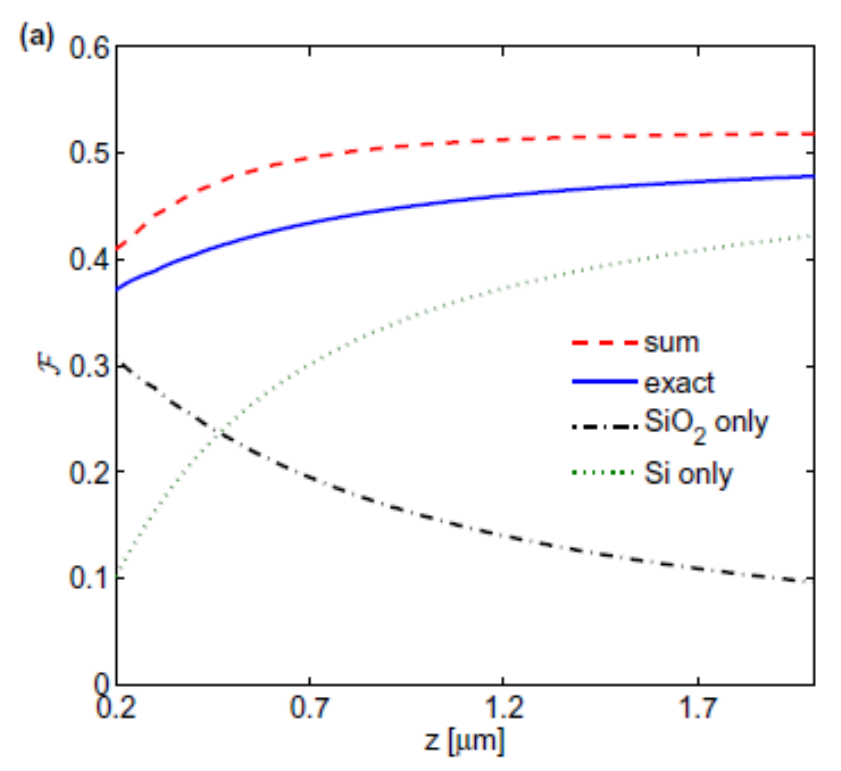}
	\includegraphics[width=0.45\columnwidth]{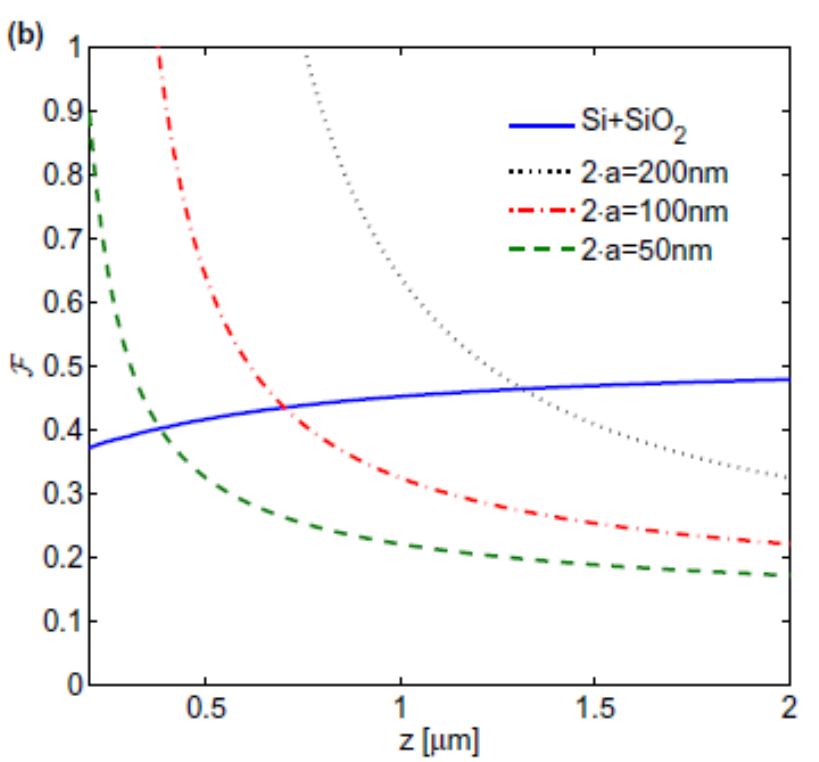}
\caption{(Color online) (a)~$\cal F$ factor ($\propto U_{CP}(z)~z^4$, see text for details) for the bilayer system of thick~Si coated by a~$\rm100$ nm layer of~SiO$_2$, similar to the system studied in~Ref.~\cite{CNT}. The exact calculation (solid curve) is compared to the sum (dashed) of two separate systems~--~the~SiO$_2$ layer alone (dashed-dotted) and the~Si layer alone (dotted). For the contribution of the~Si layer the factor~$F$ would be constant for a system of coordinates starting at its top~($z=\rm-100$ nm), but here it is rescaled to the coordinate system where~$z=0$ is at the top of the~SiO$_2$ layer (see text). The sum of the two separate contributions over-estimates the exact result by about~8-15\% over the relevant range. (b)~$\cal F$ factor (again rescaled to~$z=0$ at the top of the~SiO$_2$ layer) for the planar wafer (solid curve) reproduced from~(a) and for perfectly conducting cylindrical wires of diameters~$2a=\rm50-200$ nm (broken curves) lying on the wafer surface. Two important reasons for the differences between the wires are the different atom-wire distances~$z-2a$, which is smaller for thicker wires, and the larger solid angle subtended by the wider wire.}
\label{fig:cp}
\end{figure}

As stated above, we wish to compare contributions to the CP potential from the three different components comprising the surface: the~Si chip, the~SiO$_2$ layer of thickness~$t$, and the gold nanowire of thickness~$h$. For this comparison to be meaningful, we require a common reference for the distance variable~$z$, which we define as the distance from the top of the~SiO$_2$ surface. Then the distance from the~Si chip is~$z+\rm100$ nm and the distance from the top of the gold nanowire is~$z-h$. To factor out the strong~$z^{-4}$ dependence, we plot the quantity~${\cal F}(z)\equiv-U_{CP}(z)~\frac{2\pi}{\hbar c\alpha_0}~z^4$ in Fig.~\ref{fig:cp}(a) for the~Si+SiO$_2$ bilayer. This is compared to a sum of two models (shown separately in the figure): one where the half space for~$z<\rm-100$ nm is full of~Si while the other half is empty; and another in which only a~$\rm100$ nm thick SiO$_2$ layer exists, with empty space for~$z<\rm-100$ nm. The figure shows that simply summing the two potentials over-estimates the exact result by~8-15\% over the relevant range, but it gives the right order of magnitude.

Next let us consider the CP potential for an atom at a distance~$R$ from the center of a cylindrical conducting wire of radius~$a$ where we set~$a=h/2$. It appears that the main contribution to the integral in Eq.~(\ref{cp_general}) comes from frequencies on the order of~$\omega\sim c/R$. In our case, where~$R\rm<1\,\mu$m, the skin depth for a gold wire with resistivity~$\rho=\rm2.2\times10^{-8}\,\Omega\cdot m$ is~$\delta =\sqrt{2\rho/\mu_0\omega}\lesssim\rm10$ nm, which is much smaller than the width or thickness of the nanowires considered. We can therefore use a model where the wire is perfectly conducting (impenetrable for~EM waves), such that the~EM Green's tensor is much simpler than in the general case. The CP potential is then given by
\begin{equation}
U_{CP}(R)=-\ \frac{\hbar c\alpha_0}{2\pi}\ \frac{1}{(R-a)^4}\ F(a/R).
\end{equation}
For~$a/R>0.2$ the function~$F$ is nearly linear,~$F(a/R)\approx0.53(a/R)+0.22$, tending to~$F=\frac{3}{4}$ as~$R\rightarrow a$, where the surface of the cylinder is similar to a planar conducting surface. In the opposite limit~$a/R\ll0.1$ the function~$F$ drops to zero as~$F(a/R)\sim-\frac{2}{3\log(a/R)}$ (see appendix in \cite{Salem2010}).

Figure~\ref{fig:cp}(b) again shows the factor~$\cal F$ for the CP potential from the planar~(i.e. Si+SiO$_2$) surface in comparison to~$\cal F$ for cylindrical wires of different diameters~$2a$. It is evident that the contribution of the wire is dominant when the distance from the wire is less than~5-7 times the diameter of the wire. For larger distances the contribution of the wire falls to half or less than the contribution of the surface. Given our experience with the bilayer system~\cite{CNT}, we expect the exact calculation of the wire+surface to deviate by the same order as we observe for the bilayer, namely, less than~20\%. This degree of inaccuracy may also follow from the fact that the wires do not have circular cross-sections but square or rectangular ones. Therefore we believe that taking the sum of the two models can be expected to give at least an order of magnitude estimation of the CP potential.

\begin{figure}[ht]
	\includegraphics[width=\columnwidth]{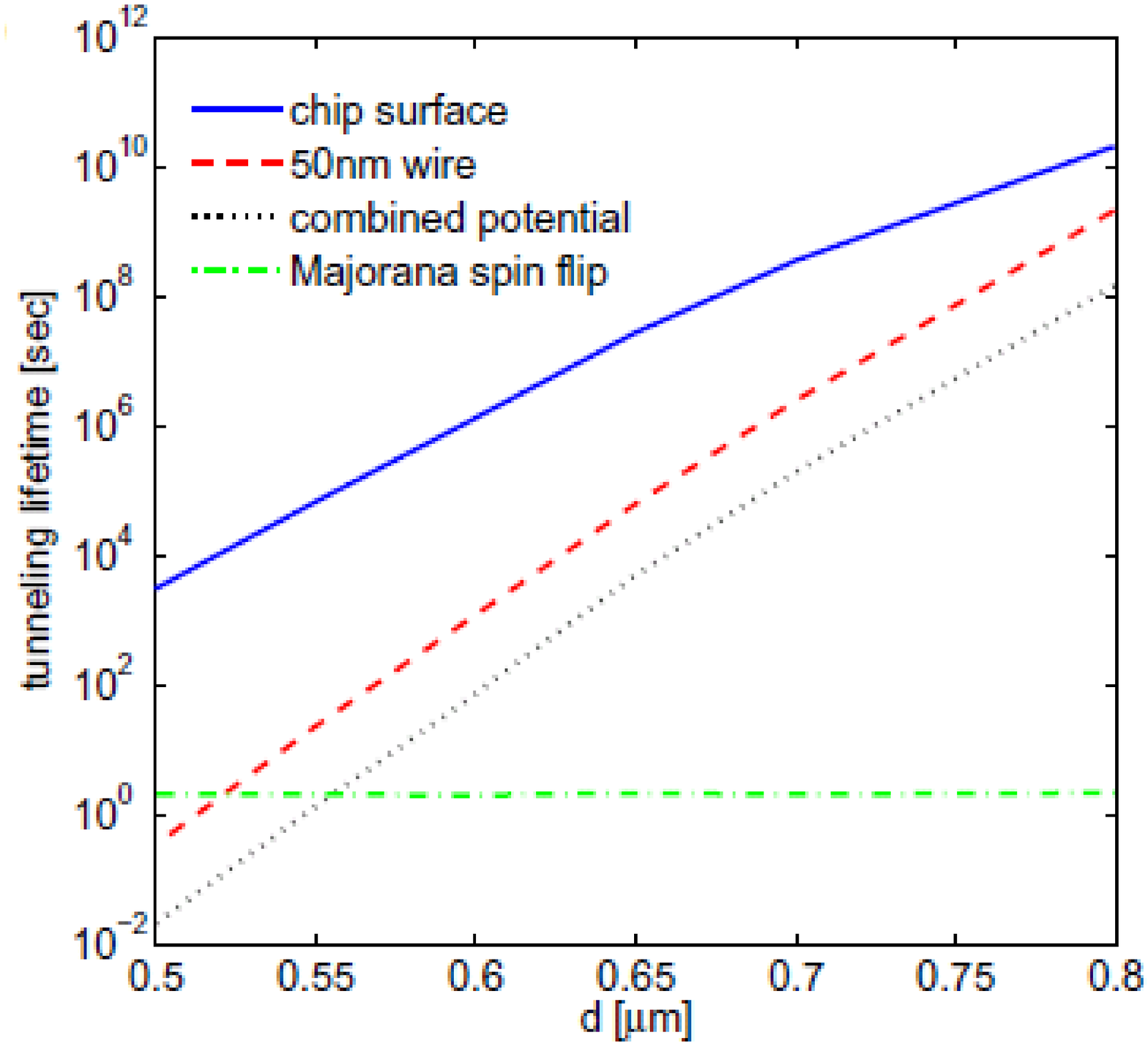}
	\caption{(Color online) Tunneling lifetimes calculated for a~BEC of~1000 atoms in traps generated at different distances~$d$, assuming a current of~$\rm40~\mu$A passing through a~$\rm50\times50$ nm trapping wire. This current is more than an order of magnitude below the maximum for such a nanowire [Fig.~\ref{fig:rho}(c)]. The solid and dashed curves are calculated assuming surface-only and wire-only contributions to the CP force respectively. Even though these CP forces are of the same order of magnitude [Fig.~\ref{fig:cp}(b)], the atomic density is much higher directly above the wire, so tunneling to the wire is much faster than tunneling to the~Si+SiO$_2$ surface; the latter tunneling proceeds mostly from the cloud edges, where the atomic density is much lower. The dotted curve is calculated for a potential combining the wire and surface CP forces; the corresponding tunneling lifetime is shorter yet because the trap barrier is reduced along the entire wire and at the cloud edges. A typical Majorana lifetime of ~$\rm2$ s is drawn for reference. Using higher currents for such wires would increase the tunneling lifetime.}
	\label{fig:lifetimea}
\end{figure}

Let us now calculate the tunneling to the surface. As a result of the CP potential, the magnetic barrier between the surface and the atoms is lowered, and atoms can tunnel through the barrier to the atomchip surface or wire. Calculated tunneling lifetimes are presented in Fig.~\ref{fig:lifetimea}, where we use a weighted average of the tunneling rate over all points in the~$(x,y)$ plane. For each point~$(x,y)$ we use the~WKB approximation for tunneling through a one-dimensional potential barrier along the~$z$ direction~\cite{CNT}:
\begin{eqnarray}\label{eq:tunnel_rate}
    \Gamma_{{\rm tunn}}=\int\int dx\ dy\ P(x,y)\ \omega_r(x,y)\nonumber \\
    \times\exp\left(-2\int_{z_{1}}^{z_{2}}dz
    \sqrt{\frac{2m}{\hbar^2}\left[U(x,y,z)-\mu\right]}
    \right),
\end{eqnarray}
where~$\mu$ is the chemical potential in the case of ensembles or the energy in the case of single atoms, and the integration over~$z$ is between the classical turning points~$z_1(x,y)$ and~$z_2(x,y)$ defined by $U(x,y,z_1)=U(x,y,z_2)=\mu$. The weighted tunneling probability appearing in the integrand is given at any point by~$P(x,y)=\frac{1}{N}\int dz\ n(x,y,z)$, where~$n(x,y,z)$ is the particle density and the transverse frequency $\omega_r(x,y)= \sqrt{\langle v_z^2\rangle}/2L(x,y)$ is the inverse of the average round-trip time for a particle moving between the turning points [$L(x,y)=z_1(x,y)-z_2(x,y)$]. In our example, these quantities are all calculated by solving the Gross-Pitaevskii equation for~1000 atoms of~$\rm^{87}Rb$. In a typical trap generated by a~Z-shaped wire, most of the tunneling occurs either at the center of the trap (where the atoms are closest to the wire) or at the trap ends (where the potential curves down towards the surface). Because of the much higher atom density directly above the wire, the lifetime is governed mostly by tunneling to the wire rather than to the surface.

\begin{figure}[ht]
	\includegraphics[width=\columnwidth]{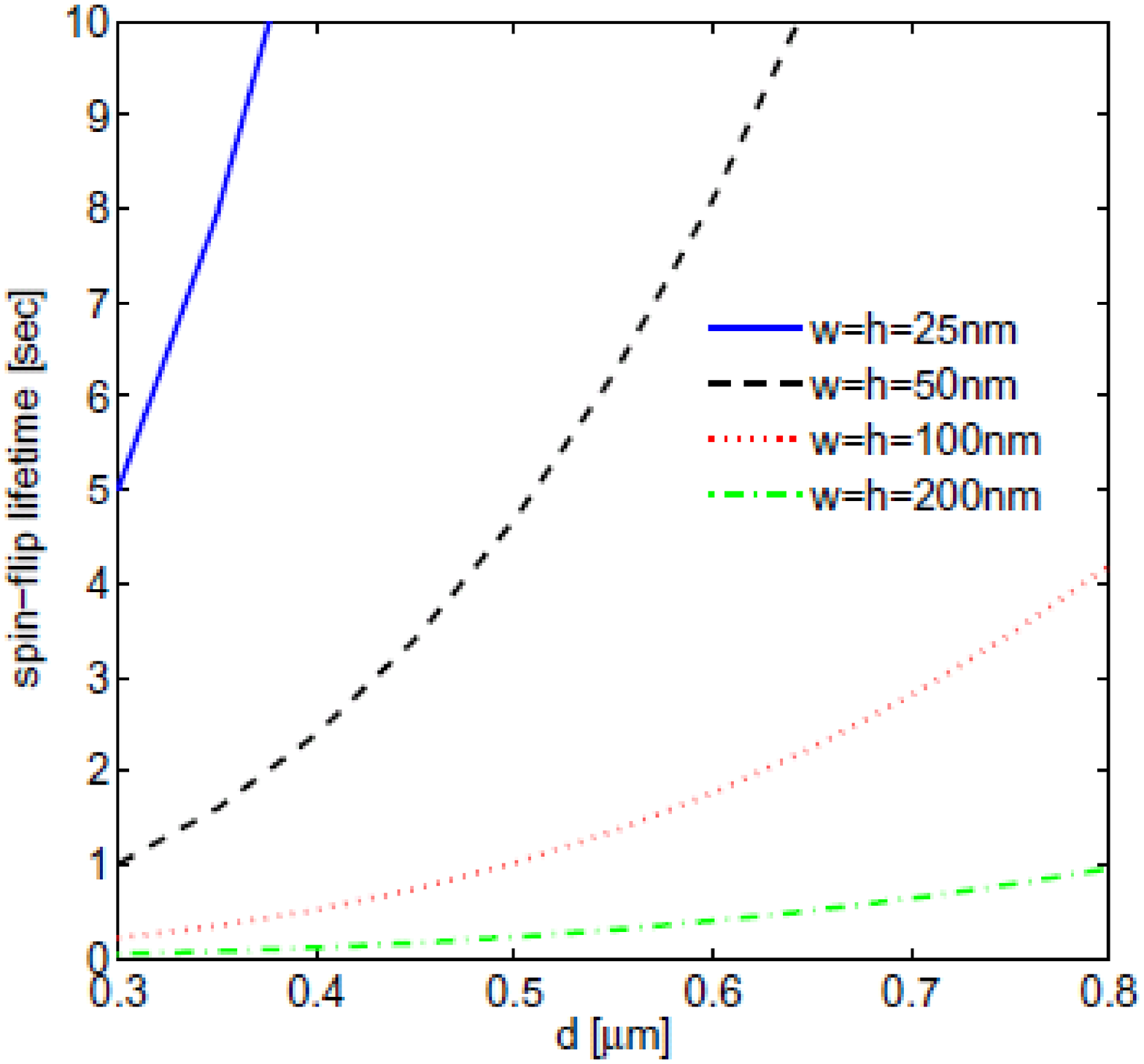}
	\caption{(Color online) Trap lifetimes due to thermal noise-induced spin-flips, calculated for atoms trapped at distances~$d$ above wires with square cross sections of~$\rm25-200$ nm. For comparison, the lifetime~$\rm1~\mu$m above a very wide wire would be $<\rm10$ ms.}
	\label{fig:lifetimeb}
\end{figure}

Next, let us also utilize the theory of Johnson noise as described previously, and present in Fig. \ref{fig:lifetimeb} the spin-flip lifetime expected in a trap formed by a nano wire. One may clearly observe the much longer spin-flip lifetimes due to the smaller cross sections.

It is quite interesting to note that narrow wires are expected to give rise to much less decoherence, not only because of the small amount of material. It has been hypothesized that narrow wires suppress thermally induced transverse currents \cite{bruder}. This is so because transverse currents charge the edges and the induced potential acts to suppress the currents. As surface induced decoherence is dominated by transverse Johnson noise currents, this may suppress decoherence by up to several orders of magnitude even at room temperature (in the same way as with anisotropic materials). For example, in Ref. \cite{bruder} a decoherence rate of $\Gamma=0.03~s^{-1}$ is calculated for a typical magnetic trap at $10~\mu$m atom-surface distance while in Fig. \ref{APB} one finds for the spin-flip rate (which is very similar to the decoherence rate) $\Gamma=3-30~s^{-1}$.

Let us also briefly mention the issue of magnetic potential corrugations. One would expect that potential corrugations produced by edge roughness would grow considerably as the edges come in close proximity to the atom. However, as shown in Fig. \ref{fig:fragmentation}, the rise is actually quite moderate (for square cross sections) as these fluctuations depend only on the variation of the center of the wire. Furthermore, the increase in potential corrugations due to edge roughness for narrower wires is somewhat balanced by the decrease of the other two sources: in Ref. \cite{YoniPRB} and in Fig. \ref{fig:fragmentation} it has been shown that narrower wires will induce less surface (or bulk inhomogeneity) induced potential corrugations. This is so due to the discreteness of the resistance wave numbers $k_x,k_y$ when the random resistance fluctuations are analyzed as waves by Fourier transforming them. For a wire with a finite length $L$ and width $W$, these wave numbers are integer products of $2\pi/L$ and $2\pi/W$. As a consequence of this discreteness, the corrugations at long wavelengths $k_xW<1$ are suppressed when the wire becomes narrower: the density of $k_y$ states becomes lower and no corresponding values of $k_y$ exist along the maximum scattering amplitude line of $k_y\sim k_x$.

Finally, let us note the possible affect of anisotropic materials, mentioned previously, in the context of nanowires. Metallic nanowires have limited maximum current densities due to the increase in their wire resistivity (Fig.~\ref{fig:rho}) from diffusive scattering at the wire surfaces. However, surface scattering for an anisotropic wire (where the surfaces are parallel to the good conductivity axis) may be significantly smaller and will have less effect on the wire resistivity, hence enabling higher current densities than discussed above. One may even speculate that at small dimensions the resistance relevant for the current density and that relevant for the Johnson noise may become decoupled.

Obviously it should be noted that nano-sized wires may have numerous limitations not presented here which need to be considered according to the specific implementation in mind. For example, introducing RF or MW currents into such wires may require more input power due to reduced impedance matching.

Let us summarize: We have discussed topics relevant to current-carrying wires including static and time dependent fluctuations and the CP force. Beyond the methods discussed above to suppress the hindering effects, one may speculate that additional methods may be found. As an outlook for future research, let us briefly give some examples of what these methods may be \cite{carsten_private}.

\begin{itemize}

\item CP force: There have been several experiments and theory papers showing that vdW and CP forces can be manipulated by choice of material and geometry, and may even change sign ~\cite{IntravaiaCP, LeonhardtCP, MundayCP, YannopapasCP, Pappakrishnan, Sambale, Milling, Lee, Tabor}.

While it is completely unclear how this could be implemented in atom chips, the wealth of work done on this topic justifies an in depth look. Could, for example, the engineering of surface plasmons or of meta-materials achieve some kind of stealth coating that would cover a broad enough range of frequencies, so as to suppress reflection by absorbtion or perfect matching of the dielectric constants?

\item Johnson noise: The issue of Johnson noise seems to have several subtleties. Some of these may be viewed in Fig. \ref{Bo}.
 \begin{figure}
\center\includegraphics [width=\columnwidth] {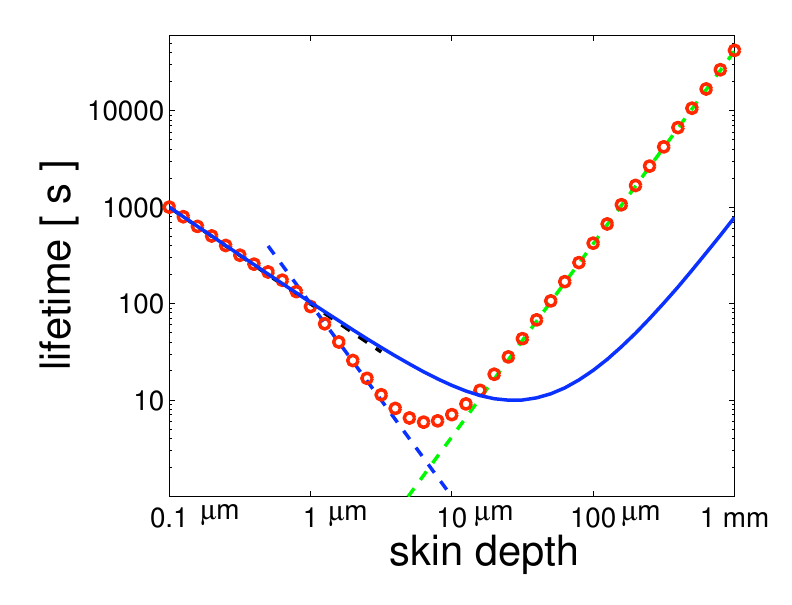}
\caption{(Color Online) Lifetime vs. skin depth. The atom-surface distance is $60~\mu$m. Solid line: bulk substrate (half space). Dashed lines and circles: 1$\mu$m thick layer. The dashed lines show approximations (Table 2.2 in \cite{Bo}). The circles are computed numerically. Taken with permission from \cite{Bo}. Similar results were obtained in \cite{scheel}. More on layer vs. half space may be found in \cite{carsten_halfspace}.} \label{Bo}
\end{figure}
At large skin depths the lifetime behaves as expected and goes up with skin depth simply because large skin depth mean small conductance and therefore a low level of noise. At medium skin depths, it is quite counterintuitive that a thin layer would enable smaller lifetimes than bulk. At small skin depths the situation is also not completely trivial. At very small skin depths, the lifetime of a thin layer behaves as a half space and goes up simply because no noise beyond a thin layer at the surface of the metal volume manages to reach the atom. The latter is hinting that if one manages to engineer a material that has a very small skin depth at the relevant frequencies for spin-flips, this could significantly suppress Johnson noise. If we cover our metal structures with such a "cut-off" low-pass filter material for which the skin depth is larger than the "filter" coating thickness for frequencies close to zero (so that DC magnetic fields or slow ramps could reach the atom), and for which the skin depth is much smaller than the "filter" coating thickness for higher frequencies (starting from say $1$ MHz, the typical Zeeman splitting in atom traps), this would indeed manage to solve the problem of Johnson noise. It remains to be seen if meta-materials may be designed with these characteristics.

In addition, what is the origin of the "kink" at a skin depth of $1~\mu$m which is also the thickness of the wire? A plausible explanation is that some effect of the bottom surface of the metallic layer comes into play. This may hint at the fact that the bottom surface, or the interface between two materials at that surface, or some interference between the top and bottom surface of the metallic wire, has significant importance. Such effects have been discussed in \cite{2surface1, 2surface2,2surface3}. Let us note that Fig. \ref{Geometry_factors} was made without including such a possible effect of the thickness. Understanding this effect may help think of ways to utilize it for a reduction of the noise.

\end{itemize}

To conclude the discussion on current-carrying wires, we may note that while conducting layers, wires and dots, offer significant potential for QC in their versatility, they also pose, as we have seen, significant challenges. It is therefore fitting to also briefly speculate upon a completely different type of atom chip which is all optical, i.e. completely made of dielectrics.

\section{Photonics}

\begin{figure}
\center\includegraphics [width=0.8\columnwidth] {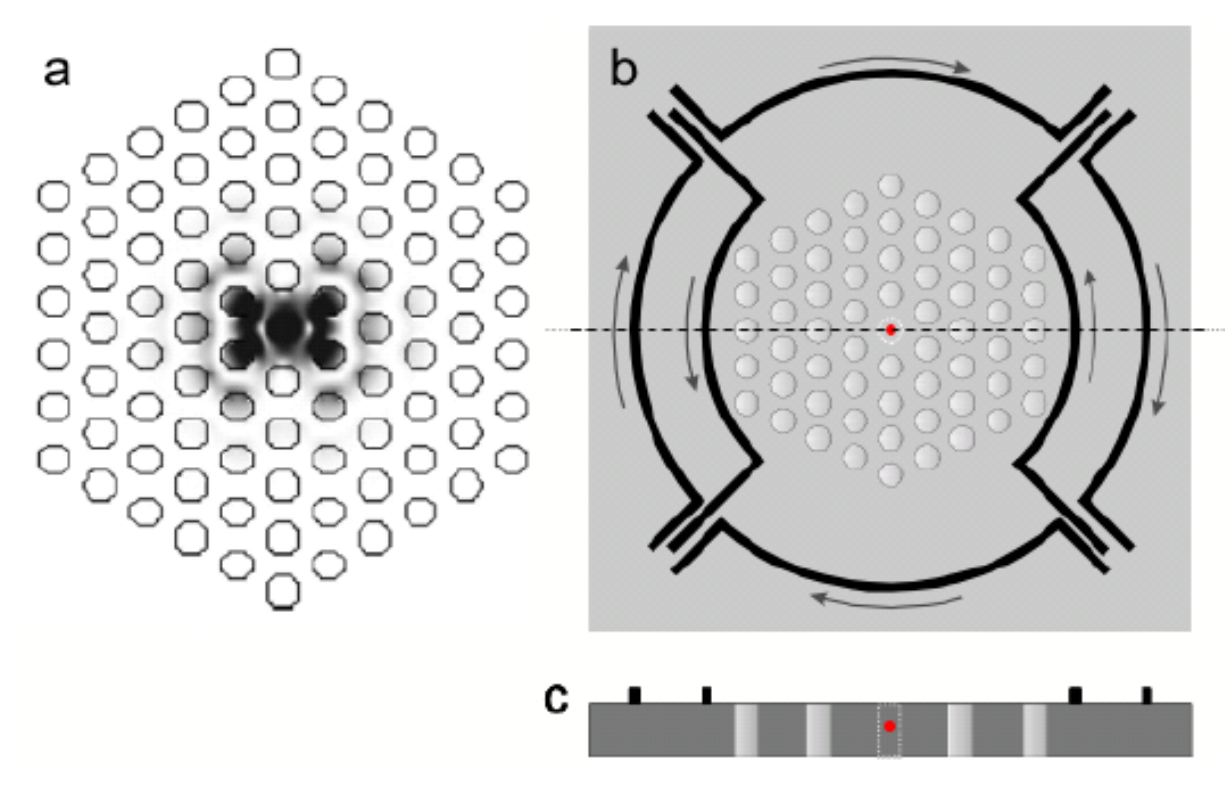}
\caption{(Color Online) An example of a combined electrical-photonic chip. a) Simulation of the cavity mode in a hexagonal lattice photonic crystal.
The cavity mode is centered and well-localized about the defect hole. b) Sketch of a
Photonic Band Gap (PBG) cavity with an integrated Ioffe microwire trap. The red dot shows the location of the trapped atom within the defect hole.
The PBG hole size is about 100 nm, and the wire diameter and width would be about 10 and 1 $\mu$m, respectively. c) Cross-section of PBG cavity and Ioffe trap through
a line intersecting the defect hole. The trapped atom depicted as a red dot in the
center of panels (b) and (c) would be located inside the cavity field maximum at
the center of the defect hole. Taken with permission from \cite{lev}.} \label{lev}
\end{figure}

Although current-carrying elements are well poised to enable further advances (at room temperature or below), one may wish to look at alternatives even further than the charged electrodes or permanent magnets, both mentioned previously. One such alternative are all optical chips. Recent advances in Cavity-QED experiments \cite{rempe2011} clearly show that the control of light and atoms in the context of high quality optical elements is advancing quickly. The technological challenge of miniaturization and monolithic fabrication, as well as integration, is not to be underestimated, but seems to have no inherent limits (beyond the fact that a minimal number of atoms is required in order to define a smooth index of refraction). Indeed, photonics is an extremely fast advancing field which will most probably be incorporated in most scenarios as a tool supplementing the main technique of a future QC chip (an example of a combined photonic-electrical chip is presented in Fig. \ref{lev}). As an in-depth analysis of the topic of photonics is beyond the scope of this paper, let us in the following briefly speculate on where this field may be headed.

Many advances have already been made. Outstanding experiments with fibers integrated on atom chips have been realized \cite{chapter,reichel_cavity}. Other experiments have trapped atoms very close to the surface of a fiber giving rise to the possibility of fiber lattices on chips \cite{fiber_lattice}. However, it is hard to imagine that fibers will enable scalability. Let us therefore focus in the following on monolithic schemes.

Monolithic schemes may deal with near and far propagating fields as well as evanescent waves. Numerous impressive works are described in \cite{chapter} and in this issue. Our own work \cite{folman_photonics} deals mainly with micro disks and toroids as high-Q devices for a strong light-atom coupling. These devices may simultaneously trap, manipulate and measure atoms and so it may indeed be feasible to construct all optical atom chips. First experiments with these devices were done in 2006 by the groups of Vahala, Kimble and Mabuchi \cite{chapter}. Other ideas for all optical chips have also been made (e.g. \cite{nakahara}). The first real integration of monolithic photonics with trapped atoms was made in the group of Ed Hinds \cite{hinds-photonics}.

\begin{figure}
\center\includegraphics [width=0.8\columnwidth] {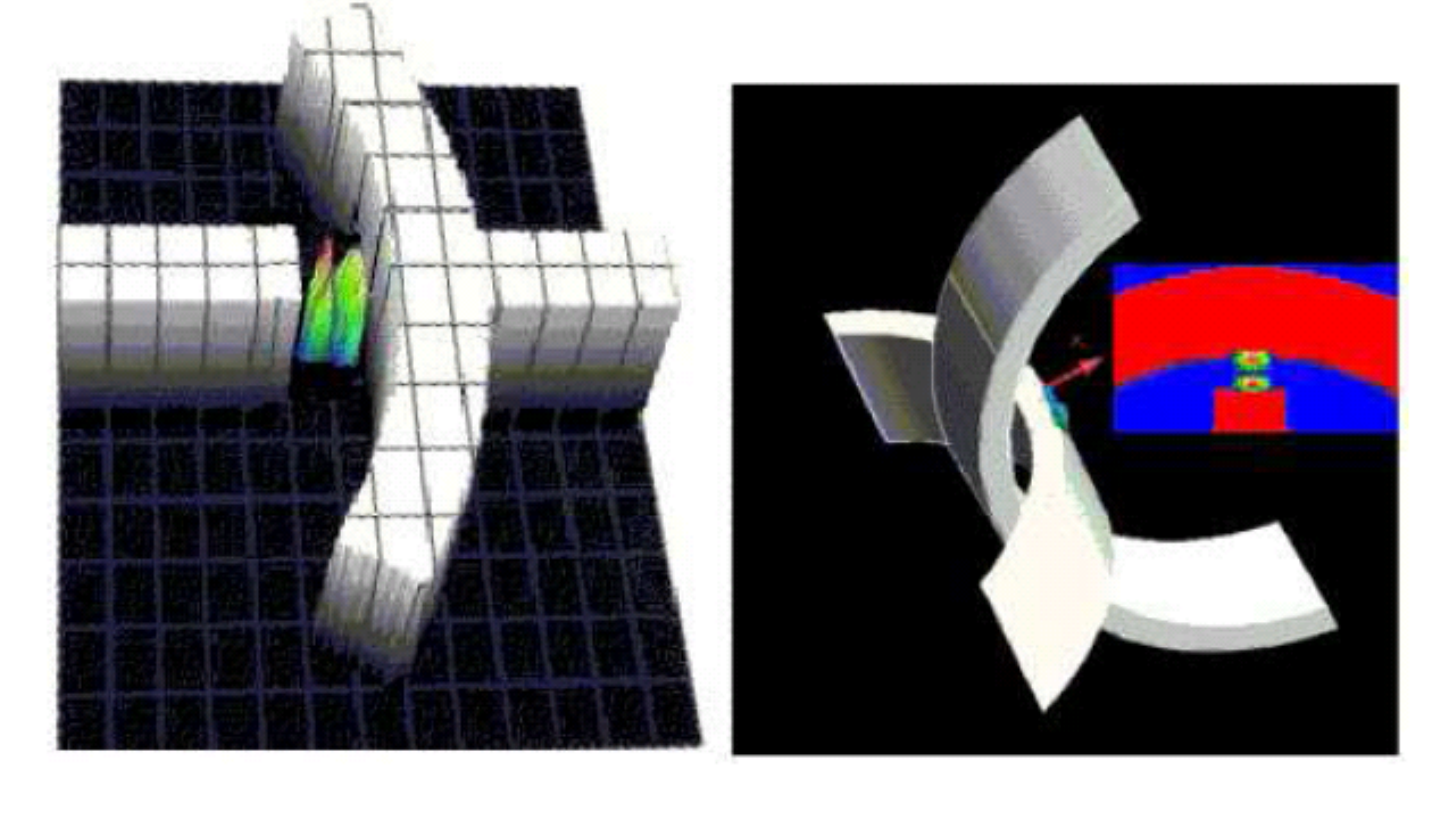}
\caption{(Color Online) An ultra low mode volume Fabry-P\'erot configuration. The stable mode is achieved with two 2D mirrors. This configuration (presented in the figure in an exaggerated manner in terms of mirror lengths) enables very close proximity between the mirrors before mirror-mirror contact is made. A clean path for atoms to enter the cavity is maintained (along the diagonal). The mode length is two optical half-wavelengths. Simulation carried out by Michael Rosenblit.} \label{Fig8}
\end{figure}

Material engineering may bring about the ability for completely novel geometries and functionalities. For example, as the ability to perform 3D fabrication advances, the presently planar chips, may enjoy 3D structures such as 3D stable mode light cavities (see Fig. \ref{Fig8}). Another futuristic element would be curved blue detuned light mirrors enabling a stable cavity mode for matter-waves \cite{wilkens}. Could this "dark" trap turn out to be the least perturbing for the qubit?

Material engineering may also bring about a photonics revolution in atom chips via sub-wavelengths optics and plasmons. The control of plasmons and plasmon-emitted/absorbed/guided light by surface material and geometry engineering is quickly advancing \cite{plasmons}. Here, very thin metallic layers or wires would control light beams with high accuracy and in small volumes. One may also think of utilizing voltage gates, as is done in directing and constricting the flow of 2D electron gases in mesoscopic physics, to dynamically control the shape of plasmons. Suggestions on how to use plasmons to trap atoms have already been made (albeit not on a flat surface) \cite{Chang2009}.

If dielectric atom chips are to be considered as a serious candidate for QC, one will of course have to return to the figure of merit discussed in the introduction, which is the ratio between the controlled and the uncontrolled coupling, or more simply put, how many logic gate operations may be done with high fidelity. Here a similar analysis to that presented above for 'electron carrying structures', would have to be done to the 'photon carrying structures'. For example, in analogy to the hindering processes described above for a current-carrying wire one would need to analyze in detail the following issues:

\begin{itemize}

\item Internal dynamic noise: In analogy to Johnson noise in current-carrying wires, changes in power, modes and polarization due to stress (e.g. due to vibrations) and temperature changes, or even frequency dependent index of refraction, are to be expected and should be analyzed. These fluctuations would create fluctuating level shifts in the qubit as well as possible excitations.

\item Internal corrugations: In analogy to magnetic potential corrugations due to static electron scattering in current-carrying wires, light scattering from surface roughness and bulk inhomogeneities will become a dominating factor. Light scattering does not only create intensity loss but changes light modes and may even reverse the light direction. For example, in \cite{folman_photonics} we have analyzed the feasibility of maintaining small light scattering.

\item Technical noise: In analogy to technical noise in current-carrying wires, laser light intensity and frequency fluctuations will have to be taken into account.

\end{itemize}

Another issue is that of tunability. If high-Q devices are required, tunability will be necessary \cite{folman_photonics}. If fast tunability is required, materials with a good Electro-Optical response need to be used. However, typically such materials have a crystalline form (e.g. $LiNbO_3$) and may not be deposited easily. More so, achieving very good surface roughness is challenging for these materials. New types of non-crystalline EO materials are being developed and should be studied in this context. Piezo materials may also be considered but again may suffer from the same problems. The state-of-the-art in tunability has been achieved in the group of Dan Stamper-Kurn and utilizes temperature control \cite{dsk}.

Most, if not all, of the above tunability implementations require some electrical connection and this requires metal structures. Here one may again be forced to deal with some of the problems discussed above concerning current-carrying wires and an effort should be made to achieve an all optical chip (e.g. by changing the index of refraction with a pump beam).

Metallic surfaces may also be needed when reflecting surfaces are required, as high reflectivity dielectric mirrors are harder to make. Such a situation may arise when some high-Q devices such as Fabry-P\'erot cavities are constructed (or even simply for a reflection Magneto-Optical Trap).
As metallic surfaces give rise to Johnson noise, one should strive to find, by way of material engineering, high reflectivity - low conductance materials. Finding such materials is not trivial especially if one requires thin layers, in which case, these materials should be suitable for deposition by some evaporation or sputter method.

To conclude, the communication revolution of the last two decades and the development of the concept of a Lab-on-a-chip in the life science field, have brought about immense capabilities in the solid state miniaturization and integration of photonic elements. This gives rise to the hope that indeed monolithic dielectric devices for QC may be formed. However, just as in the case of the metallic layers, new figures of merits will have to be formed and much higher quality standards will have to be achieved.

\section{Conclusion and outlook}

The atom chip has come to life in the beginning of the 21st century and now, about a decade later, is the topic of intensive development in dozens of laboratories and fabrication facilities around the world. Analyzing the requirements of QC, such as accuracy, integration and scalability, it is likely that any implementation of a quantum computer will require chip technology and very advanced levels of material science and engineering.

As an example of the promise of material science and engineering, the case of current-carrying wires and ground state atoms was analyzed. Specifically, I have shown that for ground state atoms, nanofabricated wires enable the following advantages at room temperature and at an atom-surface distance below one micron:

\begin{itemize}

\item coherence time of tens of seconds

\item spin-flip lifetime of tens of seconds

\item tunneling lifetime (to the surface) of tens of seconds

\item small static corrugation ($\delta B/B\sim 10^{-3}$)

\item strong immunity of magnetic lattice barriers against technical instabilities (above 10\% change of current to change tunneling probability from 0.001 to 0.1).

\end{itemize}
The above means that considering Johnson noise and spin-flips, and for mili-second gate time scales \cite{MWgate,JoergGate}, ground state atoms may be able to award us with more than the threshold of $10^4$ gate operations. It is not the purpose of this paper to promote a specific scheme for QC but rather to emphasize by way of example, the promise of material science and engineering for any type of particle or interaction used.

Extrapolating from the advances made so far with atom chips, and taking into account the new micro and nano technologies being continuously developed, one may assume that any eventual implementation of QC will indeed be possible on future atom chips. While we wait for such an optimal implementation to be realized, the coupling between quantum information processing and material engineering, promises to bring about novel insights in both quantum optics and material science.

\section{Acknowledgement}
I am sincerely thankful to Yonathan Japha and Carsten Henkel for many enlightening discussions and for their contributions over the years. I am also grateful to the members of the Ben-Gurion University atom chip group for their long standing efforts in this field. Finally, many thanks to the team of the fabrication facility for sharing my interest in building a bridge between material science and quantum optics, and for numerous chips they have so professionally fabricated for our lab and for labs around the world.

\section{Appendix A: Magnetic corrugation above a narrow wire}

Let us consider a fabricated metal wire carrying a total current~$I$. It extends along the~$\hat{x}$ direction and has a width~$w$ along~$\hat{y}$ and thickness~$h$ along~$\hat{z}$. The boundaries of the wire are located at
$y=\pm w/2+\delta y_{\pm}(x,z)$ and $z=\pm h/2+\delta z_{\pm}(x,y)$. The corrugations of the wire boundaries~$\delta y_{\pm}$ and~$\delta z_{\pm}$ can be expanded as
\begin{eqnarray}
  \delta y_{\pm}(x,z)=\sum_{n=-\infty}^{\infty}e^{2\pi i n z/h}\sum_k e^{ikx} \delta y^{\pm}_n(k)~~{\rm and}~~\nonumber \\
  \delta z_{\pm}(x,y)=\sum_{m=-\infty}^{\infty}e^{2\pi i m y/w}\sum_k e^{ikx} \delta z^{\pm}_m(k).
\end{eqnarray}
A linear theory for small corrugations predicts that the effect of each spectral component of the corrugation is responsible for a corrugation of the magnetic field near the atomic trap center with a similar wavelength~$2\pi/k$ along the~$x$ direction. However, the effect of components with wavelength much shorter than the distance~$d$ between the wire and the atomic trap (on the order of hundreds of nanometers or more) drops exponentially as~$e^{-|k|d}$ so that here we will only be interested in corrugations whose wavelengths are a few hundred nanometers or longer. We may then neglect the effect of spectral components on the order of the wire width or thickness and consider only corrugation terms with~$m=0$ and~$n=0$, namely, we may assume that~$\delta y_{\pm}$ and~$\delta z_{\pm}$ depend only on~$x$.

Corrugations of the magnetic field along the main trapping axis~$x$ above the center of a wire with geometrical perturbations are given by the Biot-Savart law as
\begin{equation}
  \delta B_x({\bf r})=\frac{\mu_0}{4\pi}\int d^3{\bf r}'\left[\delta J_y({\bf r}')
  \frac{\partial}{\partial z'}-\delta J_z({\bf r}')\frac{\partial}{\partial y'}\right]\frac{1}{
  {| \bf r- \bf r}'|}~,
  \label{eq:BS}
\end{equation}
where~$\delta J_y,~\delta J_z$ are the transverse current fluctuations in the wire. At the point exactly above the center of a nominally symmetric wire, it follows that only the symmetric components of~$\delta J_y$
[$\delta J_y(y)=\delta J_y(-y)$] and the anti-symmetric components of~$\delta J_z$
[$\delta J_z(y)=-\delta J_z(-y)$] contribute to the magnetic field. The fabrication process typically provides wires whose edge corrugations are much larger than their top or bottom surface corrugations, so that the symmetric part of~$\delta J_y$ is the major contribution to the magnetic field fluctuations.

Ohmic theory, which is adequate when the width and thickness of the wire are much larger than the electron mean free path and whose use we justify below, predicts that for wavelengths longer than the wire width or thickness the symmetric~$y$-current fluctuations in the wire have the form
\begin{equation}
  \delta J^{\rm sym}_y(x,y)=iJ^0_x \sum_{k\neq 0} k\ e^{ikx}\frac{(\delta y^+_k+\delta y^-_k)
  e^{-|k|w/2}}{1+e^{-|k|w}}\cosh(ky),
\end{equation}
such that in the limit where~$|k|w\ll 1$, $\delta J_y(x,y)\sim J_0 \partial y_c/\partial x$, where $\delta y_c=(\delta y_+ + \delta y_-)/2$ is the position of the actual center of the wire at a given point~$x$ ($y_c=y_{\rm center}$). Substituting this limit into Eq.~(\ref{eq:BS}) while assuming small deviations of the wire edges from their nominal position, and assuming that~$w\ll z$, namely, the width of the wire is much smaller than the distance of the atom to the wire, we obtain the following expression for the magnetic field corrugations above the wire
\begin{equation}
  \delta B_x(x,0,z)=\frac{iI\mu_0}{2\pi} \sum_k e^{ikx} \ k^2\ \delta y_c(k) \ {\rm K}_1(|k|z),
  \label{eq:dBx}
\end{equation}
where $I=\int dy\int dz \ J(y,z)$ is the total current in the wire and~${\rm K}_1(kz)$ is the modified Bessel function, which may be approximated by ${\rm K}_1(u)\approx (e^{-u}/u)\sqrt{1+\pi u/2}$. Our model for the fluctuation spectrum assumes that $\delta y_c(k)=\delta y_0(k_0/k)^{\alpha}e^{i\varphi}$, where~$\delta y_0$ is the edge fluctuation at some wavevector~$k_0$ and then~$\delta y_c^{rms}$ can be obtained by summing this spectrum over all~$k$. Typically,~$\alpha$ is a number between~0 (``white-noise spectrum'') and~1 (``1/f spectrum''), while~$\varphi$ is a random phase. It follows that the root-mean-square value of the field fluctuations is given by
\begin{equation}
\langle\delta B_x^2\rangle=\left(\frac{I\mu_0\delta y_0 k_0^{\alpha}}{2\pi}\right)^2
\sum_k k^{4-2\alpha} {\rm K}_1(k|z|)^2.
\end{equation}
If we assume that the distance~$|z|$ is much shorter than the length~$L$ of the measured wire,
we obtain
\begin{equation}
 \frac{\delta B_x^{rms}}{B_0}\approx
A(\alpha)\frac{\delta y_c^{rms}}{(2z)^{3/2-\alpha}},
\label{eq:fragrms}
\end{equation}
where $B_0=I\mu_0/2\pi z$ is the unperturbed magnetic field of a wire in the $y$ direction.
Here $A(\alpha)$ has units of (length)$^{1/2-\alpha}$ and is given by
\begin{equation}
A(\alpha)^2=\frac{L/\pi}{\sum_k k^{-2\alpha}}
\left[1+\frac{\pi}{4}(3-2\alpha)\right]\Gamma(3-2\alpha),
\end{equation}
where the sum is over $k$ values taking integer multiples of~$2\pi/L$ up to a cutoff~$k_{max}=2\pi/\lambda_{min}$. Typical values of this sum for~$\alpha=0$ and~$\alpha=1$ are~$\sum_k=L/\lambda_{min}$ and~$\sum_k k^{-2}=L^2/24$ respectively, when~$k_{max}\rightarrow\infty$.

The same result should be obtained if we consider diffusive surface scattering. As presented in Ref. \cite{Salem2010}, in nano-sized wires the conductivity near the boundary is reduced by diffusive surface scattering (with a typical exponential decay length $l$ from the wire edge, on the scale of the mean free path). This means that diffusive scattering is limited to a region of dimension smaller than $l$. At the same time, the corrugation wavelengths~$2\pi/k$ relevant at the atom position, e.g. similar to or larger than the atom-surface distance, induce current density directional deviations away from the edge with an exponential decay length of~$1/k$. As~$1/k \gg l$ most of the current will follow the corrugations of the boundary even in the case of surface diffusive scattering, such that the resulting $y$-current fluctuations will again generate transverse components of the current proportional to the derivative~$\partial\delta y_c/\partial x$. We thus use the ohmic theory whose general form was developed in Ref.~\cite{YoniPRB} to calculate the magnetic field corrugations above the wire.

\section{Appendix B: Johnson noise above an anisotropic material}

Turning now to magnetic field fluctuations, we use the expression for the current cross correlation
function \cite{Rytov,LifshitzRef},
\begin{eqnarray}
\lefteqn {\left\langle j_i^*\left(\vec{x}_1,\omega\right)j_j\left(\vec{x}_2,\omega '\right)\right\rangle =} \nonumber \\
      & & 4\pi \hbar \epsilon_0 \omega^2
\overline{n}\left(\omega\right)
\delta\left(\omega - \omega '\right)
\,{\rm Im}\,\epsilon_{ij}\left(\vec{x}_1; \omega \right)
\delta\left(\vec{x}_1 - \vec{x}_2\right)  \label{Eq:Lifshitz},
\end{eqnarray}
where $\overline{n}\left(\omega\right) = 1 / ({e^{\frac{\hbar\omega}{k_{\rm B} T}}-1})$
is the Bose-Einstein occupation number, and where the dielectric tensor
\begin{equation}
\epsilon_{ij}(\omega)=\frac{i \sigma_{ij}}{\epsilon_0 \omega}
\label{Eq:epsilonSigma}
\end{equation}
 is proportional to the conductivity tensor for homogeneous media.
When the anisotropic crystal axes
are aligned with the wire, the latter is diagonal,
\begin{equation}
\hat{\sigma} =
\bordermatrix{&  &  &  \cr
	 & \sigma_{xx} & 0 & 0 \cr
	 & 0 & \sigma_{yy} & 0 \cr
	 & 0 & 0 & \sigma_{zz} \cr}.\label{Eq:rhoDef}
\end{equation}

This property will be used below. We assume here that the current in the material responds locally to the electric field (Ohm's law), $j_i( {\bf x} ) = \sigma_{ij}( {\bf x} ) E_{j}( {\bf x} )$.
The vector potential and its correlation function in the quasi-static approximation are given by
\begin{eqnarray}
&\vec{A}\left(\vec{x},\omega\right) = \frac{\mu_0}{4\pi} \int {\rm d}\vec{x'}\frac{\vec{j}\left(\vec{x'},\omega\right)}{\left|\vec{x} - \vec{x'}\right|} \label{Eq:vector_potential} \\
&\left\langle A_i^*\left(\vec{x}_1,\omega\right)
A_j\left(\vec{x}_2,\omega '\right)\right\rangle
\propto \int_V {\rm d} \vec{x}'
\frac{\sigma_{ij}( \vec{x}' )
     }{ \left|\vec{x}_1 - \vec{x}'\right|\left|\vec{x}_2 - \vec{x}'\right|
     }
\label{Eq:vector_potential_correlation}.
\end{eqnarray}
Here $\vec{x}_1,\vec{x}_2$ denote the two spatial locations for which we take the correlation function, whereas $\vec{x}'$ is the integration variable, such that we sum the contribution from all points within the material volume $V$, which are at distances $\left|\vec{x}_i - \vec{x}'\right|$ ($i=1,2$) from the locations $\vec{x}_1,\vec{x}_2$. Note that this formula neglects jumps in $\vec{A}$ due to surface currents at the metal-vacuum interface. In order to calculate the correlation function of the magnetic field fluctuations, we take the curl of the vector potential correlation function, once with respect to $\vec{x}_1$ and once with respect to $\vec{x}_2$. Writing this in tensor form we get,
\begin{eqnarray}
\lefteqn{\left\langle B_i^*\left(\vec{x}_1,\omega\right) B_j\left(\vec{x}_2,\omega'\right)\right\rangle \propto}  \nonumber \\
& &\frac{1}{2}\int {\rm d}\vec{x'} \epsilon_{ilm} \epsilon_{jnp} \partial_{1,l} \partial_{2,n} \frac{\sigma_{mp}}{\left|\vec{x}_1 - \vec{x'}\right|\left|\vec{x}_2 - \vec{x'}\right|} \equiv B_{ij},\, \label{Eq:Bij1st}
\end{eqnarray}
using the Levi-Civita symbol $\epsilon_{ijk}$ and summing over repeated indices. We defined the integral holding the conductivity tensor and the geometry terms as $B_{ij}$ for convenience, and the symbol $\partial_{\alpha,l}$ ($\alpha = 1,2$) means a derivative with respect to $\vec{x_\alpha}$ in the direction of its $l$th-component. Performing the derivatives
we obtain
\begin{equation}
B_{ij} = \epsilon_{ilm}\epsilon_{jnp}\sigma_{mp}X_{ln},
\label{Eq:Bij_integral}
\end{equation}
where the geometry of the system enters through the quantity $X_{ln}$:
\begin{equation}
X_{ij} = \frac{1}{2}\int_V {\rm d} \vec{x'} \frac{\left(\vec{x}_1 - \vec{x'}\right)_i\left(\vec{x}_2 - \vec{x'}\right)_j}{\left|\vec{x}_1 - \vec{x'}\right|^3\left|\vec{x}_2 - \vec{x'}\right|^3}.
\label{Eq:Xij}
\end{equation}
We assume here a homogeneous medium (i.e., the components $\sigma_{ij}$ are
spatially constant within the material volume $V$).

Again limiting the discussion to
the 'aligned' case (\ref{Eq:rhoDef}), Eq. (\ref{Eq:Bij_integral}) is simplified, and in fact for every pair
of $i,j$ we need to sum only two integrals.  Considering the wire
geometry to be such that the atoms are located above the center of a
very long wire ($L \gg h,d$), the only non-zero elements are $B_{ii}$  ($i=x,y,z$) due
to symmetry.  Hence we obtain
\begin{eqnarray}
&B_{xx} = \sigma_{zz} X_{yy} + \sigma_{yy} X_{zz} \nonumber \\
& ~B_{yy} = \sigma_{zz} X_{xx} + \sigma_{xx} X_{zz}, \label{Eq:B_ii} \\
&B_{zz} = \sigma_{yy} X_{xx} + \sigma_{xx} X_{yy} \nonumber
\end{eqnarray}
and it is convenient to define
\begin{equation}
\tilde{Y}_{ij} \equiv B_{ij} / \sigma_{xx},
\label{Eq:Yij}
\end{equation}
assuming the good conductivity to be along $\hat{x}$.

We see that in contrast to the isotropic case, where one
has a single conductivity $\sigma_0$ for all field components $B_{ij}$,
here the conductivities $\sigma_{ii}$ give different weights to the
geometry-dependent factors $X_{ii}$.

We can now write the full expression for the power spectrum, using (\ref{Eq:spectralDensity}) and collecting all of the prefactors omitted in (\ref{Eq:vector_potential_correlation}),(\ref{Eq:Bij1st}) as
\begin{equation}
S_B^{ij}\left(\vec{x}_1,\vec{x}_2;\omega\right) = S_B^{bb}(\omega)\frac{3}{4\pi \omega/c}{\rm Im}\epsilon_{xx}\tilde{Y}_{ij},
\end{equation}
where, following \cite{HenkelPotting}, we have normalized the power spectrum to Planck's blackbody formula
\begin{equation}
S_B^{bb}(\omega)=\frac{\hbar \omega^3\bar{n}(\omega)}{3\pi \epsilon_0 c^5}.
\label{Eq:planck}
\end{equation}

As the relevant frequencies are low, the high temperature limit of the Planck function is applicable, and thus the expression for the power spectrum approaches
\begin{equation}
S_B^{ij}\left(\vec{x}_1,\vec{x}_2;\omega\right)= \frac{k_B T}{4\pi^2 \epsilon_0^2 c^4} \sigma_{xx} \tilde{Y}_{ij}.
\label{Eq:powerSpectrumFull}
\end{equation}

This expression has the same form as the result for the isotropic case \cite{HenkelPotting}, however here the tensor $\tilde{Y}_{ij}$ holds the anisotropic terms $B_{ij}$ in it.

\bibliographystyle{apsrev4-1} 

\begin{thebibliography}{10}%

\bibitem{error1} A. M. Steane, Overhead and noise threshold of fault-tolerant quantum error correction, Phys. Rev. A {\bf 68}, 042322 (2003).

\bibitem{error2} E. Knill, Scalable quantum computing in the presence of large detected-error rates, Phys. Rev. A {\bf 71}, 042322 (2005).

\bibitem{RonRev} R. Folman, P. Kr\"uger, J. Schmiedmayer, J. Denschlag and C. Henkel, Controlling cold atoms using nanofabricated surfaces: Atom Chips, Adv. At. Mol. Opt. Phys. {\bf 48}, 263 (2002).

\bibitem{ReichelRev} J. Reichel, Microchip traps and Bose–Einstein condensation, Appl. Phys. B {\bf 75}, 469 (2002).

\bibitem{fortagh} J. Fort\'agh and C. Zimmermann, Magnetic microtraps for ultracold atoms, Rev. Mod. Phys. {\bf 79}, 235 (2007).

\bibitem{baruch} C. Henkel and B. Horovitz, Noise from metallic surfaces: Effects of charge diffusion, Phys. Rev. A {\bf 78}, 042902 (2008).

\bibitem{dubessy} R. Dubessy, T. Coudreau, L. Guidoni, Electric field noise above surfaces: A model for heating-rate scaling law in ion traps, Phys. Rev. A {\bf 80}, 031402(R) (2009).

\bibitem{ion_heating} N. Daniilidis, S. Narayanan, S. A. M\"oller, R. Clark, T. E. Lee, P. J. Leek, A. Wallraff, St. Schulz, F. Schmidt-Kaler, H. H\"affner, Fabrication and heating rate study of microscopic surface electrode ion traps, New J. Phys. {\bf 13},  013032 (2011).

\bibitem{Meijer} S. A. Meek, H. Conrad, and G. Meijer, Trapping Molecules on a Chip, Science {\bf 324}, 1699 (2009).

\bibitem{Ryd_Spreeuw} A. Tauschinsky, R. M. T. Thijssen, S. Whitlock, H. B. van Linden van den Heuvell, and R. J. C. Spreeuw, Spatially resolved excitation of Rydberg atoms and surface effects on an atom chip, Phys. Rev. A {\bf 81}, 063411 (2010).

\bibitem{Ryd_Pfau} H. K\"ubler, J. P. Shaffer, T. Baluktsian, R. L\"ow, and T. Pfau, Coherent Excitation of Rydberg Atoms in Thermal Vapor Microcells, Nature Photonics {\bf 4}, 112 (2010).

\bibitem{Ryd_Scheel} J. A. Crosse, S. A. Ellingsen, K. Clements, S. Y. Buhmann, and S. Scheel, Thermal Casimir-Polder shifts in Rydberg atoms near metallic surfaces, Phys. Rev. A {\bf 82}, 010901 (2010).

\bibitem{Ryd_Martin} J. D. Carter and J. D. D. Martin,
Energy shifts of Rydberg atoms due to patch fields near metal surfaces,
Phys. Rev. A 83 (2011) 032902.

\bibitem{Ryd_Carsten} M. M. M\"uller, H. R. Haakh, T. Calarco, C. P. Koch, and C. Henkel, Prospects for fast Rydberg gates on an atom chip, paper in this issue, arXiv:1104.2739 (2011).

\bibitem{chuang} S. X. Wang, Y. Ge, J. Labaziewicz, E. Dauler, K. Berggren, Isaac L. Chuang, Superconducting microfabricated ion traps, Appl. Phys. Lett. {\bf 97}, 244102 (2010).

\bibitem{chapter} R. Folman, P. Treutlein, J. Schmiedmayer, {\it Fabrication of Atom Chips}, in "Atom Chips" (Book by Wiley-VCH), Eds. Vladan Vuletic and Jakob Reichel.

\bibitem{roman} R. Schmied, D. Leibfried, R. J. C. Spreeuw, S. Whitlock, Optimized magnetic lattices for ultracold atomic ensembles, New J. of Phys. {\bf 12}, (2010).

\bibitem{tubingen_frag} J. Fort\'agh, H. Ott, S. Kraft, A. G\"unther, C. Zimmermann, Surface effects in magnetic microtraps, Phys. Rev. A {\bf 66}, 041604 (2002).

\bibitem{Hinds_spinflip} M. P. A. Jones, C. J. Vale, D. Sahagun, B. V. Hall, E. A. Hinds, Spin Coupling between Cold Atoms and the Thermal Fluctuations of a Metal Surface, Phys. Rev. Lett. {\bf 91}, 080401 (2003).

\bibitem{orsay1} J. Est\`eve, C. Aussibal, T. Schumm, C. Fig, D. Maill, I. Bouchoul, C. I. Westbrook, A. Aspect, Role of wire imperfections in micromagnetic traps for atoms, Phys. Rev. A {\bf 70}, 043629 (2004).

\bibitem{MagScanN} S. Wildermuth, S. Hofferberth, I. Lesanovsky, E. Haller, L. M. Andersson, S. Groth, I. Bar-Joseph, P. Kr\"uger, J. Schmiedmayer, Bose-Einstein condensates:  Microscopic magnetic-field imaging, Nature {\bf 435}, 440 (2005).

\bibitem{simonSc} S. Aigner, L. Della Pietra, Y. Japha, O. Entin-Wohlman, T. David, R. Salem, R. Folman, J. Schmiedmayer, Long-Range Order in Electronic Transport Through Disordered Metal Films, Science {\bf 319}, 1226 (2008).

\bibitem{emmert} A. Emmert, A. Lupa{\c s}cu, G. Nogues, M. Brune, J.-M. Raimond, S. Haroche, Measurement of the trapping lifetime close to a cold metallic surface on a cryogenic atom-chip, Eur. Phys. J. D {\bf 51}, 173 (2009).

\bibitem{Valery} V. Dikovsky, Y. Japha, C. Henkel, R. Folman, Reduction of Magnetic Noise in Atom Chips by Material Optimization, Eur. Phys. J. D {\bf 35}, 87 (2005).

\bibitem{OrsayFragSup} J.-B. Trebbia, C. L. Garrido Alzar, R. Cornelussen, C. I. Westbrook, I. Bouchoule, Roughness Suppression via Rapid Current Modulation on an Atom Chip, Phys. Rev. Lett. {\bf 98}, 263201 (2007).

\bibitem{fermani} R. Fermani, S. Scheel, P. L. Knight, Trapping cold atoms near carbon nanotubes: Thermal spin flips and Casimir-Polder potential, Phys. Rev. A {\bf 75}, 062905 (2007).

\bibitem{YoniPRB} Y. Japha, O. Entin-Wohlman, T. David, R. Salem, S. Aigner, J. Schmiedmayer, R. Folman, Model for Organized Current Patterns in Disordered Conductors, Phys. Rev. B {\bf 77}, 201407(R) (2008).

\bibitem{TalAni} T. David, Y. Japha, V. Dikovsky, R. Salem, C. Henkel, R. Folman, Magnetic interactions of cold atoms with anisotropic conductors, Eur. Phys. J. D {\bf 48}, 321 (2008).

\bibitem{peter} G. Sinuco-Le\'on, B. Kaczmarek, P. Kr\"uger, and T. M. Fromhold, Atom chips with two-dimensional electron gases: Theory of near-surface trapping and ultracold-atom microscopy of quantum electronic systems, Phys. Rev. A {\bf 83}, 021401(R) (2011).

\bibitem{Milton} K. A. Milton, Resource Letter VWCPF-1: Van der Waals and Casimir-Polder forces, arXiv:1101.2238v2 (2011).

\bibitem{VuleticCP} Y. J. Lin, I. Teper, C. Chin, V. Vuleti\'c, Impact of the Casimir-Polder Potential and Johnson Noise on Bose-Einstein Condensate Stability Near Surfaces, Phys. Rev. Lett. {\bf 92}, 050404 (2004).

\bibitem{cornell} J. M. Obrecht and R. J. Wild and M. Antezza and L. P. Pitaevskii and S. Stringari and E. A. Cornell, Measurement of the Temperature Dependence of the Casimir-Polder Force, Phys. Rev. Lett. {\bf 98}, 063201 (2007).

\bibitem{hinds_vdw} V. Sandoghdar, C. I. Sukenik, E. A. Hinds, Direct measurement of the van der Waals interaction between an atom and its images in a micron-sized cavity, Phys. Rev. Lett. {\bf 68}, 3432 (1992); C. I. Sukenik, M. G. Boshier, D. Cho, V. Sandoghdar, E. A. Hinds, Measurement of the Casimir-Polder force, Phys. Rev. Lett. {\bf 70}, 560 (1993).

\bibitem{vigue} S. Lepoutre, V. P. A. Lonij, H. Jelassi, G. Tr\'enec, M. B\"uchner, A. D. Cronin and J. Vigu\'e, Atom interferometry measurement of the atom-surface van der Waals interaction, Eur. Phys. J. D (2011).

\bibitem{atom-probe} M. Gierling, P. Schneeweiss, G. Visanescu, P. Federsel, M. H\"affner, D. P. Kern, T. E. Judd, A. G\"unther, and J. Fort\'agh, Cold-atom scanning probe microscopy, Nature Nanotechnology {\bf 6}, 446 (2011).

\bibitem{carsten_probe} H. R. Haakh and C. Henkel,
Magnetic near fields as a probe of charge transport in spatially
dispersive conductors,	arXiv:1107.2268v1 (2011).

\bibitem{Ovchinnikov} Yu. B. Ovchinnikov, S. V. Shulga, V. I. Balykin, An atomic trap based on evanescent light waves, J. Phys. B {\bf 24}, 3173 (1991).

\bibitem{Schmiedmayer} J. Schmiedmayer, Quantum wires and quantum dots for neutral atoms, Eur. Phys. J. D {\bf 4}, 57 (1998).

\bibitem{Shevchenko} A. Shevchenko, T. Lindvall, I. Tittonen and M. Kaivola, Microscopic electro-optical atom trap on an evanescent-wave mirror, Eur. Phys. J. D {\bf 28}, 273 (2004).

\bibitem{rosenblit1} M. Rosenblit and Y. Japha and P. Horak, R. Folman, Simultaneous optical trapping and detection of atoms by microdisk resonators, Phys. Rev. A {\bf 73}, 063805 (2006).

\bibitem{Ricci} L. Ricci and D. Bassi, A. Bertoldi, Combined static potentials for confinement of neutral species, Phys. Rev. A {\bf 76}, 023428 (2007).

\bibitem{Bender} H. Bender, P. Courteille, C. Zimmermann, S. Slama, Towards surface quantum optics with Bose-Einstein condensates in evanescent waves, Appl. Phys. B {\bf 96}, 275 (2009).

\bibitem{Gillen} J. I. Gillen, W. S. Bakr, A. Peng, P. Unterwaditzer, S. F\"olling, M. Greiner, Two-dimensional quantum gas in a hybrid surface trap, Phys. Rev. A {\bf 80}, 021602(R) (2009).

\bibitem{Chang2009} D. E. Chang, J. D. Thompson, H. Park, V. Vuleti\'c, A. S. Zibrov, P. Zoller, and M. D. Lukin, Trapping and Manipulation of Isolated Atoms Using Nanoscale Plasmonic Structures, Phys. Rev. Lett. {\bf 103}, 123004 (2009).


\bibitem{ReichelIFM} W. H\"ansel, J. Reichel, P. Hommelhoff, T. W. H\"ansch, Trapped-atom interferometer in a magnetic microtrap, Phys. Rev. A {\bf 64}, 063607 (2001).

\bibitem{Horikoshi} M. Horikoshi and K. Nakagawa, Atom chip based fast production of Bose-Einstein condensate, Appl. Phys. B {\bf 82}, 363 (2006).

\bibitem{Anderson-poster} D. Z. Anderson, private communication (2009).

\bibitem{Salem2010} R. Salem, Y. Japha, J. Chab\'e, B. Hadad, M. Keil, K. A. Milton and R. Folman, Nanowire atomchip traps for sub-micron atom-surface distances, New J. Phys. {\bf 12}, 023039 (2010).

\bibitem{JoergIFM} T. Schumm, S. Hofferberth, L. M. Andersson, S. Wildermuth, S. Groth, I. Bar-Joseph, J. Schmiedmayer, P. Kr\"uger, Matter-wave interferometry in a double well on an atom chip, Nature Physics {\bf 1}, 57 (2005).

\bibitem{Ketterle_int} G. B. Jo, J.-H. Choi, C. A. Christensen, Y. R. Lee, T. A. Pasquini, W. Ketterle, D. E. Pritchard, Matter-Wave Interferometry with Phase Fluctuating Bose-Einstein Condensates, Phys. Rev. Lett. {\bf 99}, 240406 (2007).

\bibitem{MWgate} P. Treutlein, T. W. H\"ansch, and J. Reichel, A. Negretti, M. A. Cirone, and T. Calarco, Microwave potentials and optimal control for robust quantum gates on an atom chip, Phys. Rev. A {\bf 74}, 022312 (2006).

\bibitem{Treulein_MW} P. B\"ohi, M. F. Riedel, J. Hoffrogge, J. Reichel, T. W. H\"ansch, and P. Treutlein, Coherent manipulation of Bose–Einstein condensates with state-dependent microwave potentials on an atom chip, Nature Physics {\bf 5}, 592 (2009).

\bibitem{groth} S. Groth, P. Krueger, S. Wildermuth, R. Folman, T. Fernholz, D. Mahalu, I. Bar-Joseph and J. Schmiedmayer, Atom Chips: Fabrication and Thermal Properties, Appl. Phys. Lett. 85, 2980 (2004).

\bibitem{lukin-frag} D. W. Wang, M. D. Lukin, and E. Demler, Disordered Bose-Einstein Condensates in Quasi-One-Dimensional Magnetic Microtraps, Phys. Rev. Lett. {\bf 92}, 076802 (2004).

\bibitem{hinds-frag} M. P. A. Jones, C. J. Vale, D. Sahagun, B. V. Hall, C. C. Eberlein, B. E. Sauer, K. Furusawa, D. Richardson, and E. A. Hinds, Cold atoms probe the magnetic field near a wire, J. Phys B {\bf 37}, L15 (2004).

\bibitem{Whi06} S. Whitlock, B. V. Hall, T. Roach, R. Anderson, M. Volk, P. Hannaford, A. I. Sidorov, Effect of magnetization inhomogeneity on magnetic microtraps for atoms, Phys. Rev. A {\bf 75}, 043602 (2007).

\bibitem{FAB:Kruger07} P. Kr\"uger, L. M. Andersson, S. Wildermuth, S. Hofferberth, E. Haller, S. Aigner, S. Groth, I. Bar-Joseph, J. Schmiedmayer, Potential Roughness near Lithographically Fabricated Atom Chips, Phys. Rev. A {\bf 76}, 063621 (2007).

\bibitem{Krueger2004c} P. Kr{\"u}ger, L. M. Andersson, S. Wildermuth, S. Hofferberth, E. Haller, S. Aigner, S. Groth, I. Bar-Joseph, J. Schmiedmayer, Disorder Potentials near Lithographically Fabricated Atom Chips, eprint arXiv:cond-mat/0504686 (2004).

\bibitem{Krueger2004} P. Kr{\"u}ger, PhD thesis, Coherent matter waves near surfaces, University of Heidelberg (2004).

\bibitem{Carsten_thermal} C. Henkel and M. Wilkens, Heating of trapped atoms near thermal surfaces, Europhys. Lett. {\bf 47}, 414 (1999).

\bibitem{Sidles} J. A. Sidles, J. L. Garbini, W. M. Dougherty, and S. H. Chao,
The classical and quantum theory of thermal magnetic noise, with applications
   in spintronics and quantum microscopy,
Proc. IEEE {\bf 91}, 799 (2003), preprint quant-ph/0004106.

\bibitem{HenkelFundamentalLimits}
C. Henkel, P. Kr\"uger, R. Folman, and J. Schmiedmayer, Fundamental limits for coherent manipulation on atom chips, Appl. Phys. B {\bf 76}, 174 (2003).

\bibitem{HenkelPotting} C. Henkel and S. P\"otting, Coherent transport of matter waves, Appl. Phys. B {\bf 72}, 73 (2001).

\bibitem{HenkelPottingWilkens} C. Henkel, S. P\"otting, and M. Wilkens, Loss and heating of particles in small and noisy traps, Appl. Phys. B {\bf 69}, 379 (1999).

\bibitem{ScheelSpatialDecoherence} R. Fermani, S. Scheel, and P. L. Knight, Spatial decoherence near metallic surfaces, Phys. Rev. A {\bf 73}, 032902 (2006).

\bibitem{Zhang07} B. Zhang and C. Henkel, Magnetic noise around metallic microstructures, J. Appl. Phys. {\bf 102}, 084907 (2007).

\bibitem{LandauLifshitz}
L. D. Landau and E. M. Lifshitz, {\it Quantum Mechanics: Non-Relativistic Theory}, 3rd ed., Pergamon (1977).

\bibitem{Mandel}
L. Mandel and E. Wolf, {\it Optical coherence and quantum optics}, Cambridge (1995).

\bibitem{Note:TrapInhomogeneity}
It should be noted that considering the noise only at the trap center (as usually done) may not be sufficient, as by doing so one neglects the fact that the trap is commonly spatially inhomogeneous (e.g. harmonic). The atoms are distributed in the trap with a certain density profile, and move as they have finite temperature. Taking this into account introduces corrections to the theory which in some cases may be important.

\bibitem{Cornell2}
D. M. Harber, J. M. McGuirk, J. M. Obrecht, and E. A. Cornell, Thermally induced losses in ultra-cold atoms magnetically trapped near room-temperature surfaces, J. Low Temp. Phys. {\bf 133}, 229 (2003).

\bibitem{TubingenNoise}
J. Fort\'agh, H. Ott, S. Kraft, A. G\"unther, and C. Zimmermann, Surface effects in magnetic microtraps, Phys. Rev. A {\bf 66}, 041604 (2002).

\bibitem{ZhangNoise}
B. Zhang, C. Henkel, E. Haller, S. Wildermuth, S. Hofferberth, P. Kr{\"u}ger,
  and J. Schmiedmayer, Relevance of sub-surface chip layers for the lifetime of magnetically trapped atoms, Eur. Phys. J. D {\bf 35}, 97 (2005).

\bibitem{ReichelClock} P. Treutlein, P. Hommelhoff, T. Steinmetz, T. W. H\"ansch, J. Reichel, Coherence in Microchip Traps, Phys. Rev. Lett. {\bf 92}, 203005 (2004).

\bibitem{ReichelRephasing} C. Deutsch, F. Ramirez-Martinez, C. Lacro\^ute, F. Reinhard, T. Schneider, J. N. Fuchs, F. Pi\'echon, F. Lalo\"e, J. Reichel, and P. Rosenbusch, Spin Self-Rephasing and Very Long Coherence Times in a Trapped Atomic Ensemble, Phys. Rev. Lett. {\bf 105}, 020401 (2010).

\bibitem{Reichel2010} K. Maussang, G. E. Marti, T. Schneider, P. Treutlein, Y. Li, A. Sinatra, R. Long, J. Est\'eve, and J. Reichel, Enhanced and Reduced Atom Number Fluctuations in a BEC Splitter, Phys. Rev. Lett. {\bf 105}, 080403 (2010).

\bibitem{Nqubit1} N.G. van Kampen, A Soluble Model for Quantum Mechanical Dissipation, J. Stat. Phys. {\bf 78}, 299 (1995).

\bibitem{Nqubit2} G. M. Palma, K.-A. Suominen, A. K. Ekert, Quantum computers and dissipation, Proc. R. Soc. Lon. A {\bf 452}, 567 (1996).

\bibitem{Nqubit3} B. J. Dalton, Scaling of decoherence effects in quantum computers, J. mod. Optics {\bf 50}, 951 (2003).

\bibitem{Nqubit4} R. Doll, M. Wubs, P. H\"anggi and S. Kohler, Limitation of entanglement due to spatial qubit separation, Europhys. Lett. {\bf 76}, 547 (2006).

\bibitem{Malkov}  M. P. Malkov, I. B. Danilov, A. G. Zeldovich, and A. B. Fradkov. {\it Handbook on Physical and Technical Basis of Cryogenics}, Energiya, Moskwa (1973).

\bibitem{LinPrivate} Yu-Ju Lin, Private communications.

\bibitem{Rytov} S. M. Rytov, Y. A. Kravtsov, and V. I. Tatarskii, {\it Principles of Statistical Radiophysics III: Elements of Random Fields}, Springer, Berlin (1989).

\bibitem{LifshitzRef} E. M. Lifshitz, Sov. Phys. JETP {\bf 2}, 73 (1956) ; J. Exp. Theor. Phys. USSR {\bf 29}, 94 (1955).

\bibitem{TalPhD} Tal David, PhD thesis, Ben-Gurion University (2009).

\bibitem{CNT} P. G. Petrov, S. Machluf, S. Younis, R. Macaluso, T. David, B. Hadad, Y. Japha, M. Keil, E. Joselevich and R. Folman, Trapping cold atoms using surface-grown carbon nanotubes, Phys. Rev. A {\bf 79}, 043403 (2009).

\bibitem{peano} V. Peano, M. Thorwart, A. Kasper and R. Egger, Nanoscale atomic waveguides with suspended carbon nanotubes, Appl. Phys. B {\bf 81}, 1075 (2005).

\bibitem{fortaghCNT} B. Gr{\"u}ner, M. Jag, A. Stibor, G. Visanescu,M. H{\"a}ffner, D. Kern, A. G{\"u}nther and J. Fort{\'a}gh, Integrated atom detector based on field ionization near carbon nanotubes, Phys. Rev. A {\bf 80}, 063422 (2009).

\bibitem{Hau} Brian Murphy and Lene Vestergaard Hau, Electro-Optical Nanotraps for Neutral Atoms, Phys. Rev. Lett. {\bf 102}, 033003 (2009).

\bibitem{roux} C. Roux, A. Emmert, A. Lupa{\c s}cu, T. Nirrengarten, G. Nogues, M. Brune, J.-M. Raimond and S. Haroche, Bose Einstein condensation on a superconducting atom chip, Europhys. Lett. {\bf 81}, 56004 (2008).

\bibitem{valery2} V. Dikovsky, V. Sokolovsky, B. Zhang, C. Henkel and R. Folman, Superconducting atom chips: advantages and challenges, Eur. Phys. J. D {\bf 51}, 247 (2009).

\bibitem{mukai} T. Mukai, C. Hufnagel, A. Kasper, T. Meno, A. Tsukada, K. Semba and F. Shimizu, Persistent supercurrent atom chip, Phys. Rev. Lett. {\bf 98}, 260407 (2007).

\bibitem{cano} D. Cano, B. Kasch, H. Hattermann, R. Kleiner, C. Zimmermann, D. Koelle and J. Fort\'agh, Meissner Effect in Superconducting Microtraps, Phys. Rev. Lett. {\bf 101}, 183006 (2008).

\bibitem{fortagh2010} B. Kasch, H. Hattermann, D. Cano, T. E. Judd, S. Scheel, C. Zimmermann, R. Kleiner, D. Koelle, and J. Fort\'agh,  New J. Phys. {\bf 12}, 065024 (2010).

\bibitem{fortagh2011} D. Cano, H. Hattermann, B. Kasch, C. Zimmermann, R. Kleiner, D. Koelle and J. Fort\'agh, Experimental system for research on ultracold atomic gases near superconducting microstructures, Eur. Phys. J. D, DOI 10.1140/epjd/e2011-10680-8 (2011).

\bibitem{ZhangSC} B. Zhang, R. Fermani, T. M\"uller, M. J. Lim, and R. Dumke, Design of magnetic traps for neutral atoms with vortices in type-II superconducting microstructures, Phys. Rev. A {\bf 81}, 063408 (2010).

\bibitem{muller} T. M\"uller, B. Zhang, R. Fermani, K. S. Chan, Z. W. Wang, C. B. Zhang, M. J. Lim, and R. Dumke, Trapping of ultra-cold atoms with the magnetic field of vortices in a thin-film superconducting micro-structure, New J. Phys. {\bf 12}, 043016 (2010).

\bibitem{hinds_permanent} C. D. J. Sinclair, E. A. Curtis, I. Llorente Garcia, J. A. Retter, B. V. Hall, S. Eriksson, B. E. Sauer, E. A. Hind, Bose-Einstein condensation on a permanent-magnet atom chip, Phys. Rev. A {\bf 72}, 031603(R) (2005).

\bibitem{amsterdam} S. Whitlock, R. Gerritsma, T. Fernholz, R. J. C. Spreeuw, Two-dimensional array of microtraps with atomic shift register on a chip, New J. Phys. {\bf 11}, 023021 (2009).

\bibitem{hannaford} S. Ghanbari, P. B. Blakie, P. Hannaford, T. D. Kieu, Superfluid to Mott insulator quantum phase transition in a 2D permanent magnetic lattice, Eur. Phys. J. B {\bf 70}, 305 (2009).

\bibitem{folman2003} P. Krueger, X. Luo, M. W. Klein, K. Brugger, A. Haase, S. Wildermuth, S. Groth, I. Bar-Joseph, R. Folman, J. Schmiedmayer, Trapping and manipulating neutral atoms with electrostatic fields, Phys. Rev. Lett. {\bf 91}, 233201 (2003).

\bibitem{joerg1998} J. Schmiedmayer, Quantum wires and quantum dots for neutral atoms, Eur. Phys. J. D {\bf 4}, 57 (1998).

\bibitem{Shevchenko2004} A. Shevchenko T. Lindvall, I. Tittonen and M. Kaivola, Microscopic electro-optical atom trap on an evanescent-wave mirror, Eur. Phys. J. D {\bf 28}, 273 (2004).

\bibitem{ketterle_noise} A. Leanhardt, Y. Shin, A. Chikkatur, D. Kielpinski, W. Ketterle, and D. Pritchard, Bose-Einstein Condensates near a Microfabricated Surface, Phys. Rev. Lett. {\bf 90}, 100404, (2003).

\bibitem{shimi} S. Machluf, J. Coslovsky, P. G. Petrov, Y. Japha and R. Folman, Coupling between internal spin dynamics and external degrees of freedom in the presence of colored noise, Phys. Rev. Lett. {\bf 105}, 203002 (2010).

\bibitem{casimir} H. B. G. Casimir and D. Polder, The Influence of Retardation on the London-van der Waals Forces, Phys. Rev. {\bf 73}, 360 (1948).

\bibitem{bruder} C. Schroll, W. Belzig, and C. Bruder, Decoherence of cold atomic gases in magnetic microtraps, Phys. Rev. A {\bf 68}, 043618 (2003).

\bibitem{carsten_private} I am grateful to Carsten Henkel for many brainstorming encounters, whether in person or by e-mail.

\bibitem{IntravaiaCP} F. Intravaia and C. Henkel and A. Lambrecht, Role of surface plasmons in the Casimir effect, Phys. Rev. A {\bf 76}, 033820 (2007).

\bibitem{LeonhardtCP} U. Leonhardt and T. G. Philbin, Quantum levitation by laft-handed metamaterials, New J. Phys. {\bf 9}, 254 (2007).

\bibitem{MundayCP} J. N. Munday, F. Capasso and V. A. Parsegian, Measured long-range repulsive Casimir-Lifshitz forces, Nature {\bf 457}, 170 (2009).

\bibitem{YannopapasCP} V. Yannopapas and N. V. Vitanov, First-Principles Study of Casimir Repulsion in Metamaterials, Phys. Rev. Lett. {\bf 103}, 120401 (2009).

\bibitem{Pappakrishnan} V. Pappakrishnan and D. Genov, Casimir-Polder Force Reversal with Metamaterials, http://meetings.aps.org/Meeting/SES10/Event/134690.

\bibitem{Sambale} A. Sambale, S. Y. Buhmann, H. T. Dung, D. G. Welsch, Resonant Casimir-Polder forces in planar meta-materials, Phys Scripta T {\bf 135}, 014019 (2009).

\bibitem{Milling} A. Milling, P. Mulvaney, I. Larson, Direct measurement of repulsive van der Waals interactions, Journal of Colloid and Interface Science {\bf 180}, 460 (1996).

\bibitem{Lee} S. Lee, W. M. Sigmund, Repulsive van der Waals forces for silica and alumina, Journal of Colloid and Interface Science {\bf 243}, 365 (2001).

\bibitem{Tabor} R. F. Tabor, R. Manica, D. Y. C. Chan, F. Grieser, and R. R. Dagastine, Repulsive van der Waals forces in soft matter: why bubbles
do not stick to walls, Phys. Rev. Lett. {\bf 106}, 064501 (2011).

\bibitem{Bo} Bo Zhang, Ph.D. Thesis, Magnetic Fields near micro structured surfaces: application to atom chips, Potsdam (2008).

\bibitem{scheel} S. Scheel, P. K. Rekdal, P. L. Knight and E. A. Hinds, Atomic spin decoherence near conducting and superconducting films, Phys. Rev. A {\bf 72}, 042901 (2005).

\bibitem{carsten_halfspace} C. Henkel, Magnetostatic field noise near metallic surfaces, Eur. Phys. J. D {\bf 35}, 59 (2005).

\bibitem{2surface1} S. Bauer, Optical properties of a metal film and its application as an infrared absorber and as a beam splitter, Am. J. Phys. {\bf 60}, 257 (1992).

\bibitem{2surface2} S. A. Biehs, D. Reddig, and M. Holthaus, Thermal radiation and near-field energy density of thin metallic films, Eur. Phys. J. B {\ bf 55}, 237 (2007).

\bibitem{2surface3} S. A. Biehs, Thermal heat radiation, near-field energy density and near-field radiative heat transfer of coated materials, Eur. Phys. J. B {\bf 58}, 423 (2007).

\bibitem{rempe2011} H.P. Specht, C. N\"olleke, A. Reiserer, M. Uphoff, E. Figueroa, S. Ritter, G. Rempe, A Single-Atom Quantum Memory, ArXiv 1103.1528 (2011).

\bibitem{lev} Benjamin Lev, Ph.D. Thesis, Magnetic Microtraps for Cavity QED, Bose-Einstein Condensates, and Atom Optics, Caltech (2005).

\bibitem{reichel_cavity} J. Volz, R. Gehr, G. Dubois, J. Est\'eve and J. Reichel, Measurement of the internal state of a single atom without energy exchange, Nature
{\bf 475}, 210 (2011).

\bibitem{fiber_lattice} E. Vetsch, D. Reitz, G. Sague', R. Schmidt, S. T. Dawkins, and A. Rauschenbeutel, Optical Interface Created by Laser-Cooled Atoms Trapped in the Evanescent Field Surrounding an Optical Nanofiber, Phys. Rev. Lett. {\bf 104}, 203603 (2010).

\bibitem{folman_photonics}M. Rosenblit, P. Horak, S. Helsby and R. Folman, Single-atom detection using whispering gallery modes of microdisk resonators, Phys. Rev. A {\bf 70}, 053808 (2004); M. Rosenblit, Y. Japha, P. Horak and R. Folman, Simultaneous optical trapping and detection of atoms by microdisk resonators, Phys. Rev. A {\bf 73}, 063805 (2006); M. Rosenblit, P. Horak, E. Fleminger, Y. Japha and R. Folman, Design of microcavity resonators for single-atom detection, J. Nanophoton. {\bf 1}, 011670 (2007), Special issue; US Patent 7,466,889B1 (2008).

\bibitem{nakahara} E. H. Lapasar, K. Kasamatsu, Y. Kondo, M. Nakahara, and T. Ohmi, Selective Application of Two-Qubit Gate in Neutral Atom Quantum Computer, http://arxiv.org/abs/1101.4300

\bibitem{hinds-photonics} M. Kohnen, M. Succo, P. G. Petrov, R. A. Nyman, M. Trupke and E. A. Hinds, An array of integrated atom-photon junctions, Nature Photonics {\bf 5}, 35–38 (2011).

\bibitem{wilkens} M. Wilkens, E. Goldstein, B. Taylor, P. Meystre, Fabry-P\'erot interferometer for atoms, Phys. Rev. A {\bf 47}, 2366 (1993).

\bibitem{plasmons}  J. B. Pendry, Mart\'in-Moreno, F. J. Garcia-Vidal,
Mimicking Surface Plasmons with Structured Surfaces, Science {\bf 305}, 847 (2004).

\bibitem{dsk} T. P. Purdy and D. M. Stamper-Kurn, Integrating cavity quantum electrodynamics and ultracold-atom chips with on-chip dielectric mirrors and temperature stabilization, Appl. Phys. B {\bf 90}, 401 (2008).

\bibitem{JoergGate} E. Charron, M. A. Cirone, A. Negretti, J. Schmiedmayer, and T. Calarco, Theoretical analysis of a realistic atom-chip quantum gate, Phys. Rev. A {\bf 74}, 012308 (2006).



\end{thebibliography}

%

\end{document}